\documentclass[twocolumn,twocolappendix]{aastex631}

\usepackage{comment}
\usepackage{soul}
\usepackage{hyperref}

\received{June 28, 2023}
\revised{September 22, 2023}
\accepted{October 13, 2023}

\submitjournal{ApJ}

\shorttitle{SMACS J0723 Lens Model}
\shortauthors{Chow et al.}
\graphicspath{{./}{figures/}}
\usepackage{amsmath}
\usepackage{siunitx}
\usepackage{pifont}
\usepackage{longtable}
\usepackage{multirow}

\begin{document}

\title{A Lens Finder Map to check claimed High-z Galaxies behind SMACS J0723.3–7327}

\author[0000-0003-3878-9118]{Alex Chow}
\affiliation{Department of Physics, The University of Hong Kong, Pokfulam Road, Hong Kong}

\author[0000-0002-4490-7304]{Sung Kei Li}
\affiliation{Department of Physics, The University of Hong Kong, Pokfulam Road, Hong Kong}

\author[0000-0002-8785-8979]{Tom Broadhurst}
\affiliation{Department of Theoretical Physics, University of the Basque Country UPV/EHU, Bilbao, Spain}
\affiliation{Donostia International Physics Center (DIPC), 20018 Donostia, Spain}
\affiliation{IKERBASQUE, Basque Foundation for Science, Bilbao, Spain}

\author[0000-0003-4220-2404]{Jeremy Lim}
\affiliation{Department of Physics, The University of Hong Kong, Pokfulam Road, Hong Kong}

\author[0000-0002-0000-9895]{Man Cheung Alex Li}
\affiliation{Department of Physics, The University of Hong Kong, Pokfulam Road, Hong Kong}

\author[0000-0001-6985-2939]{James Nianias}
\affiliation{Department of Physics, The University of Hong Kong, Pokfulam Road, Hong Kong}

\author[0000-0002-7265-7920]{Jake Summers}
\affiliation{School of Earth and Space Exploration, Arizona State University, Tempe, AZ 85287-1404, USA}

\author[0000-0001-8156-6281]{Rogier Windhorst}
\affiliation{School of Earth and Space Exploration, Arizona State University, Tempe,
AZ 85287-1404, USA}


\begin{abstract}
The first science image released by the JWST reveals numerous galaxies in the distant background of the galaxy cluster SMACS\,J0723.3–7327.  Some have claimed redshifts of up to $z \simeq 20$, challenging standard cosmological models for structure formation.  
Here, we present a lens model for SMACS\,J0723.3–7327 anchored on five spectroscopically-confirmed systems at $1.38 \leq z \leq 2.21$ that are multiply lensed, along with twelve other systems with proposed image counterparts sharing common colours, spectral energy distributions, and morphological features, but having unknown redshifts.  Constrained only by their image positions and, where available, redshifts, our lens model correctly reproduces the positions and correctly predicts the morphologies and relative brightnesses of all these image counterparts, as well as providing geometrically-determined redshifts spanning $1.4 \lesssim z \lesssim 6.7$ for the twelve candidate multiply-lensed galaxies lacking spectroscopic measurements.  From this lens model, 
we create a lens finder map that defines regions over which galaxies beyond a certain redshift are predicted to be multiply lensed.  Applying this map to three galaxies claimed to be at $10 \lesssim z \lesssim 20$, we find no image counterparts at locations (with an uncertainty of $\sim$$0\farcs5$) where they ought to be sufficiently magnified to be detectable -- suggesting instead that these galaxies lie at $z \lesssim 1.7$--3.2.  In lieu of spectroscopy, the creation of reliable lens finder maps for cluster fields are urgently needed to test and constrain redshifts inferred from photometry for a rapidly increasing number of candidate high-$z$ galaxies found with the JWST.
\end{abstract}



\section{Introduction} \label{sec:intro}
Gravitational lensing by galaxy clusters has enabled the discovery of galaxies at high redshifts that would otherwise have been too dim to be detectable or have escaped notice -- thus allowing us to probe fainter galaxy populations in the early universe or to discover galaxies at higher redshifts than would otherwise have been possible.  Measurements of the galaxy luminosity function at, in particular, $z \gtrsim 10$ will provide stringent tests of cosmological models for structure formation, along with the very nature of Dark Matter (DM).  For example, models incorporating cold yet ultra-light particles (e.g., axions) for DM predict a suppression of low-mass DM halos owing to the quantum interference of the constituent particles that act on macroscopic scales as waves \citep[][]{schive2014,Harko2019,Hui2021}. In this model, galaxies are expected to arise later in the universe than for models incorporating cold and ultra-massive particles (e.g., WIMPs) for DM, in which DM halos with masses down to the Jeans limit are predicted.  

Prior to the era of the James Webb Space Telescope (JWST), the most distant galaxy discovered through gravitational lensing was MACS\,0647-JD \citep{Coe2013,Chan2017}, which is multiply lensed by the galaxy cluster MACS\,J0647.7+7015 as seen in images taken with the Hubble Space Telescope (HST) under the The Cluster Lensing and Supernova Survey with Hubble (CLASH) program \citep{Postman2012} (Program ID: 12459, PI: M. Postman). Based on synthetic stellar populations fitted to its spectral energy distribution (SED), MACS\,0647-JD was inferred to most likely have a redshift (thus determined photometrically) of $z_{\rm phot} \simeq {10.7}_{-0.4}^{+0.6}$ \citep{Coe2013}.  This photometrically-determined redshift was subsequently supported by measurements for its geometric redshift of $z_{\rm geo} \simeq {10.8}_{-0.4}^{+0.3}$ \citep{Chan2017}.  The latter measurement relies only on the geometry of gravitational lensing, whereby image counterparts of multiply-lensed systems at higher redshifts form configurations have larger angular separations in the sky.  In the work of \citet{Chan2017}, the geometric redshift of MACS\,0647-JD was determined based on a lens model for the foreground lensing galaxy cluster constructed using the positions and photometrically-determined redshifts of nine multiply-lensed systems (at redshifts significantly lower than that of MACS\,0647-JD) as constraints, as none of the multiply-lensed systems had at the time redshifts measured spectroscopically.  Both the photometrically- and geometrically-determined redshifts of MACS\,0647-JD have since been confirmed through spectroscopy with the JWST, placing this galaxy at $z_{\rm spec} = 10.17$ \citep{Hsiao2023}.

Near-infrared images of the galaxy cluster SMACS\,J0723.3-7327 released by the JWST on 2022 July 11, the first science images made publicly available from the telescope, has stimulated a number of searches for galaxies in the early universe lying behind the cluster.  This cluster was originally discovered by \citet{Repp2018} in a visual inspection of optical images from the Digital Sky Survey (DSS), conducted as the southern extension of the Massive Cluster Survey \citep{Ebeling2001},  towards bright X-ray sources contained in the ROSAT All-Sky Survey (RASS) Bright Source Catalogue \citep{Voges1999}.  It was one of the clusters subsequently imaged by the HST in the Reionization Lensing Cluster Survey (RELICs) program \citep{Coe2019}.  Based on the JWST images for this cluster, four papers have been published reporting galaxy candidates at $z_{\rm phot} \gtrsim 10$, all based on the putative detection of a Lyman break in the SEDs of these galaxies.  The first paper to appear in the published literature (Advance Access publication in 2022 Nov) reported four candidates at $10 \lesssim z_{\rm phot} \lesssim 11.5$ \citep{Adams2023}.  This paper was quickly followed by two others (in 2023 Jan), one reporting two candidates at $10 \lesssim z_{\rm phot} \lesssim 13$ \citep[along with eighteen others at $8 \lesssim z_{\rm phot} \lesssim 10$, all supplemented by mid-infrared images taken also with the JWST;][]{Rodighiero2023}, and the other reporting an astonishing eighty-seven candidates at $11 \lesssim z_{\rm phot} \lesssim 20$ \citep{Yan2023}.  Another paper appeared soon thereafter (in 2023 Feb) reporting fifteen candidates at $10 \lesssim z_{\rm phot} \lesssim 15$ \citep{Atek2023}.  Pushing back on some of these claims, from a recalibration of the JWST images and revised model fits to the SEDs of galaxy candidates at $z_{\rm phot} \gtrsim 10$, \citet{Adams2023} argued that the putative Lyman breaks reported by \citet{Yan2023} and \citet{Atek2023} may be better explained by a drop in brightness towards shorter wavelengths owing to dust extinction.  This explanation would place the galaxies at much lower redshifts than their claimed redshifts of $z_{\rm phot} \gtrsim 10$.

Here, we test the putative redshifts of selected galaxies at $z_{\rm phot} \gtrsim 10$ behind SMACS\,J0723.3-7327 using an entirely different approach: by determining whether these galaxies, at their claimed redshifts, ought to be multiply lensed by the foreground galaxy cluster.  By searching for their predicted lensed counterparts where sufficiently magnified to be detectable, we can either affirm these galaxies to lie at their claimed redshifts, or place upper limits on their redshifts purely through the geometry of gravitational lensing.  The veracity of this approach depends on the reliability of a lens model for SMACS\,J0723.3-7327.  In our work, we constructed a lens model for the cluster based on five multiply-lensed systems having redshifts of $1.38 \leq z_{\rm spec} \leq 2.21$, supplemented by twelve other multiply-lensed systems identified by the common colours (for which we confirm to also have common SEDs) and shared morphologies of their proposed image counterparts but of unknown redshifts.  Constrained only by the positions and, where available, redshifts of these image counterparts, we test the reliability of the lens model thus constructed by assessing its ability to reproduce the positions (a test of internal consistency) as well as both the relative brightnesses and lensed morphologies (tests of predictability) of all the image counterparts.  From this lens model, we determine precise geometric redshifts spanning $1.4 \lesssim z_{\rm geo} \lesssim 6.7$ for the twelve multiply-lensed systems with no spectroscopic redshift determinations.  We also create a lens finder map to assess the veracity of claimed high-$z$ galaxies discovered behind SMACS\,J0723.3-7327 -- and, where refuted to be at their claimed redshifts, to place upper limits on the actual redshifts of these galaxies.

The remainder of this paper is oganized as follows.  Section \ref{sec:obs} provides a succinct description of the data on which our lens model for SMACS\,J0723.3-7327 is based.  Readers interested only in how we constructed the lens model can skip ahead to Section \ref{sec:methodology}.  In Section \ref{sec:results}, we present the parameters of our lens model for SMACS\,J0723.3-7327, as well as the various tests conducted to assess the internal consistency and predictability of the lens model.  We also provide precise geometric redshifts for the twelve proposed multiply-lensed systems used (in concert with the five others having spectroscopically-determined redshifts) to constrain the lens model despite not having known redshifts.  Based on this lens model, we present in Section \ref{sec:discussion} a lens finder map specifying regions over which galaxies beyond a certain redshift ought to be multiply lensed.  We then demonstrate the utility of this map for assessing the claimed redshifts of purported high-$z$ galaxies, and for placing upper limits on their actual redshifts when their predicted image counterparts are not seen.  Finally, in Section \ref{sec:summary}, we provide a concise summary of the main points of this paper.  We adopt throughout a concordance $\Lambda$CDM cosmology with $\Omega_M=0.3$, $\Omega_\Lambda=0.7$, and $H_0=70\:\mathrm{km\:s^{-1}\:Mpc^{-1}}$; values quoted from other publications have been converted to this cosmology where necessary.





\section{Data} \label{sec:obs}

\subsection{JWST}\label{subsec:JWST}
SMACS\,J0723.3–7327 was observed in six filters with the Near-Infrared Camera (NIRCam) on the JWST. We found the background of the level 3 products from the MAST archive to be not subtracted satisfactorily. We therefore recalibrated the images from the corresponding lower level products using version 1.7.2 of the JWST calibration pipeline developed by \citet{Bushouse2023} (hereafter JWST pipeline), along with the \texttt{jwst\_0995.pmap} to calibrate the absolute fluxes.  In this procedure, the images were first processed through the Stage 1 (\texttt{detector1pipeline}) of the JWST pipeline, along with flagging and subtraction of snowballs with the algorithm developed by Chris Willot\footnote{\url{https://github.com/chriswillott/jwst}}.  Next, we subtracted wisps present in the F150W and F200W images using the wisp templates from the NRCA3, NRCA4, NRCB3, and NRCB4 detectors.  We then processed the wisp-subtracted images with the default setting of the Stage 2 JWST pipeline (\texttt{calwebbimage2}), before proceeding to the $1/f$ noise correction and sky subtraction using the package \texttt{Profound} \citep{Robotham2018}.  Finally, we complete the image calibration by drizzling the images with a pixel size $0.^{\prime\prime}06/\rm pix$ and aligning the World Coordinate System (WCS) using the JWST Stage 3 pipeline (\texttt{calwebbimage3}).


\subsection{HST}
SMACS\,J0723.3–7327 was one of the targets of the Reionization Lensing Cluster Survey \citep[RELICs;][]{Coe2019,Salmon2020}, in which fourty-one galaxy clusters were observed with the Hubble Space Telescope (HST).  SMACS\,J0723.3–7327 was imaged in seven broadband filters in this program, spanning the optical (F435W being the shortest-wavelength filter) to the near-infrared (F160W being longest-wavelength filter).  For our purposes, these images provide a means for identifying cluster members for inclusion in our lens model.  To make such identifications possible, we retrieved science-ready images of SMACS\,J0723.3–7327, along with photometric and redshift catalogs as published by \citet{Coe2019}, from the RELICs archive\footnote{\url{https://archive.stsci.edu/hlsp/relics}}.   Photometry of objects was conducted using \texttt{SExtractor}.  Redshifts were determined by fitting synthetic stellar populations to the SEDs of the detected objects where deemed sufficiently bright, providing a probability distribution in $z$ using a Bayesian method as implemented in the algorithm \texttt{BPZ} \citep{BPZ2000,BPZ2004,BPZ2006}.

\subsection{MUSE} \label{sec:MUSE}
Prior to the public release of the images taken by the JWST, spectroscopic observations of SMACS\,J0723.3-7327 had been made with the Multi Unit Spectroscopic Explorer (MUSE) on the Very Large Telescope (VLT) (Program ID: 0102.A-0718(A), PI: A. Edge).  We downloaded the reduced data from the ESO archive\footnote{\url{https://archive.eso.org/downloadportal/b98dd1c1-3bc7-4f41-894d-be15de5dccfc}} to aid in the identification of cluster members as described in Section\,\ref{sec:clusmem}. This data had previously been used by \citet{Golubchik2022}, \citet{Caminha2022}, and \citet{Mahler2023} to identify multiply-lensed systems behind SMACS\,J0723.3–7327 and to determine their redshifts. 

The observation of SMACS\,J0723.3-7327 by the MUSE-VLT covered a field of $1^{\prime} \times 1^{\prime}$, compared with a field of $2\farcm35 \times 2\farcm35$ imaged by the JWST of this cluster.  The wavelength coverage of the MUSE-VLT spectroscopy spanned 4750.31--$9350.31\,\mathrm{\AA}$ at a spectral resolutions $\sim$$26.5\,\mathrm{\AA}$, corresponding to velocity resolutions ranging from $176.3\:\rm km\:s^{-1}$ at 4750\,\AA\ downwards to $84.5\:\rm km\:s^{-1}$ at 9350\,\AA.  The measured spectra of each object contains numerous sky lines, which we removed to the extent possible using the skyline cleaning program \texttt{ZAP} \citep{Soto2016}.  The point-spread function (PSF) of the telescope was measured by fitting a Moffat function to a star at $\alpha = 07^{\rm h} \, 23^{\rm m} \, 13.06063\fs$ and $\delta = -73^{\circ} \, 27^{\prime} \, 05\farcs908$, for which we obtained a full-width half-maximum (FWHM) of $0\farcs91$.

\section{Lens Model Construction}\label{sec:methodology}
We used the parametric lens modelling algorithm \texttt{glafic} \citep{oguri2010,Oguri2021} to construct a lens model for SMACS\,J0723.3–7327.  This model was constrained using five multiply-lensed systems having redshifts determined spectroscopically ($z_{\rm spec}$), along with twelve other multiply-lensed systems mostly identified by others based on the common colours and anticipated locations of candidate image counterparts; and for which we checked their proposed image counterparts to have common SEDs (Section \ref{constraints}).  The ingredients of our lens models are cluster members identified by their colours or, where available, $z_{\rm spec}$, along with a cluster-scale Dark Matter halo (Section \ref{ingredients}).

\texttt{glafic} solves for a best-fit lens model by minimising the chi-square between the observed properties of the multiply-lensed lensed systems used as constraints -- in the case here, their positions -- and the same properties predicted by the lens model for these systems.  The minimisation can be carried out in either the image plane (i.e., at the redshift of the lensing object, in this case SMACS\,J0723.3–7327 at $z = 0.39$) or the source plane (i.e., at the redshift, either $z_{\rm spec}$ or $z_{\rm geo}$, of a given multiply-lensed system).  When minimising in the image plane, the best-fit source positions along with the best-fit lens model are computed together so as to most closely reproduce the measured positions of the image counterparts used as constraints.  When minimising in the source plane, the best-fit source positions along with the best-fit lens model are computed together so as to most closely reproduce the de-lensed source positions for a given set of image counterparts used as constraints.
When constructing lens models involving large fields of view (as is necessarily the case for galaxy clusters) and constrained by relatively large numbers of multiply-lensed systems, minimising in the image plane requires very long computational times so as to make this approach impractical; we therefore carried out the minimisation in the source plane.  In this case, \texttt{glafic} minimises:
\begin{equation}
    \chi^2 = \sum_{i}\frac{(\vec{\beta}_{i,\mathrm{obs}}-\vec{\beta}_{\mathrm{pred}})^{\mathrm{T}}\mathrm{M}_{i}^{2}(\vec{\beta}_{i,\mathrm{obs}}-\vec{\beta}_{\mathrm{pred}})}{\sigma_{\theta_i}^2},
    \label{eq:chisq_src}
\end{equation}
where $\vec{\beta}_{i,\mathrm{obs}}$ is the de-lensed source position of the $i^{\rm th}$ lensed image, $\vec{\beta}_{\mathrm{pred}}$ is the model-predicted source position, $\sigma_{\theta_i}$ is the measurement uncertainty of the image position in the image plane, and $\rm{M}_{i}$ is the magnification tensor.

When solving for the lens model, we allowed \texttt{glafic} to treat the geometric redshift, $z_{\mathrm{geo}}$, of the systems without $z_{\rm spec}$ as a free parameter (otherwise, $z_{\mathrm{geo}} \equiv z_{\rm spec}$), thus allowing us to derive their geometric redshifts.  As mentioned earlier, solving for $z_{\mathrm{geo}}$ relies purely on the geometry of gravitational lensing, whereby multiply-lensed images of objects at higher redshifts form configurations having larger angular separations in the sky.  To determine the tolerances of the parameters characterising our lens model, we conducted a Markov-Chain Monte-Carlo (MCMC) simulation (Section \ref{sec:MCMC}) as implemented in \texttt{glafic}.  Tolerances in our lens model translate to systematic uncertainties in the predicted image positions, image brightnesses, and individual $z_{\mathrm{geo}}$ of the multiply-lensed systems used as constraints.  

\subsection{Model Constraints}\label{constraints}
\begin{figure*}
    \centering
    \includegraphics[width=\textwidth]{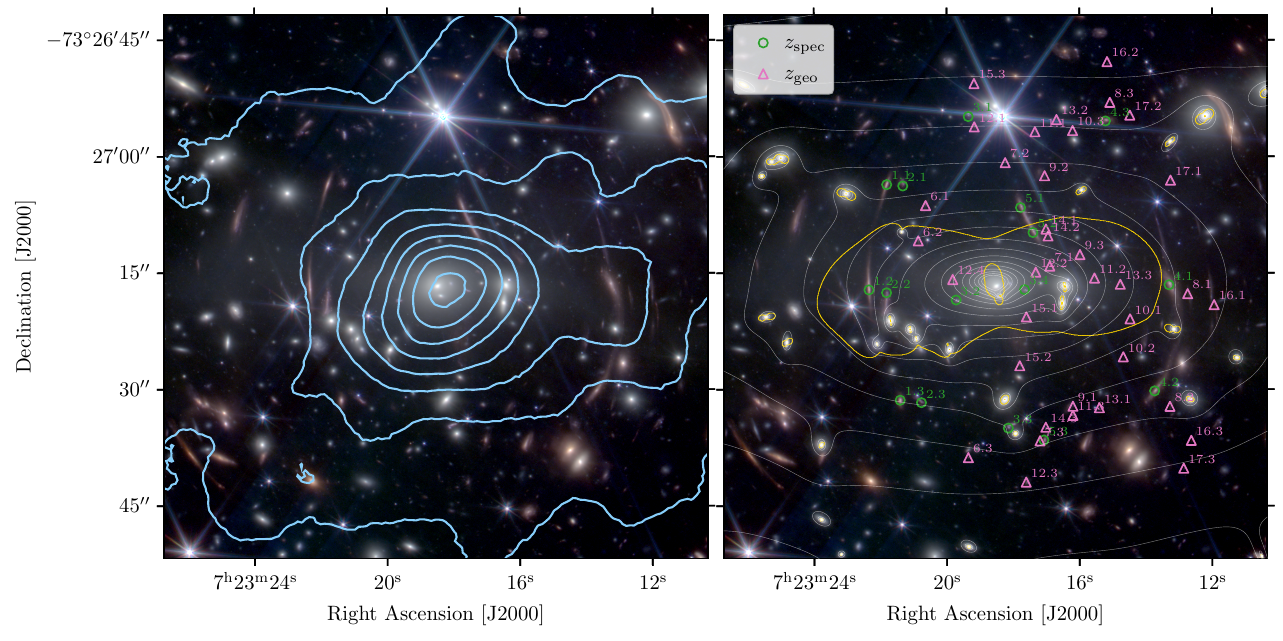}
    \caption{Composite colour image of SMACS\,J0723.3–7327 made using JWST images in F090W and F150W for blue, F200W and F277W for green, together with F356W and F444W for red.  \textit{Left}: Cyan iso-contours of the X-ray emission as observed with the Chandra X-ray Observatory. \textit{Right}: Green circles enclose five sets of multiply-lensed systems confirmed via their spectroscopic redshifts ($z_{\rm spec}$), whereas pink triangles enclose twelve sets of proposed multiply-lensed systems for which we derive geometrical redshifts ($z_{\rm geo}$) simultaneously with our lens model; the same symbols are adopted in the following relevant figures in this paper.  Identification numbers follow convention whereby whole number represents a particular system and decimal part its individual image counterparts. White iso-contours delineate projected mass distribution of the cluster inferred from our lens model.  Orange curves delineates the critical curve at $z=1.45$, the redshift of system 1, based on our lens model.}
    \label{fig:RGB_lens_img}
\end{figure*}

\subsubsection{Identification of Image Counterparts}\label{sec:counterparts}
Five multiply-lensed systems had previously been identified towards SMACS\,J0723.3–7327 from images obtained in the RELICs program, with four having redshifts measured spectroscopically from the aforementioned MUSE-VLT observations (Section \ref{sec:MUSE}).  The image counterparts of these systems are enclosed in green circles and labelled 1--5 in Figure \ref{fig:RGB_lens_img}, such that the whole number corresponds to a particular system and the decimal part the corresponding individual image counterparts for each system.  \citet{Golubchik2022} determined the redshifts of systems 1, 2, and 5, and \citet{Mahler2023} the redshift of system 3, from MUSE-VLT spectra.  \citet{Noirot2022} measured the redshift of system 4 from spectroscopic observations with the Near Infrared Spectrograph (NIRSpec) on the JWST. The image positions (defined in the manner described below) and redshifts of all these multiply-lensed systems are listed in Table \ref{tab:lens_img}.


In addition to the five systems having $z_{\rm spec}$, we used eleven other candidate multiply-lensed systems without spectroscopic redshifts to constrain our lens model for SMACS\,J0723.3–7327.  These candidates were either identified or used also by \citet{Caminha2022}, \citet{Pascale2022}, and \citet{Mahler2023} in their lens model for SMACS\,J0723.3–7327 based on the common colours, shared morphological features, and best guesses for the positions of their proposed image counterparts.  The latter are enclosed by pink triangles and labelled 6--16 in Figure \ref{fig:RGB_lens_img}.  We have examined the SEDs of the proposed image counterparts to verify that they do look alike for each system (see Section \ref{sec:relens}).  Furthermore, we identified another candidate multiply-lensed system, labelled 17 and having image counterparts enclosed also by pink triangles in Figure \ref{fig:RGB_lens_img}, based on the same criteria.  The image positions (again defined in the manner described below) of all these candidate multiply-lensed systems are listed also in Table \ref{tab:lens_img}, following those for the confirmed multiply-lensed systems.  When constructing our lens model, the redshifts of all these systems were left unspecified, except for system 16, for their $z_{\rm geo}$ to be freely solved by \texttt{glafic}.  For system 16, colour images constructed from the F277W, F356W, and F444W filters of the JWST indicate one or more bright emission lines in the F356W filter.  If associated with the [OIII] line, for which spectroscopy with NIRSpec on the JWST show to be the brightest line in galaxies at $z > 5$ \citep{carnall2023}, then system\,16 must have a redshift of $5.15 \leq z \leq 7.16$.  We therefore constrained the redshift of system\,6 accordingly when solving for the lens model (from which we obtained $z_{\rm geo} = 6.7$). 

Among the candidate multiply-lensed systems identified or used by \citet{Caminha2022}, \citet{Pascale2022}, and \citet{Mahler2023} in their lens model for SMACS\,J0723.3–7327, we omitted five such systems.  Three of these systems have all their image counterparts surrounding individual cluster members, making the lens model particularly dependent on the adopted mass models for these cluster members. Two other systems constitute dim and compact features located next to two corresponding bright candidate multiply-lensed systems (that we also used in our lens model): they are too dim to yield confident associations based on their colours or SEDs, and do not show appreciable morphologies.  The same is true for the remaining multiply-lensed system, which also is dim and compact, not used to constrain our lens model for SMACS\,J0723.3–7327.

As can be seen in the right panel of Figure \ref{fig:RGB_lens_img}, whereas the image counterparts of the five multiply-lensed systems having spectroscopically-confirmed redshifts (enclosed by the green circles) are quite evenly distributed around the cluster centre, the twelve candidate multiply-lensed systems have most of their image counterparts (enclosed by the pink triangles) located on the western half of the cluster.  Overall, the sky distribution of the (confirmed and proposed) image counterparts suggests that SMACS\,J0723.3–7327 has a lopsided mass distribution, such that there is more mass on the western compared with the eastern half of the cluster so as to generate more multiply-lensed images on the western half.  Indeed, an X-ray map of SMACS\,J0723.3–7327 by the Chandra X-ray Observatory (cyan contours in the left panel of Fig.\,\ref{fig:RGB_lens_img}) reveals its intracluster medium to also be lopsided, having a higher surface brightness on the western compared to the eastern side of the cluster \citep{Mahler2023}; indeed, the lens model for SMACS\,J0723.3–7327 constructed by \citet{Mahler2023} requires slightly more mass on the western compared with the eastern side of the cluster.  As we will show, and as demonstrated by other lensing analyses of SMACS\,J0723.3–7327 mentioned later in Section\,\ref{sec:bestfit_para}, the major axis of the cluster mass distribution is projected along an east-west direction, evenly straddled to the north and south by both the confirmed and proposed multiply-lensed image counterparts.  The lens model we construct for SMACS\,J0723.3–7327 is therefore well constrained over the inner region of the cluster populated quite evenly over all azimuthal angles by multiply-lensed images, although farther away from the cluster centre only on the western half well populated by multiply-lensed image counterparts.

\subsubsection{Defining Positional Centres of Image Counterparts}\label{sec:image positions}
To define the positions of the image counterparts used as constraints on the lens model, we first looked for a compact and relatively bright feature (hereafter knot) in each image counterpart.  Where detectable, we used the brightest pixel of the knot in the F444W image as the position of the image counterpart (except for systems 2 and 4, both of which have multiple knots).  As a conservative estimate of the uncertainty in the positions ($\sigma_{\theta_i}$ in Eq.\ref{eq:chisq_src}) thus defined, we adopt a position uncertainty of $3\sigma_{\rm Gauss} = 0\farcs24$, where $\sigma_{\rm Gauss}$ is the standard deviation of the Gaussian function representing the point spread function (PSF) of the JWST in the F444W image, for which the full-width half-maximum is 0\farcs19.
Where no knot is seen (and for systems 2 and 4 with multiple knots), we simply used the brightest pixel close the visually-determined centroid of the image counterpart in the F444W image, for which we now set an even more conservative positional uncertainty of $5\sigma_{\rm Gauss} = 0\farcs39$.  Both the positions and positional uncertainties thus defined were then used as constraints on the lens model we constructed for SMACS\,J0723.3–7327.


\startlongtable
\begin{deluxetable*}{cccccc}
\tablewidth{0pt}
\tablecaption{Multiply-lensed systems used to constrain our lens model.}
\tablehead{ \colhead{I.D.$^{(1)}$} & \multicolumn{2}{c}{Position$^{(2)}$}  & \colhead{Knot$^{(3)}$} & \colhead{$z$$^{(4)}$} & \colhead{$\mu$$^{(5)}$} \\
\colhead{} & \colhead{$\Delta \alpha$ ($\arcsec$)} & \colhead{$\Delta \delta$ ($\arcsec$)} & \colhead{} & \colhead{} & \colhead{} }
\label{tab:lens_img}
\startdata
1.1 & -14.0104 & 13.156 & $\times$ & 1.45 & $10.7\pm 0.28$\\
1.2 & -16.29 & -0.4018 & $\times$ & -- & $-18.5\pm 0.75$\\
1.3 & -12.2683 & -14.6064 & $\times$ & -- & $8.24\pm 0.12$\\
2.1 & -11.9397 & 12.9767 & $\times$ & 1.38 & $9.26\pm 0.18$\\
2.2 & -14.025 & -0.7693 & $\times$ & -- & $-19.7\pm 0.82$\\
2.3 & -9.5294 & -14.9199 & $\times$ & -- & $7.43\pm 0.10$\\
3.1 & -3.5121 & 21.9192 & $\checkmark$ & 1.99 & $5.40\pm 0.05$\\
3.2 & -5.0729 & -1.7047 & $\checkmark$ & -- & $-6.04\pm 0.12$\\
3.3 & 1.6561 & -18.1979 & $\checkmark$ & -- & $10.7\pm 0.40$\\
3.4 & 3.7448 & -0.3654 & $\checkmark$ & -- & $-5.68\pm 0.26$\\
4.1 & 22.3441 & 0.2659 & $\times$ & 2.21 & $-19.6\pm 0.52$\\
4.2 & 20.4952 & -13.3991 & $\times$ & -- & $12.2\pm 0.20$\\
4.3 & 14.1577 & 21.3515 & $\times$ & -- & $8.24\pm 0.14$\\
5.1 & 3.2263 & 10.1987 & $\times$ & 1.43 & $37.6\pm 1.34$\\
5.2 & 4.8269 & 6.9845 & $\times$ & -- & $-36.1\pm 1.32$\\
5.3 & 6.2351 & -19.732 & $\times$ & -- & $5.00\pm 0.05$\\
\hline
6.1 & -8.9971 & 10.4607 & $\checkmark$ & $1.63\pm0.02$ & $22.3\pm 0.41$\\
6.2 & -9.9391 & 5.9154 & $\checkmark$ & -- & $-20.4\pm 0.35$\\
6.3 & -3.4876 & -21.9947 & $\checkmark$ & -- & $5.11\pm 0.07$\\
7.1 & 6.9995 & 2.6689 & $\checkmark$ & $1.66\pm0.02$ & $-8.37\pm 0.22$\\
7.2 & 1.2475 & 15.9861 & $\checkmark$ & -- & $8.22\pm 0.10$\\
7.3 & 5.744 & -19.8259 & $\checkmark$ & -- & $5.66\pm 0.06$\\
8.1 & 24.7513 & -0.8807 & $\checkmark$ & $2.94\pm0.05$ & $-19.4\pm 0.51$\\
8.2 & 22.4728 & -15.3835 & $\checkmark$ & -- & $12.4\pm 0.36$\\
8.3 & 14.7255 & 23.7538 & $\checkmark$ & -- & $7.16\pm 0.10$\\
9.1 & 9.9759 & -15.3767 & $\checkmark$ & $1.43\pm0.02$ & $6.12\pm 0.08$\\
9.2 & 6.3246 & 14.3062 & $\checkmark$ & -- & $10.1\pm 0.18$\\
9.3 & 10.8768 & 4.1576 & $\checkmark$ & -- & $-10.4\pm 0.24$\\
10.1 & 17.3312 & -4.1303 & $\times$ & $1.7\pm0.035$ & $-24.4\pm 0.52$\\
10.2 & 16.4567 & -9.0091 & $\times$ & -- & $23.9\pm 0.43$\\
10.3 & 9.93 & 20.13 & $\times$ & -- & $6.35\pm 0.12$\\
11.1 & 5.0785 & 19.9741 & $\checkmark$ & $1.78\pm0.02$ & $6.39\pm 0.07$\\
11.2 & 12.7489 & 1.1319 & $\checkmark$ & -- & $-6.46\pm 0.11$\\
11.3 & 9.9778 & -16.5454 & $\checkmark$ & -- & $7.00\pm 0.09$\\
12.1 & -2.7388 & 20.5891 & $\times$ & $2.58\pm0.08$ & $7.16\pm 0.09$\\
12.2 & 5.1863 & 1.8872 & $\times$ & -- & $-6.32\pm 0.25$\\
12.3 & 3.9615 & -25.1461 & $\times$ & -- & $5.31\pm 0.07$\\
12.4 & -5.5048 & 0.9368 & $\times$ & -- & $-6.46\pm 0.131$\\
13.1 & 13.3798 & -15.5725 & $\times$ & $1.96\pm0.055$ & $8.04\pm 0.12$\\
13.2 & 7.8456 & 21.541 & $\times$ & -- & $6.30\pm 0.09$\\
13.3 & 16.0719 & 0.3286 & $\times$ & -- & $-8.53\pm 0.17$\\
14.1 & 6.5056 & 7.439 & $\times$ & $1.37\pm0.02$ & $-57.8\pm 12.7$\\
14.2 & 6.7741 & 6.5474 & $\times$ & -- & $58.2\pm 12.7$\\
14.3 & 6.4844 & -18.0781 & $\times$ & -- & $5.17\pm 0.07$\\
15.1 & 3.9813 & -3.8778 & $\checkmark$ & $1.93\pm0.03$ & $-19.7\pm 0.78$\\
15.2 & 3.137 & -10.1656 & $\checkmark$ & -- & $24.74\pm 0.63$\\
15.3 & -2.7663 & 26.1761 & $\checkmark$ & -- & $4.39\pm 0.04$\\
16.1 & 28.1565 & -2.2967 & $\checkmark$ & $6.71\pm0.18$ & $-15.1\pm 0.42$\\
16.2 & 14.3595 & 28.9998 & $\checkmark$ & -- & $5.94\pm 0.07$\\
16.3 & 25.2218 & -19.7616 & $\checkmark$ & -- & $9.62\pm 0.16$\\
17.1 & 22.5351 & 13.7418 & $\checkmark$ & $5.08\pm0.245$ & $-41.6\pm 1.80$\\
17.2 & 17.2959 & 22.1135 & $\checkmark$ & -- & $12.8\pm 0.24$\\
17.3 & 24.2396 & -23.3562 & $\checkmark$ & -- & $6.05\pm 0.09$\\
\enddata
\tablecomments{(1) Whole number for the system and decimal part its individual image counterparts having positions as shown in Fig.\,\ref{fig:RGB_lens_img}.  (2) Relative to $\rm R.A. = 07^{\rm h}23^{\rm m}18\fs5239$ and $\delta = -73^{\circ}27^{\prime}16\farcs750$. (3) Whether image has compact knots.  (4) Spectroscopic redshifts for I.D. 1--5, geometric redshifts for I.D. 6--17 as inferred from our lens model.  (5) Lensing magnification, whereby negative sign indicates negative parity.}
\end{deluxetable*}

\subsection{Model Ingredients}\label{ingredients}

\subsubsection{Cluster Members}\label{sec:clusmem}
\begin{figure}
    \centering
    \includegraphics[width=0.45\textwidth]{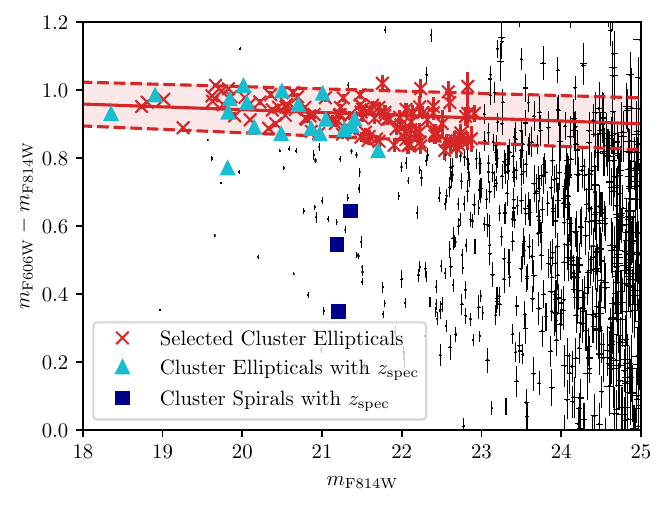}
    \caption{Colour versus apparent magnitude for selected galaxies (see text) towards SMACS\,J0723.3–7327.  Colours derived from HST images in filters straddling the 4000\,\AA\ break at the cluster redshift.  Cyan triangles are early-type galaxies that define a red sequence, and blue squares are late-type galaxies, selected as cluster members based on their $z_{\rm spec}$ as determined from spectra taken by the MUSE-VLT.  Red line is a linear fit to the colour-magnitude dependence of the red-sequence cluster members, with dashed lines corresponding to the $\pm 1\sigma$ uncertainty of this fit.  Among remaining galaxies without $z_{\rm spec}$, red crosses indicate those lying in the red band (within their $\pm 1\sigma$ measurement uncertainties) and hence selected also as cluster members.}
    \label{fig:CMD}
\end{figure}

We selected cluster members using two methods.  First, for objects detected in the MUSE-VLT observations, which span a field-of-view smaller than the JWST observations (see Section \ref{sec:MUSE}), we extracted spectra for all the objects compiled in the RELICs catalog.  Then, for each object, we derived its $z_{\rm spec}$ by fitting spectral templates to its measured spectrum using the program \texttt{MarZ} \citep{Hinton2016}.  Plotting the number of galaxies versus their $z_{\rm spec}$, we then fitted a Gaussian function to the highest peak in galaxy numbers to infer a cluster redshift, $z_{\rm cluster}$, and an associated standard deviation, $\sigma_{z, \rm cluster}$, of $z_{\rm cluster} \pm \sigma_{z, \rm cluster} = 0.389 \pm 0.0172$.  Having a corresponding velocity dispersion $\sigma_{v, \rm cluster}=1669\:\rm km\:s^{-1}$, we then selected all objects within $z_{\rm cluster} \pm 2\sigma_{z, \rm cluster}$ as cluster members.

For galaxies having unreliable $z_{\rm spec}$, spectra too noisy to be useful, or those lying beyond the field-of-view of the MUSE-VLT observation, we derived their colours from the HST images in the F606W and F814W filters.  
Figure \ref{fig:CMD} shows the colours thus measured against the apparent magnitudes of these objects in the F814W filter ($m_{\rm F814W}$), where for comparison the cyan trangles and dark blue squares are cluster members selected by their $z_{\rm spec}$.  The filters used for measuring colours straddle the 4000\,\AA\ break of galaxies at the redshift of SMACS\,J0723.3–7327, and therefore selectively picks out cluster members by their relatively red colours as measured in these two filters.  As can be seen, a red sequence is clearly defined by the concentration of relatively bright and red galaxies denoted by the cyan triangles, corresponding to early-type cluster members having $z_{\rm spec} = z_{\rm cluster} \pm 2\sigma_{z, \rm cluster}$.  The red line passing through these cluster members is a linear fit to their colour-magnitude dependence, with the dashed lines corresponding to the $\pm 1\sigma$ uncertainty of the fit.  We define cluster members without $z_{\rm spec}$ as those having colours that, within their $\pm 1 \sigma$ measurement uncertainties, lie in the red band of Figure \ref{fig:CMD}, and which are brighter than $m_{\rm F814W} = 23 \rm \, mag$; the galaxies thus selected are indicated by red crosses. Three relatively bright spiral galaxies with $z_{\rm spec}$ indicating that they are cluster members are indicated by the dark blue squares.  In this way, we efficiently select the brightest cluster members -- those that can create the most appreciable local perturbations in lensing magnification -- for inclusion in the lens model.

When constructing the lens model, we parametrise the mass distributions of each cluster member using a modified Jaffe profile \citep{Keeton2001}:
\begin{equation}
    \rho(r) = \frac{\rho_{0}}{r^2(r^2+r_{\mathrm{trun}}^2)} \, \, ,
\end{equation}
where $\rho_{0}$ is the scaling constant, $r$ the radius from the centre, and $r_{\mathrm{trun}}$ the truncation radius of the galaxy.  To avoid having many more free parameters than there are constraints, we scale the truncation radius of each galaxy to its luminosity such that:
\begin{equation}
\label{eq:fj_2}
    \frac{r_{\mathrm{trun}}}{r_{\mathrm{trun, G02}}} = \left(\frac{L}{L_{\mathrm{G02}}}\right)^{\frac{1}{2}} \, \, ,
\end{equation}
where $L_{\mathrm{G02}}$ is the luminosity of the 2nd-brightest cluster member.  Moreover, we scale the mass of each cluster member as characterised by its velocity dispersion, $\sigma_v$, to its luminosity according to the Faber-Jackson relation \citep{faber1976} such that:
\begin{equation}
\label{eq:fj_1}
    \frac{\sigma_v}{\sigma_{v,\mathrm{G02}}} = \left(\frac{L}{L_{\mathrm{G02}}}\right)^{\frac{1}{4}}
\end{equation}
Finally, we fix their ellipticities ($e$) and position angles ($\theta$) for their major axes to the values compiled in the RELICs catalog.  We note that the exact parameterisation of individual cluster members (except for the central giant elliptical galaxy) has little effect on the derived parameters of the main component -- the cluster-scale Dark Matter (DM) halo -- that dominates the cluster mass and hence our lens model.  Cluster members, however, can have an appreciable effect on lensed images that happen to lie nearby in projection.

As is typical when constructing lens models for galaxy clusters, we treat the cluster central giant elliptical galaxy -- the Brightest Cluster Galaxy (BCG) -- differently than the other cluster members, as it oftentimes contributes significantly to lensed images formed around the cluster centre.  As the presence of any galaxy-scale DM halo around the BCG is usually degenerate with the cluster-scale DM halo \citep[e.g.,][]{Chow2023}, we only consider the stellar component of the BCG.  From its SED as retrieved from the photometric catalog, we fit a synthetic stellar population having an exponentially decaying star-formation rate to deduce a stellar mass for the BCG of $M_*=1.25^{+0.02}_{-0.02}\times10^{11}M_{\odot}$, with stars first forming at a lookback time of $6.86^{+0.06}_{-0.07}\:\rm Gyr$ and having a metallicity of $2.38^{+0.04}_{-0.04}\:Z_{\odot}$.  We note this fit is not meant to provide an accurate representation of its star-formation history in the distant past, but to provide a reasonably accurate estimate of its total stellar mass (an estimate that is relatively immune to the formation history of old stars).  The mass profile of the BCG is parameterised by fitting a S\'ersic function \citep{sersic1963} to its light profile in the F606W filter.




\subsubsection{Cluster-Scale DM Halo}
X-ray images of SMACS\,J0723.3-7327 show an elongated morphology that is quite sharply peaked on the BCG.  Analyses of the X-ray morphology of the intracluster medium by \citet{Mahler2023} and \citet{Liu2023} suggest this cluster to be reasonably close to being dynamically relaxed, although betraying evidence for global disturbances.  Indeed, the presence of a radio halo suggests interactions with another galaxy group or cluster in the recent past, such that the cluster has not yet fully recovered to become dynamically relaxed.  Nonetheless, for simplicity, we model its DM halo with an axially symmetric Navarro-Frenk-White (NFW) profile \citep{Navarro1996}.  We note our lens models does not include a specific mass component for the intracluster medium (ICM), which has a total mass far exceeding the cluster members combined.  Studies find the mass profile of the ICM, in the case of closely-relaxed clusters, to closely follow the total DM-dominated mass profile of the cluster \citep[e.g.,][]{Donahue2014}; in such cases, the mass contributed by the ICM is absorbed into the mass component representing the cluster-scale DM halo.


\subsubsection{External Shear}
To allow for the possibility of a non-uniform gravitational field external to the cluster, specifically mass concentrated along a particular direction, we allow an additional contribution to the lensing potential as parametrised by
\begin{equation}
    \phi = \frac{1}{2}r^2 \gamma \cos{[2(\theta-\theta_{\gamma})]} \,\, ,
\end{equation}
where $r$ and $\theta$ are the usual spherical coordinates, $\gamma$ the shear strength, and $\theta_{\gamma}$ the position angle of the external shear.

\subsection{Lens Model Parameter Tolerances}\label{sec:MCMC}
Uncertainties in the positions of the image counterparts used as constraints (Section \ref{constraints}) translate to uncertainties in the parameters of the best-fit lens model.  To assess the tolerances of the lens model parameters, we performed a Markov Chain Monte Carlo (MCMC) simulation as implemented in \texttt{glafic} to sample the parameter space around the best-fit parameters of the lens model, as well as the derived values of $z_{\rm geo}$ for the proposed multiply-lensed systems without $z_{\rm spec}$.  In this way, we estimate the precision of the derived $z_{\rm geo}$ (Section \ref{sec:fk}), and gain an understanding of how uncertainties in our lens model parameters affect the lens finder map for high-$z$ galaxies (Section \ref{sec:finder_map}).

\section{Parameters and Reliability of Lens Model}\label{sec:results}

\begin{deluxetable}{cccc}
\tablewidth{0pt}
\tablecaption{Parameters of Lens Model.}
\tablehead{\colhead{} & \colhead{} & \colhead{Unit} & \colhead{Best-fit}}
\label{tab:LM_para}
\startdata
Cluster & $x$ & arcsec & $-0.344^{+0.17}_{-0.17}$ \\
DM & $y$& arcsec & $0.728^{+0.05}_{-0.05}$ \\
 & $e$ & -- & $0.495^{+0.01}_{-0.01}$ \\
 & $\theta$ & deg & $90.3^{+0.07}_{-0.17}$ \\
 & $M_{200}$& $10^{15}M_{\odot}$ & $3.60^{+0.12}_{-0.04}$ \\
 & $c_{200}$ & & $2.44^{+0.02}_{-0.01}$  \\
\hline
BCG & $M$ & $10^{11}M_{\odot}$ & $[1.25]$\\
Stellar & $R_e$ & $\mathrm{kpc}$ & $[25.1]$\\
 & $e$& --& $[0.242]$ \\
 & $\theta$& deg & $[97.1]$ \\
 & $n$ & -- & $[2.73]$ \\
\hline
Cluster & $\sigma_{v,\mathrm{G02}}$ &$\mathrm{km\:s^{-1}}$ & $156^{+0.11}_{-0.19}$ \\
Members & $r_{\mathrm{trun,G02}}$ &$\mathrm{kpc}$ & $343^{+11.7}_{-17.9}$ \\
\hline
External & $x$ & arcsec & $-32.1^{+0.05}_{-0.05}$ \\
Shear & $y$ & arcsec & $-9.06^{+0.12}_{-0.12}$\\
 & $\gamma$ & -- & $0.065^{+0.003}_{-0.003}$ \\
& $\theta_{\gamma}$ & deg & $-18.2^{+0.07}_{-0.06}$ 
\enddata
\tablecomments{Positions $(x,y)$ with respect to $\rm R.A. = 7^{\rm h}23^{\rm m}18\fs.5239$ and $\delta = -73^{\circ}27^{\prime}16\farcs750$.  $e$ is ellipticity and $\theta$ is position angle for major axis of individual mass components.  $M_{200}$ and $c_{200}$ are mass and concentration, respectively, of the NFW halo.  $M_*$ is stellar mass and $R_e$ effective (i.e., half-light) radius of BCG, for which $n$ is its fitted S\'ersic index.  $\sigma_v$ and $r_{\rm trun}$ are velocity dispersion and truncation radius of cluster members, all referenced with respect to the second-brightest galaxy G02 (see text).  $\gamma$ and $\theta_{\gamma}$ are, respectively, shear strength and position angle of external shear.}
\end{deluxetable}

\subsection{Best-fit Parameters} \label{sec:bestfit_para}

Table \ref{tab:LM_para} lists the best-fit parameters of the lens model constructed for SMACS\,J0723.3-7327, along with their uncertainties as estimated from an MCMC simulation.  The redshifts of the twelve multiply-lensed systems lacking spectroscopic measurements, determined geometrically along with the lens model, are listed in Table \ref{tab:lens_img}.  The projected surface mass density of the cluster implied by the lens model is indicated by contours in Figure \ref{fig:RGB_lens_img}.  This mass distribution is particularly well constrained in regions densely populated by the multiply-lensed images used to constrain the lens model (enclosed by green circles for systems with, and pink triangles for systems without, $z_{\rm spec}$), but less so at increasing radial distances away from the available constraints.  We find a total mass within the Einstein ring (for a lensed source approaching infinity, with a corresponding radius for the Einstein ring of 20\farcs6 or 109.0\,kpc) for this cluster of $8.90^{+0.18}_{-0.1}\times 10^{13}M_{\odot}$.  

Using also a parametric approach, \citet{Mahler2023} derive a total mass of $7.63^{+2.0}_{-2.0}\times 10^{13}M_{\odot}$ within a radius of 128\,kpc from the cluster centre, corresponding to the maximal separation of the image counterparts of System\,4 from the cluster centre.  At this same radius, the enclosed mass that we derive is $8.90^{+0.18}_{-0.1}\times 10^{13}M_{\odot}$, similar to within the inferred uncertainties as that derived by \citet{Mahler2023}.  Also using a parametric approach, \citet{Caminha2022} find a total mass $8.6^{+0.2}_{-0.2}\times 10^{13}M_{\odot}$ within a radius of 128\,kpc of the cluster centre, again similar to that we derived within the inferred uncertainties.
On the other hand, the cluster mass that we derived is significantly (nearly 20\%) higher than that derived by \citet{Diego2023} of $7.28^{+0.03}_{-0.13}\times 10^{13}M_{\odot}$ over a radius of 128\,kpc, derived using a free-form approach (rather than an analytical function) for the cluster-scale DM halo.
Reassuringly, all these lens models find the mass distribution to be elongated at a position angle of about $90^{\circ}$, similar to that inferred in our lens model.

Extrapolating the cluster mass to a radius of $\sim$3.2\,Mpc -- corresponding to 200 times the critical density of the universe at the cluster redshift -- so as to compare with the cluster mass inferred from weak-lensing measurements, we infer a cluster mass of $M_{200} = 3.6^{+0.12}_{-0.04} \times 10^{15} M_{\odot}$.  This value is about $50\%$ higher than that inferred by \citet{Finner2023} based on a weak-lensing analysis of $2.33\pm0.3\times 10^{15}M_{\odot}$.  Whereas we allow both $M_{200}$ and $c_{200}$, the latter corresponding to the concentration parameter of the NFW halo, to freely vary, \citet{Finner2023} force $M_{200}$ and $c_{200}$ to follow the $c$--$M$ relation relation proposed by \citet{Diemer2019}.  We derive $c_{200} = 2.44^{+0.02}_{-0.01}$, which is significantly smaller than the corresponding value derived by \citet{Finner2023} of $c_{200} = 4$. Nonetheless, reassuringly, the mass distribution that \citet{Finner2023} infer from their weak-lensing analysis also is elongated at a position angle of about $90^{\circ}$, similar to that inferred from all the strong-lensing analyses described above.


\subsection{Internal Consistency}\label{sec:consistency}
Internal consistency refers to the ability of a lens model to reproduce the image properties used to constrain that lens model, a test that any lens model must pass to be considered viable.   Here, we consider two metrics for internal consistency.

\subsubsection{Image Positions}\label{sec:pred_img_pos}
\begin{figure}
    \centering
    \includegraphics[width=0.45\textwidth]{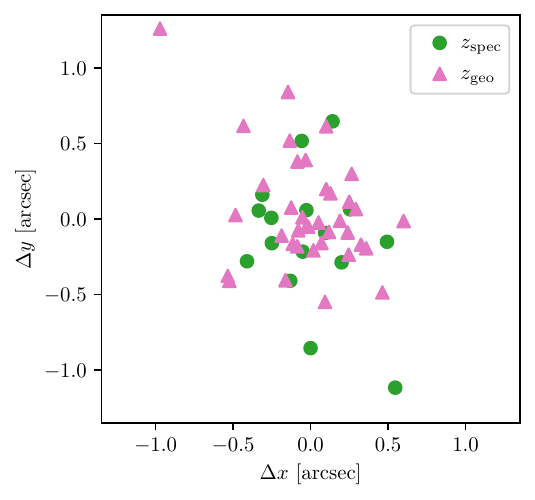}
    \caption{Model-predicted positions relative to observed positions of all image counterparts used to constrain our lens model.  Observed positions are all placed at the origin.
    }
    \label{fig:rms_scatter}
\end{figure}
Figure \ref{fig:rms_scatter} shows the predicted position of a given image counterpart relative to its observed position, which is placed at the origin of this plot for all the image counterparts used as constraints.  The root-mean-square (rms) dispersion between the predicted and observed positions of the multiply-lensed counterparts with $z_{\rm spec}$ is $0\farcs492$, compared with that for the multiply-lensed counterparts without $z_{\rm spec}$ (but for which we derive $z_{\rm geo}$) of $0\farcs404$.  These positional dispersions are, at worst, a factor of 2 larger than the positional uncertainties adopted for the image counterparts used as constraints; as mentioned in Section \ref{constraints}, we set a positional uncertainty of $0\farcs24$ for the image counterparts displaying knots, and $0\farcs39$ for the image counterparts without knots.  Furthermore, tolerances in the lens model parameters contribute an additional rms uncertainty of 0\farcs086 to the predicted positions of image counterparts.  We therefore consider the level of reproducibility in positions to be satisfactory, 
especially given the relative simplicity of our lens model in terms of the number of components employed (Section \ref{ingredients}).


\subsubsection{Adopted Cosmology} \label{sec:fk}
Translating the image positions and, where available, redshifts of multiply-lensed images into actual constraints for a lens model requires a cosmological model, for which we adopt the concordant $\Lambda$CDM cosmological parameters of $\Omega_M=0.3$, $\Omega_\Lambda=0.7$, and $H_0=70\:\mathrm{km\:s^{-1}\:Mpc^{-1}}$.  To check whether the observed image positions and input redshifts ($z_{\rm spec}$) or model-determined output redshifts ($z_{\rm geo}$) are in agreement with the adopted cosmology, we employ the quantity $f_k$, where the suffix $k$ refers to the system $k$, as introduced by \citet{Broadhurst2005} and subsequently modified by \citet{Chan2017}.  The latter defines this quantity as:
\begin{equation} \label{eq:fk_cosmo}
    f_k(z) = \frac{D_{\mathrm{ls}_k}(z)}{D_{\mathrm{s}_k}(z)} \bigg/ \frac{D_{\mathrm{ls}_0}(z_0)}{D_{\mathrm{s}_0}(z_0)},
\end{equation}
where $D_{\mathrm{ls}_k}$ is the angular diameter distance between the lens and system $k$ (at its actual sky position and known or determined redshift), and $D_{\mathrm{s}_k}$ the angular diameter distance between system $k$ and the observer.  $z_0$ is a fiducial redshift, for which we chose $z_0=1.45$, the spectroscopically-determined redshift of system 1.  Thus defined, $f_k(z)$ depends solely on the adopted cosmology and can be rewritten in terms of the observables as:
\begin{equation} \label{eq:fk_obs}
    f_k = \frac{2}{n_k(n_k-1)}\sum_{i,j;i>1}\frac{(\vec{\theta}_{k,i}-\vec{\theta}_{k,j})\cdot[\vec{\hat{\alpha}}(\vec{\theta}_{k,i})-\vec{\hat{\alpha}}(\vec{\theta}_{k,j})]}{[\vec{\hat{\alpha}}(\vec{\theta}_{k,i})-\vec{\hat{\alpha}}(\vec{\theta}_{k,j})]^2} \, \, ,
\end{equation}
where $n_k$ is the number of lensed counterparts for system $k$, $\vec{\theta}_{k,i}$ is the observed image position, and $\vec{\hat{\alpha}}(\vec{\theta}_{k,i})$ is the reduced deflection angle at $\vec{\theta}_{k,i}$ at $z_0$. 

\begin{figure}
    \centering
    \includegraphics[width=0.45\textwidth]{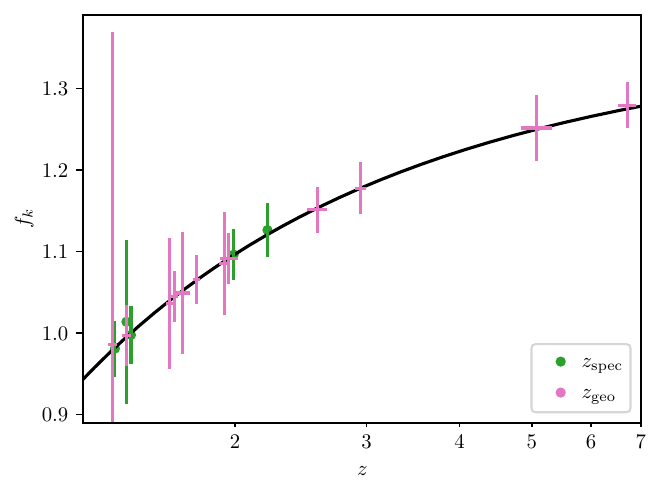}
    \caption{Observed $f_k$ (computed from Eq.\,\ref{eq:fk_obs}) versus $z$ ($z_{\rm spec}$ where available, and where not $z_{\rm geo}$) for the multiply-lensed systems that serve as constraints on our lens model.  Theoretical $f_k$ (computed from Eq.\,\ref{eq:fk_cosmo}) indicated by black curve based on the concordance cosmological parameters of the $\Lambda$CDM model adopted (see text).}
    \label{fig:fk_1_halo}
\end{figure}

Figure \ref{fig:fk_1_halo} shows the $f_k$ values for all the multiply-lensed systems plotted against their redshifts as determined spectroscopically where available, or where not available then geometrically with uncertainties as obtained from the MCMC simulation.  The curve indicates the same dependence as predicted by the adopted cosmology.   Deviations from this curve reflect differences in the deflection fields at the model-predicted and observed positions of a given image counterpart; we therefore computed the rms uncertainty in the individual values $f_k$ by evaluating the rms dispersion in the deflection field over an error circle of radius 0\farcs45 (corresponding to the rms dispersion between the model-predicted and observed positions of all the image counterparts used to constrain the lens model, as described earlier in Section \ref{sec:pred_img_pos}) centred on the observed position of a given image counterpart.  As can be seen, the $f_k$ values of all the multiply-lensed systems used to constrain the lens model are in good agreement with the adopted cosmology, demonstrating the internal consistency of our lens model.

\subsection{Predictability}\label{sec:predictability}
The ability of a lens model to correctly predict image properties not used as constraints on the lens model provides a measure of its reliability.  Proof that our lens model for SMACS\,J0723.3-7327 is reliable provides confidence for using this model to test or constrain the redshifts of claimed high-$z$ galaxies behind the cluster.  Here, we consider two metrics for predictability.

\subsubsection{Morphology of Multiply-Lensed Images}\label{sec:relens}
\begin{figure}
    \centering
    \gridline{\hspace{-1em} \fig{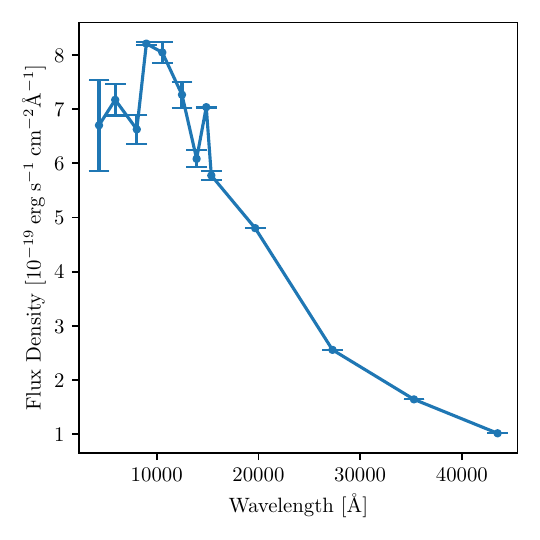}{0.14\textwidth}{} \hspace{-3.75em}
          \fig{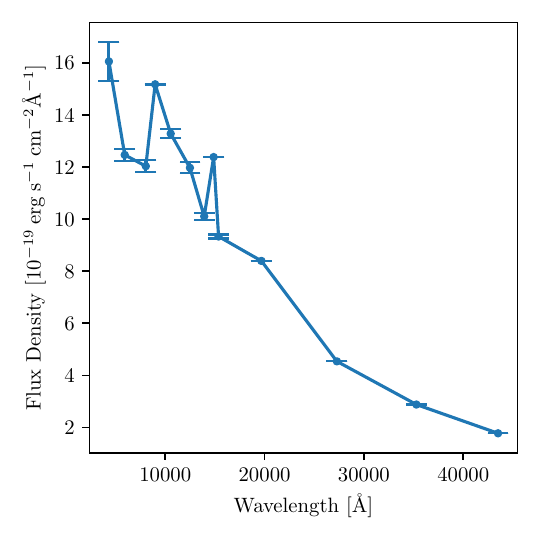}{0.14\textwidth}{} \hspace{-3.75em}
          \fig{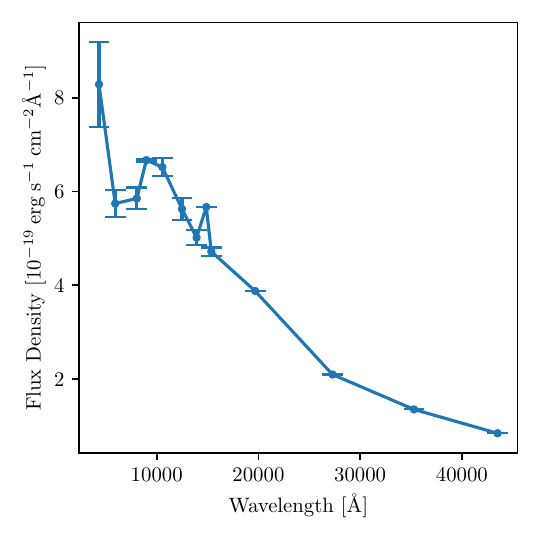}{0.14\textwidth}{}
          }
    \vspace{-3em}
    \gridline{\fig{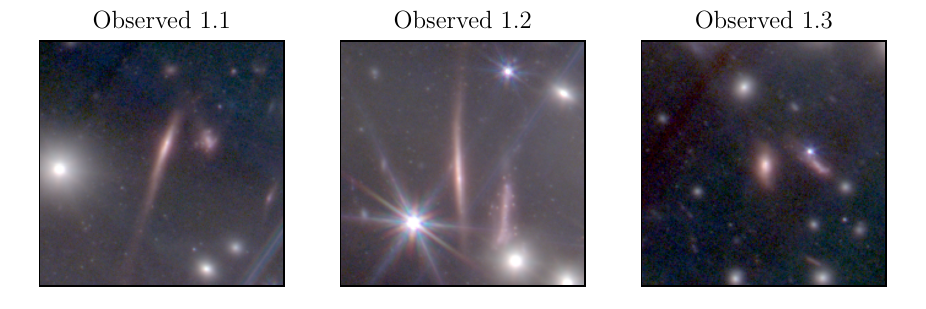}{0.45\textwidth}{}}
    \vspace{-3em}
    \gridline{\fig{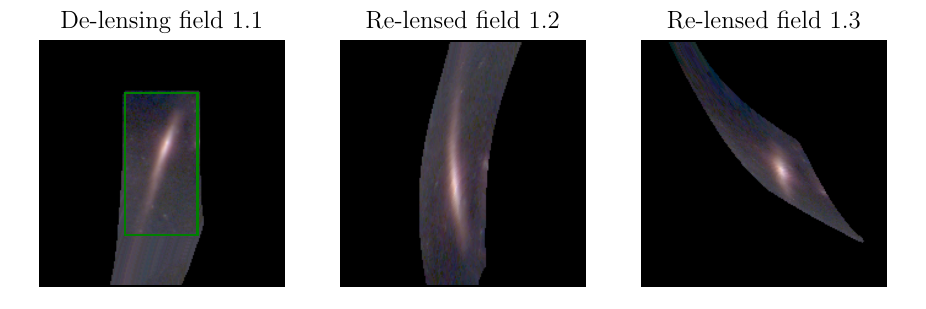}{0.45\textwidth}{}}
    \vspace{-3em}
    \gridline{\fig{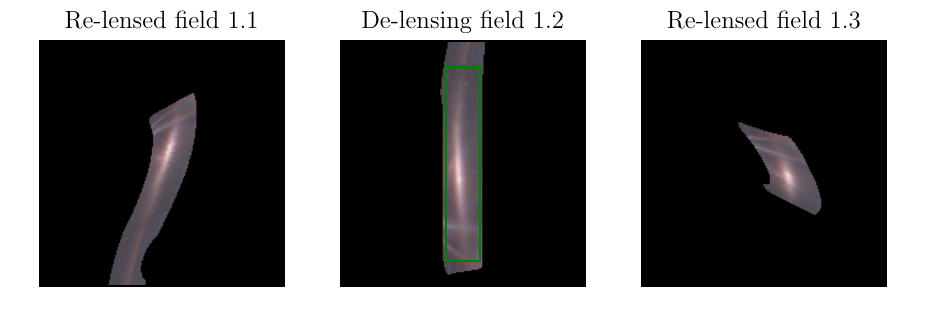}{0.45\textwidth}{}}
    \vspace{-3em}
    \gridline{\fig{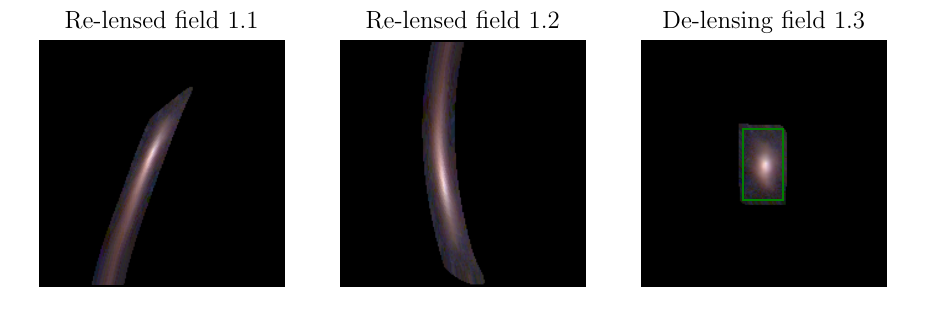}{0.45\textwidth}{}}
    \vspace{-6mm}
    \caption{First row showing SEDs of the individual image counterparts  of system\,1 shown in second row. Each of the remaining rows show field around each image counterpart, enclosed in green boxes and chosen so as to avoid neighbouring unrelated objects (panels labeled de-lensing field 1.x, where x is the decimal part corresponding to a given image counterpart), that is de-lensed and then re-lensed to locations of its other image counterparts (panels labeled re-lensed field 1.x). The de-lensed and then re-lensed images bear a close resemblance to their observed image counterpart at the corresponding location, testifying to the ability of our lens model to make reliable predictions.  Corresponding results for all remaining multiply-lensed systems used to constrain our lens model are shown in Appendix\,\ref{sec:appen_A}.}
    \label{fig:relens_sys1}
\end{figure}

To assess the ability of our lens model to reproduce the observed morphologies of the image counterparts for each multiply-lensed system, we first de-lens the image of a given counterpart to the source plane, and then re-lens it back to the image plane at the position of a different counterpart. Figure\,\ref{fig:relens_sys1} shows examples of re-lensed images for every image counterpart in system\,1.  The second row of this figure shows all of its observed image counterparts.  The third row shows the predicted image counterparts 1.2 and 1.3 after de-lensing and re-lensing the area enclosed by the green box around the image counterpart 1.1.  The fourth and fifth rows show, respectively, the corresponding situations by de-lensing and re-lensing the image counterparts 1.2 and 1.3.  To check their reliability as image counterparts, the top row of Figure\,\ref{fig:relens_sys1} shows the SEDs for each image counterpart of this system. The re-lensed images for each of the other systems can be found in Appendix\,\ref{sec:appen_A}. In every case, the morphology of a given image counterpart as predicted from de-lensing and the re-lensing is in good agreement with its observed morphology, bestowing confidence on the reliability of our lens model.  In the top row of Figure\,\ref{fig:relens_sys1} and each of the figures shown in Appendix\,\ref{sec:appen_A}, we also plot the SED of each image counterpart; as mentioned in Section\,\ref{sec:counterparts}, we checked that each set of proposed image counterparts do indeed share the same SEDs.



\subsection{Relative Brightnesses of Multiply-Lensed Images} \label{sec:rela_bri}
\begin{figure}
    \centering
    \includegraphics[width=0.45\textwidth]{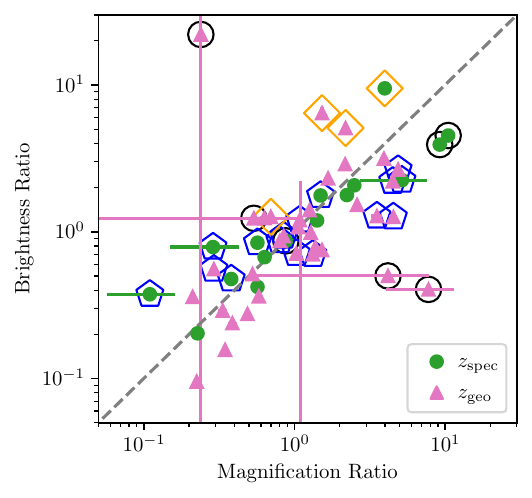}
    \caption{Observed brightness ratio versus predicted magnification ratio for every pair of image counterparts in each multiply-lensed system used to constrain our lens model.  The brightness of each image counterpart was measured using individually tailored photometry, having measurement uncertainties as reflected by the $\pm 1\sigma$ error bars in brightness ratios.  The magnification ratio is computed from the lens model magnification at the defined position of each image counterpart (see Section\,\ref{sec:image positions}), with $\pm 1\sigma$ error bars reflecting tolerances in the lens model parameters and hence its magnification at any given location as determined from an MCMC simulation (see Section\,\ref{sec:MCMC}).  Lack of visible error bars are for cases where the corresponding error bars are smaller than the symbol size.  In reality, the lens model magnification can vary considerably over the area spanned by the individual images.  Black circles enclose those for which at least one image counterpart in a pair combination is lensed into an arc, and hence for which there are large variations in magnification along one or both image counterparts thus making their magnification ratio ill defined.  Orange diamonds enclose those near bright stars or galaxies and blue pentagons enclose those projected against bright intracluster light, thus compromising the photometry of these image counterparts.  Dashed diagonal line corresponds to a perfect one-to-one match.}
    \label{fig:flux_mag_ratio}
\end{figure}
Although the relative brightnesses along with positions of multiply-lensed images can be used as constraints on lens models, the former is seldom (if ever) employed for galaxy clusters owing to the difficulty in performing photometry of strongly-lensed images having complex morphologies (oftentimes, lensed into arcs), sometimes located near bright cluster members (or other bright objects), or confused with intracluster light.  With this in mind, here we compare the relative brightnesses of the image counterparts for each multiply-lensed system as predicted by our lens model against those observed.  The brightness of each image counterpart was determined through individually tailored photometry, and measured in the filter where the image counterparts of a given multiply-lensed system have the highest signal-to-noise (S/N) ratios.  We use the ratio in lensing magnification predicted by our lens model at the position defined for each image counterpart (see Section \ref{constraints} and Table \ref{tab:lens_img}) as a stand-in for their relative brightnesses.  The uncertainties plotted for this ratio reflect those of the relevant pair of lensing magnifications as evaluated from our MCMC simulation. 

The results for every possible pair combination of image counterparts for each multiply-lensed system is shown in Figure \ref{fig:flux_mag_ratio}.  As can be seen, there is reasonably good agreement between the predicted and observed relative brightnesses for the majority of image counterparts.  The image counterparts enclosed in different shapes, for which this agreement is relatively poor, correspond to those for which at least one image counterpart is: (i) lensed into an arc exhibiting large variations in magnifications along its length (enclosed by black circles in Fig.\,\ref{fig:flux_mag_ratio}); (ii) lying adjacent to bright stars or galaxies (orange diamonds), complicating its photometry; and (iii) projected against relatively bright intracluster light (blue pentagons), again complicating its photometry.

\section{Geometric Applications of Lens Model}\label{sec:discussion}
Our lens model, which we go to great pains to ensure is both internally consistent (Section \ref{sec:consistency}) and make verifiable predictions (Section \ref{sec:predictability}), provide geometrically-determined redshifts for multiply-lensed galaxies lacking spectroscopic measurements.  In turn, provided the position and redshift of an image, our lens model can be used to predict whether that image belongs to a multiply-lensed system, and if so the positions of its lensed counterparts at its claimed redshift (Section \ref{sec:finder_map}).  The lack of lensed counterparts would then imply that the redshift of the galaxy has been wrongly determined, and furthermore imposes an upper limit on its redshift (Section \ref{sec:Yan_highz}).  Until spectroscopic measurements become available, this approach for testing the validity of claimed redshifts provides a sobering assessment of purported galaxies in the very early universe, and just as importantly insights into how even model fits to SEDs from the HST and JWST combined can still lead to highly erroneous redshifts.

\subsection{Lens Finder Map for High-$z$ Galaxies}\label{sec:finder_map}
\begin{figure*}
    \centering
    \includegraphics[width=0.85\textwidth]{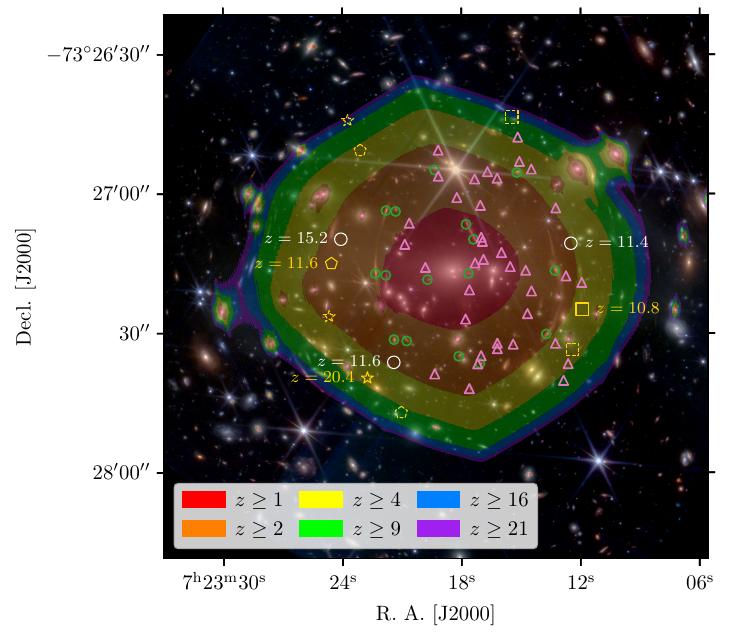}
    \caption{Lens finder map for SMACS\,J0723.3-7327, whereby outermost boundary of a given colour enclose regions over which galaxies having claimed redshifts beyond the lower limit specified for a given colour are predicted by our lens model to be multiply lensed.  Green circles and pink triangles are the same as Fig.\,\ref{fig:RGB_lens_img}.  Solid white and yellow shapes, both with adjacent redshift labels, enclose candidate high-$z$ galaxies found by \citet{Yan2023}, for which our lens model predicts should have image counterparts.  These image counterparts are predicted at locations enclosed by corresponding dashed white and yellow shapes: although those enclosed by dashed white shapes have predicted magnifications making them too dim to be detectable, those enclosed by dashed yellow shapes have predicted magnifications making them sufficiently bright to be detectable -- yet none are seen, as is more clearly shown in Figs.\,\ref{fig:F150DB-040}--\ref{fig:F200DB-045}.}
    \label{fig:SL_region}
\end{figure*}

The outermost boundary of the individual colour-shaded regions in Figure \ref{fig:SL_region} enclose sight-lines over which images of galaxies beyond a certain redshift (see legend for colour coding of redshifts) are predicted to have lensed counterparts.  As can be seen, the radii of these regions increase with increasing redshift, such that at $z \gtrsim 16$, all galaxies that form images within $\sim$$43^{\prime\prime}$ of the cluster centre should have lensed counterparts.  The increasing radii of these regions with increasing redshifts again reflects the geometry of gravitational lensing, whereby multiply-lensed images of galaxies at higher redshifts form configurations have larger angular separations in the sky.  These regions therefore also define locations whereby galaxies beyond a certain redshift are the most strongly lensed, making them both the largest and the most easily detectable.  Tolerances in the parameters of our lens model (given the measurement uncertainties in positions of the image counterparts used as constraints) translate to uncertainties in the boundary locations indicated in Figure \ref{fig:SL_region}, but these uncertainties are small -- amounting to an rms uncertainty of just $\sim$$0\farcs5$ no matter the redshift involved.


\subsection{Testing Claims for High-z Candidates}\label{sec:Yan_highz}
Six of the eighty-seven galaxies that \citet{Yan2023} propose may lie over the redshift range $10 \lesssim z_{\rm phot} \lesssim 20$ behind SMACS\,J0723.3-7327 have image locations in our lens finder map of Figure \ref{fig:SL_region} where they should be accompanied by lensed counterparts.  These galaxies are enclosed by either yellow shapes or white circles in Figure \ref{fig:SL_region}, and have redshifts determined from model fits to their SEDs ranging from $z_{\rm phot} \simeq 10.8$ to $z_{\rm phot} \simeq 20.4$.  As we will show, the three galaxies enclosed by yellow shapes, having $10.8 \lesssim z_{\rm phot} \lesssim 20.4$, are predicted by our lens model to have image counterparts sufficiently bright to be detectable above the (local) detection threshold (sometimes highly elevated in the close vicinity of a relatively bright foreground star or a galaxy).  Although image counterparts also are predicted for the three galaxies enclosed by white circles, their predicted lensing magnifications are either too low to make them detectable (if these galaxies are indeed multiply lensed) or their predicted image counterparts are hidden by the glare of a bright galaxy.

\begin{figure*}
    \centering
    \gridline{\fig{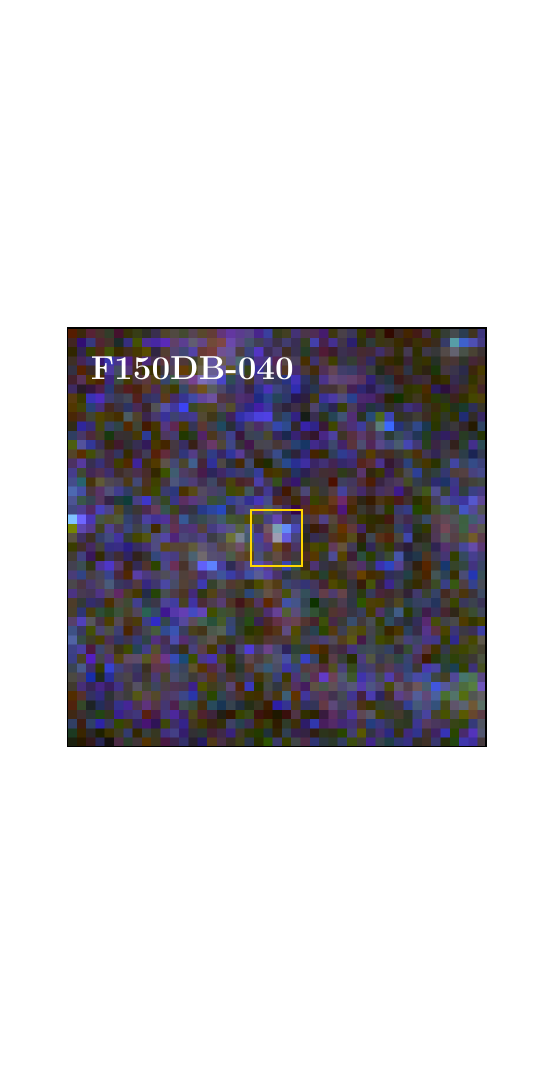}{0.18\textwidth}{}
          \fig{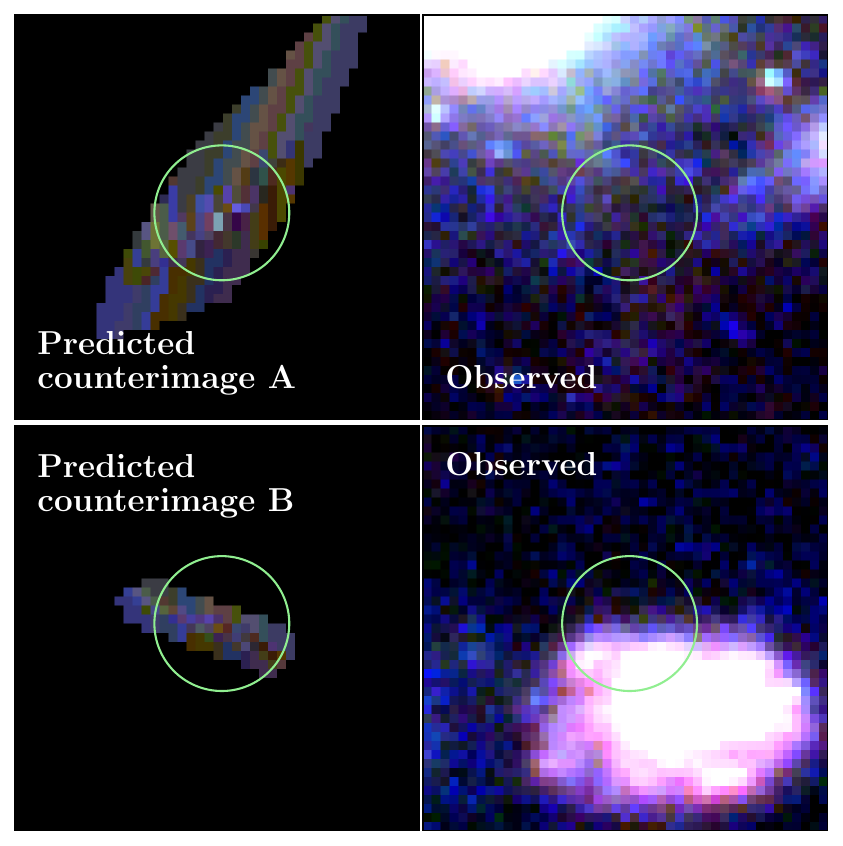}{0.36\textwidth}{}
          \fig{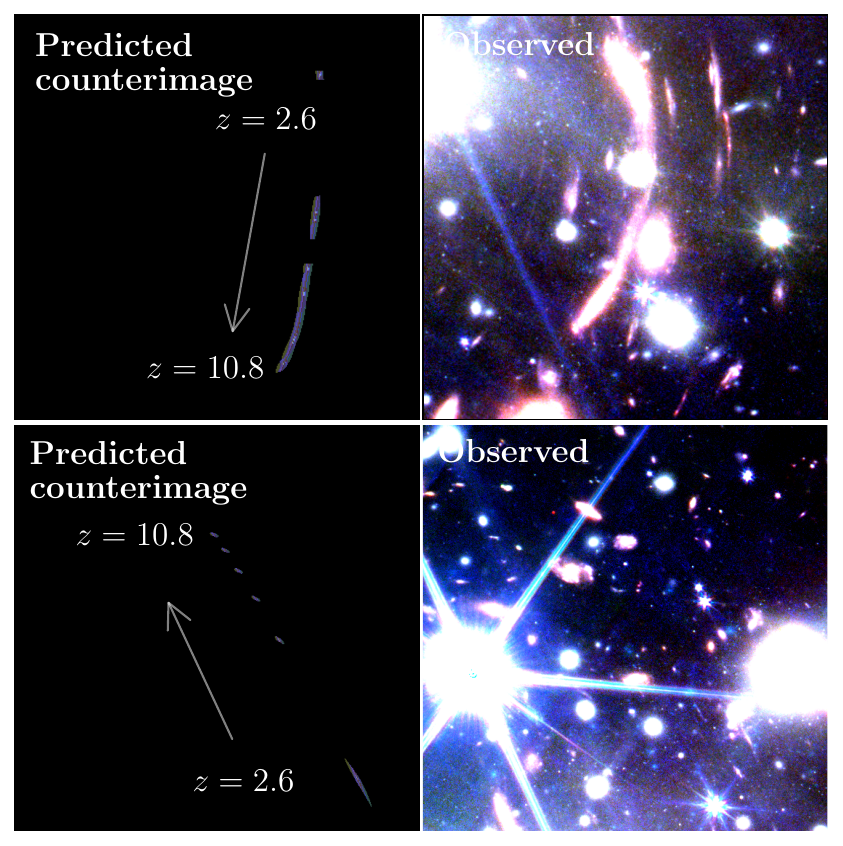}{0.36\textwidth}{}
          }
    \caption{Colour images constructed from the same JWST filters as in Fig.\,\ref{fig:RGB_lens_img}, displayed such that all panels have the same contrast.  Panel in first column shows F150DB-040 having $z_{\rm phot}=10.8$ as inferred by \citet{Yan2023}.  Orange box encloses region de-lensed and then re-lensed according to our lens model to generate image counterparts shown by panels in second column.  Panels in third column show corresponding observed fields, revealing none of the predicted image counterparts.  Green circle has a radius of $0\farcs45$, corresponding to the rms uncertainty in the predicted positions of image counterparts.  Panels in fourth column show predicted image counterparts at representative redshifts of $z = 2.6,3.5,4.6,6.1,8.1,10.8$, to be compared with panels in fifth column showing corresponding observed fields.  Once again no image counterparts are detected, implying that F150DB-040 is not multiply lensed and hence has $z_{\rm geo} \lesssim 2.5$ according to our lens model.}
    \label{fig:F150DB-040}
\end{figure*}

Figure \ref{fig:F150DB-040} shows the galaxy proposed by \citet{Yan2023} to be at $z_{\rm phot} \simeq 10.8$.  The panel in the first column shows its observed image, and the panels in the next two columns the model-predicted lensed counterparts and the corresponding observed fields, respectively, given its nominal redshift.  While the image counterpart with the lower-predicted lensing magnification (lower row) may escape detectability owing to its location next to a relatively bright galaxy resulting in a S/N ratio of only $2.0\sigma$, the image counterpart with the higher-predicted lensing magnification (upper row) ought to be easily detectable at a significance level of $29.3\sigma$.

\begin{figure*}
    \centering
    \gridline{\fig{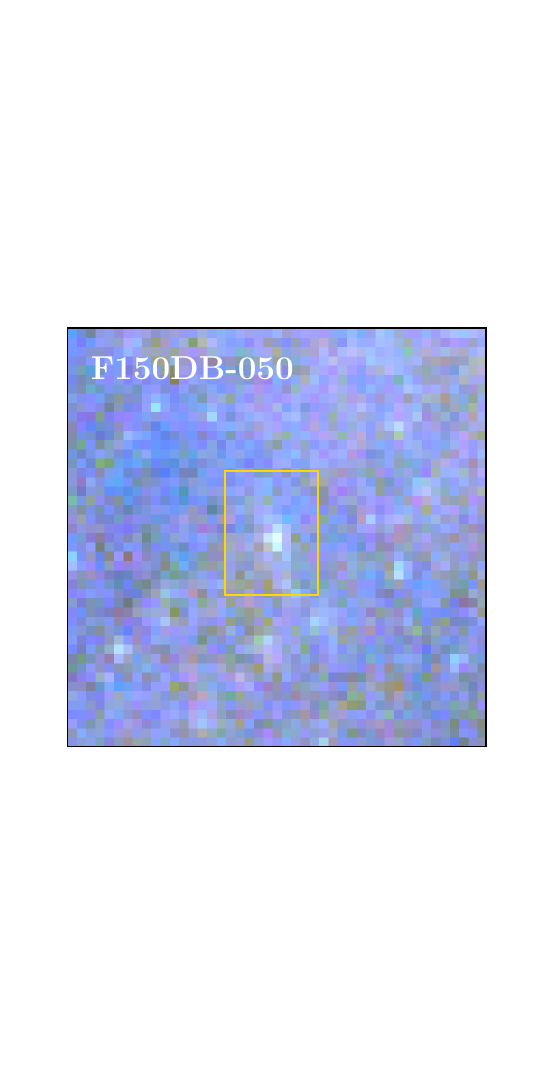}{0.18\textwidth}{}
          \fig{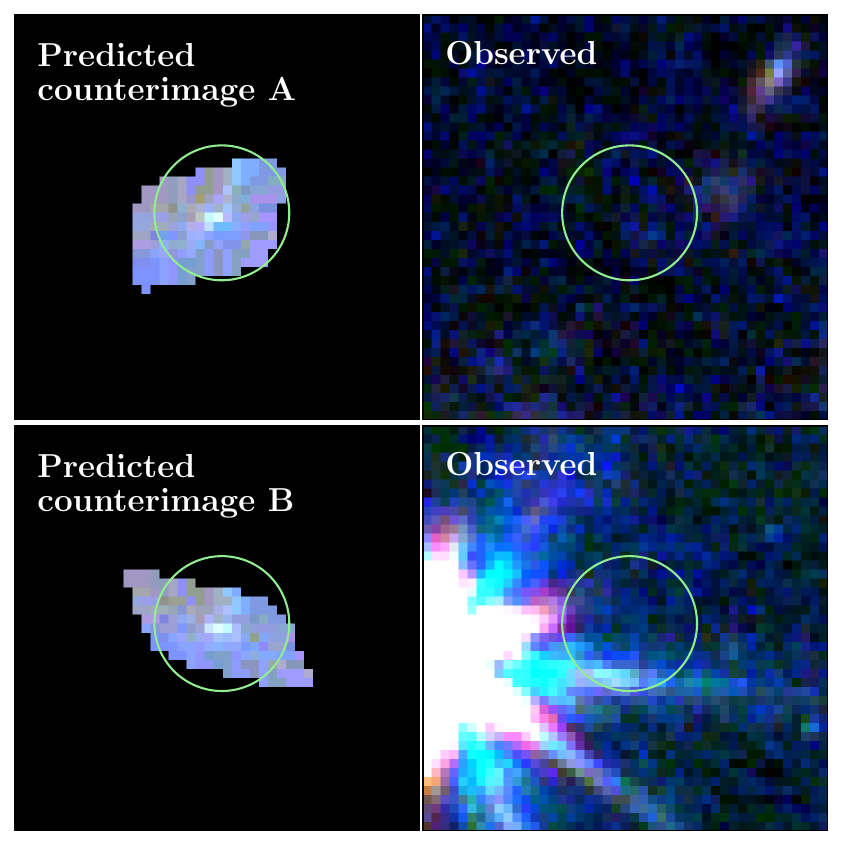}{0.36\textwidth}{}
          \fig{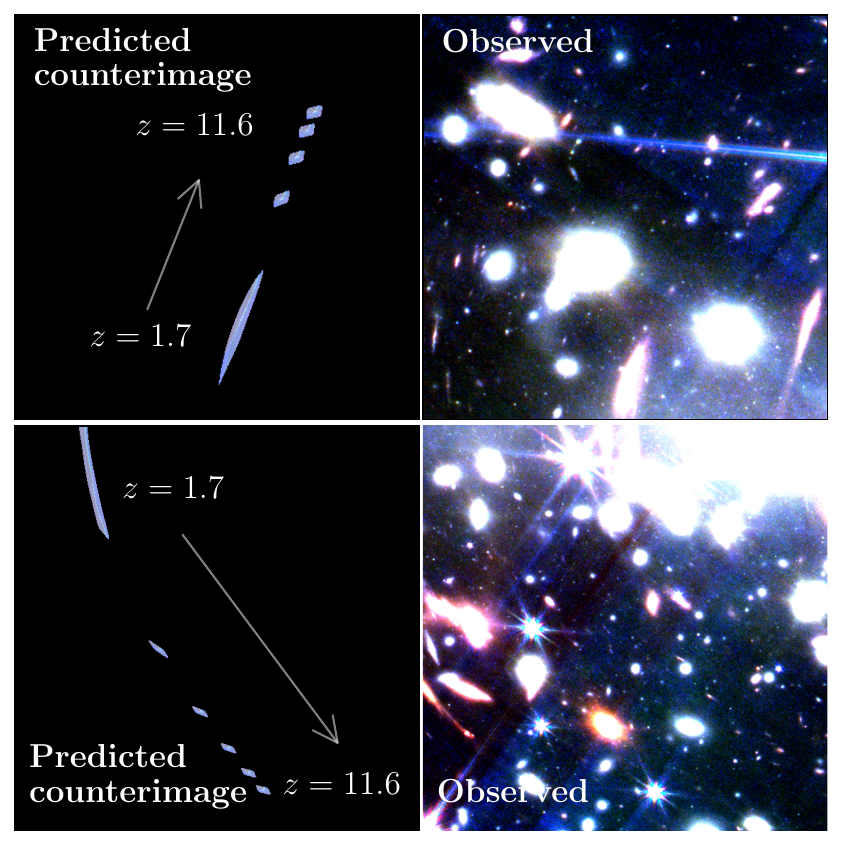}{0.36\textwidth}{}
          }
    \caption{Same as Fig.\,\ref{fig:F150DB-040} and with the same contrast, but now for F150DB-050 having $z_{\rm phot}=11.6$ as inferred by \citet{Yan2023}.  Panels in fourth column show predicted image counterparts at representative redshifts of $z = 1.7,2.5,3.7,5.4,7.9,11.6$.  Once again none of the predicted image counterparts are seen at its claimed $z_{\rm phot}$ or any of the $z$ considered as indicated by the panels in the fourth column.  The lack of image counterparts implies that F150DB-050 is not multiply lensed, and hence has $z_{\rm geo} \lesssim 1.6$ according to our lens model.} 
    \label{fig:F150DB-050}
\end{figure*}

\begin{figure*}
    \centering
    \gridline{\fig{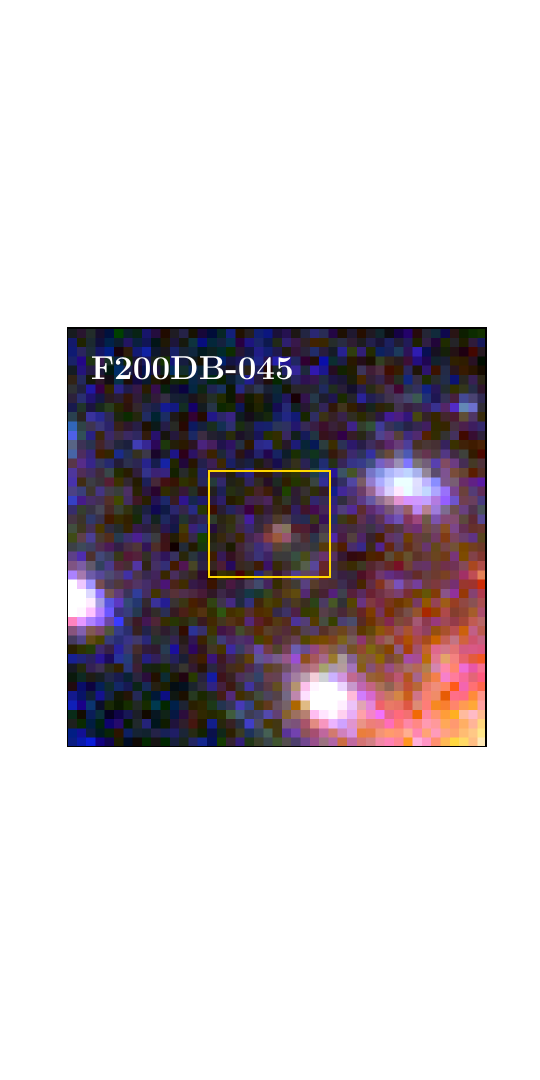}{0.18\textwidth}{}
          \fig{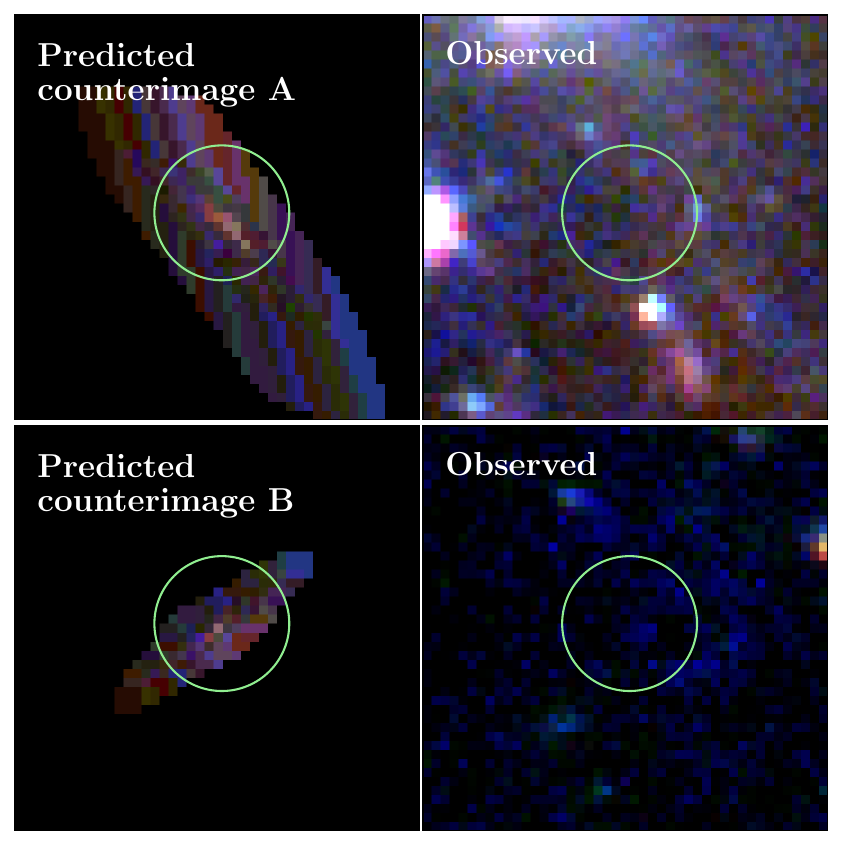}{0.36\textwidth}{}
          \fig{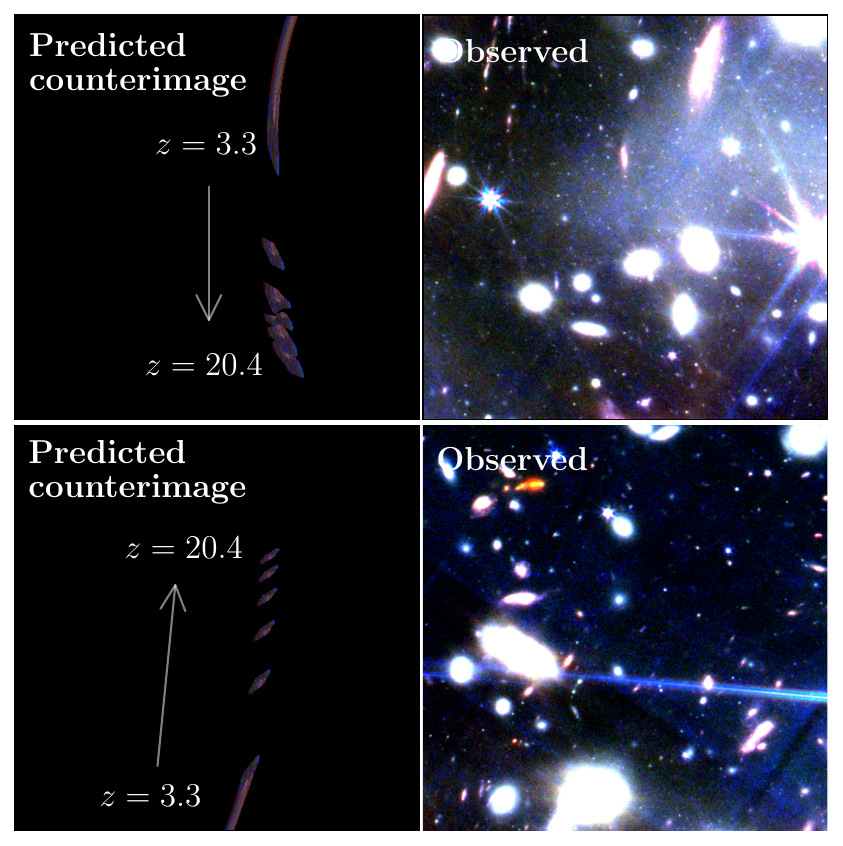}{0.36\textwidth}{}
          }
    \caption{Same as Fig.\,\ref{fig:F150DB-040} and with the same contrast, but now for F200DB-045 having $z_{\rm phot}=20.4$ as determined by \citet{Yan2023}.  Panels in fourth column show predicted image counterparts at representative redshifts of $z = 3.3,4.8,6.8,9.8,14.2,20.4$.  Yet again none of the predicted image counterparts are seen at its claimed $z_{\rm phot}$ or any of the $z$ considered as indicated by the panels in the fourth column.  The lack of image counterparts implies that F150DB-050 is not multiply lensed, and hence has $z_{\rm geo} \lesssim 3.2$ according to our lens model.}
    \label{fig:F200DB-045}
\end{figure*}

Figure \ref{fig:F150DB-050} shows the corresponding situation for the galaxy inferred by \citet{Yan2023} to be at $z_{\rm phot} \simeq 11.6$.  Although one of the model-predicted image counterparts lies close to a relatively bright star (lower row), both ought to be easily detectable at significance levels of $13.9\sigma$ and $19.3 \sigma$.  Figure \ref{fig:F200DB-045} shows the corresponding situation for the galaxy proposed by \citet{Yan2023} to be at $z_{\rm phot} \simeq 20.4$.  In this case also, both its image counterparts ought to be easily detectable at significance levels of $36.3\sigma$ and $40.0\sigma$.  Table\,\ref{tab:Yan_counimg} lists the predicted image counterpart positions, magnifications, and apparent magnitude in the F356W band for all three aforementioned galaxies if at the redshifts claimed by \citet{Yan2023}.

\begin{deluxetable*}{ccccccc}
\tablewidth{0pt}
\tablecaption{Predicted Properties of Image Counterparts.}
\tablehead{\colhead{Object$^{(1)}$} & \colhead{$z_{\rm phot}^{(2)}$} & \colhead{Predicted$^{(3)}$} & \multicolumn{2}{c}{Position$^{(4)}$} & \colhead{$\mu$$^{(5)}$} & \colhead{$m_{\rm F356W}$$^{(6)}$} \\
\colhead{} & \colhead{} & \colhead{Counterimage} & \colhead{$\Delta \alpha$ ($\arcsec$)} & \colhead{$\Delta \delta$ ($\arcsec$)}} 
\startdata
\multirow{2}{*}{F150DB-040} & \multirow{2}{*}{$10.8$} & A & $7^{\rm h}23^{\rm m}12\fs4335$ & $-73^{\circ}27^{\prime}33\farcs687$  & $22.84\pm2.03$  &  $ 27.28 \pm 0.04$\\
 & & B & $7^{\rm h}23^{\rm m}15\fs4849$ & $-73^{\circ}26^{\prime}43\farcs618$  & $5.22\pm0.05$  &  $ 28.88 \pm  0.54$\\
 \hline
\multirow{2}{*}{F150DB-050} & \multirow{2}{*}{$11.6$} & A & $7^{\rm h}23^{\rm m}23\fs1242$ & $-73^{\circ}26^{\prime}50\farcs779$  & $6.92\pm0.11$  &  $ 27.7 \pm 0.06$\\
 & & B & $7^{\rm h}23^{\rm m}21\fs0651$ & $-73^{\circ}27^{\prime}47\farcs191$  & $6.04\pm0.07$  & $ 27.84 \pm 0.08 $ \\
 \hline
\multirow{2}{*}{F200DB-045} & \multirow{2}{*}{$20.4$}& A & $7^{\rm h}23^{\rm m}24\fs6798$ & $-73^{\circ}27^{\prime}26\farcs592$  & $18.82\pm0.66$  &  $ 26.47 \pm 0.03 $\\
 &  & B & $7^{\rm h}23^{\rm m}23\fs7545$ & $-73^{\circ}26^{\prime}44\farcs352$  & $5.08\pm0.04$  &  $ 27.89 \pm  0.03$
\enddata
\label{tab:Yan_counimg}
\tablecomments{(1) I.D. of the high-$z$ candidates reported by \citet{Yan2023}. (2) Photometric redshift derived by \citet{Yan2023}. (3) Labelled according to Figs.\,\ref{fig:F150DB-040}--\ref{fig:F200DB-045}. (4) Sky position. (5) Lensing magnification. (6) Apparent magnitude in F356W band.}
\end{deluxetable*}

The lack of detectable image counterparts at the locations predicted by our lens model implies that, provided our lens model is reliable, these galaxies cannot be at their claimed redshifts. To allow for errors in their redshifts, we have checked for image counterparts at different redshifts to the lowest values for which these galaxies are predicted to be multiply lensed.  The panels in the fourth and fifth columns of Figs.\,\ref{fig:F150DB-040}--\ref{fig:F200DB-045} show, respectively, the model-predicted lensed counterparts and the corresponding observed fields at the redshifts indicated and as specified in the captions for these figures.  As is clearly evident, no image counterparts are seen despite being predicted to be sufficiently magnified to be detectable.

Our lens finder map can be used to place upper limits on the redshifts of these galaxies, for which we find form only single images.  For F150DB-040 (claimed to be at $z_{\rm phot} \simeq 10.8$) as shown in Figure \ref{fig:F150DB-040}, our lens model constrains this galaxy to be at $z_{\rm geo} \lesssim 2.5$ so as not to produce multiple images.  For F150DB-050 (claimed to be at $z_{\rm phot} \simeq 11.4$) as shown in Figure \ref{fig:F150DB-050}, the same consideration requires this galaxy to be at $z_{\rm geo} \lesssim 1.6$.  For F200DB-045 (claimed to be at $z_{\rm phot} \simeq 20.4$) as shown in Figure \ref{fig:F200DB-045}, this consideration places the galaxy at $z_{\rm geo} \lesssim 3.2$.  \citet{Adams2023} is able to successfully fit the measured SED of F200DB-045 with a dusty stellar population at $z_{\rm phot}=0.7$, consistent with the upper limit inferred geometrically from our lens model.

\begin{figure*}
    \centering
    \includegraphics[width=\textwidth]{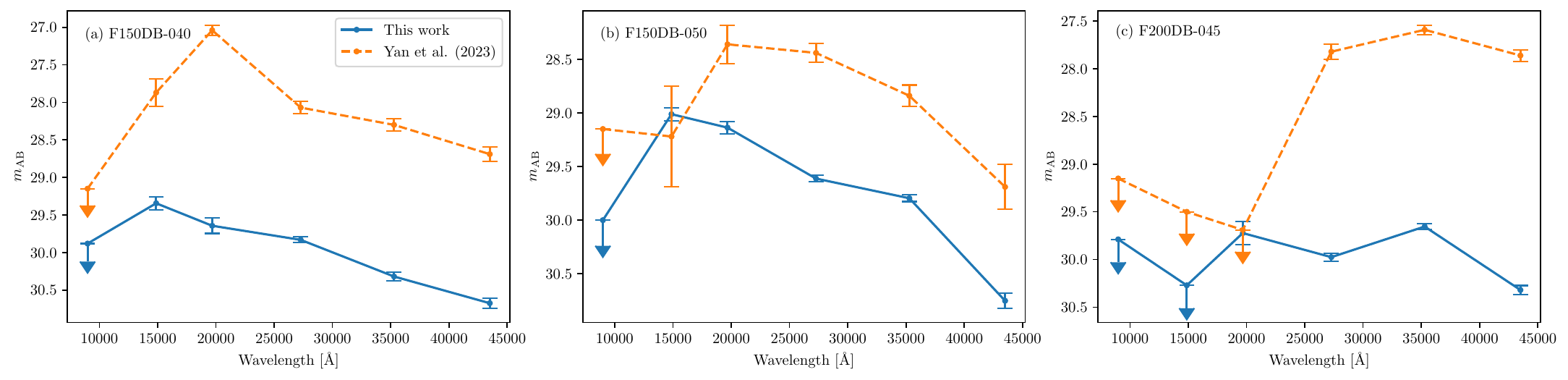}
    \caption{SEDs from the JWST for (a) F150DB-040 ($z_{\rm phot} = 10.8$), (b) F150DB-050 ($z_{\rm phot} = 11.6$), and (c) F200DB-045 ($z_{\rm phot} = 20.4$) as retrieved from \citet{Yan2023} (orange) and as we determined based on our reprocessed images (blue).  In our reprocessed images, neither F150DB-040 nor F200DB-045 show a statistically significant drop in brightness towards shorter wavelengths in filters where they are not detectable, therefore providing no direct evidence for a Lyman break.}
    \label{fig:Yan_SED}
\end{figure*}

Apart from \citet{Adams2023}, a number of other studies have pointed to the degeneracy between dust extinction in relatively low-$z$ galaxies and the Lyman-break of purported high-$z$ galaxies when fitting synthetic stellar populations to SEDs having relatively poor S/N ratios \citep{Zavala2023, Finkelstein2022, Donnan2023, Castellano2022}.  Figure \ref{fig:Yan_SED} shows the measured SEDs of the three galaxies that are the subjects of Figures\,\ref{fig:F150DB-040}--\ref{fig:F200DB-045}.  We plot in orange their SEDs as measured by \citet{Yan2023}, and in blue their SEDs as measured from the reprocessed images employed in our work (see Section \ref{subsec:JWST}).  In our reprocessed images, the measured SED of F150DB-050 displays at best a relatively weak drop in brightness of $0.99 \pm 0.06$\,mag between the second-shortest wavelength filter (at which the object is detectable) and the shortest-wavelength filter (at which the object is no longer detectable) in which images were taken, hardly strong enough to be convincingly attributed to the Lyman break.  For the other two galaxies, no statistically significant change in brightness can be claimed between the shortest-wavelength filter in which the object is detected and the next shortest wavelength filter(s) in which the object is not -- and hence no evidence for a Lyman break.  For completeness, we note that we fitted model SEDs to the measured SEDs of all three galaxies, but did not find these fits to provide meaningful measures of either the stellar properties or redshifts of these galaxies given the quality of their measured SEDs.  For all these galaxies, the lack of lensed image counterparts provides the strongest evidence that they cannot lie at $z_{\rm phot} \gtrsim 10$, but instead must lie at $z_{\rm geo} \lesssim 1.6$--3.2.

\section{Summary and Conclusions}\label{sec:summary}
The explosive growth of candidate high-$z$ galaxies discovered in images taken by the JWST, less than a year after the first scientific image from this telescope was released, stresses cosmological models for structure formation if these galaxies have correctly inferred stellar masses at their purported redshifts.  Spectroscopic measurements are required to confirm their redshifts, but such measurements are unlikely to keep up with the number of candidate high-$z$ galaxies discovered, and may not be practical for very faint galaxies given the competition for observing time on the JWST.   Other methods need to be developed to reliably weed out likely low-$z$ interlopers as well as to identify the best candidates for follow-up spectroscopy.

In this paper, we described a method to test whether candidate high-$z$ galaxies lying behind galaxy clusters are actually at their claimed redshifts, and if not then determine upper limits on their redshifts.  This method relies on constructing a reliable lens model (corresponding to the projected surface mass density) for the galaxy cluster in the foreground of the candidate high-$z$ galaxies -- in this case SMACS\,J0723.3–7327, the subject of the first scientific image released from the JWST.  Our lens model for SMACS\,J0723.3–7327 is anchored on five spectroscopically-confirmed systems at $1.38 \leq z_{\rm spec} \leq 2.21$ that are multiply lensed, along with twelve other systems with proposed image counterparts having common colours and spectral energy distributions, shared morphologies, but unknown redshifts.  Constrained only by their image positions and, where available, redshifts, we solve for a lens model simultaneously with the source positions of all these systems together with the redshifts of the twelve systems with unknown redshifts.  For the latter, we determine precise geometric redshifts (which depend only on the lensing geometry) spanning the range $1.4 \lesssim z_{\rm geo} \lesssim 6.7$. 

We show that our lens model is able to reproduce the positions of the multiply-lensed images used as constraints to an rms dispersion of $0\farcs5$, a test of its internal consistency.  As a second test of its internal consistency, we show that the model-predicted configurations of the multiply-lensed images at their spectroscopic or geometrically-determined redshifts ($z_{\rm geo}$) are consistent with those predicted given the adopted cosmology.   As tests of its predicability, our lens model is able to correctly reproduce the relative brightnesses and individual morphologies of the multiply-lensed images used as constraints.  Such tests are crucial to establish the reliability of lens models, and yet are not always conducted in their entirety in published lens model.

From this lens model, we create a lens finder map that defines regions over which galaxies beyond a certain redshift are predicted to be multiply lensed.  Applying this map to three galaxies claimed to be at $10 \lesssim z_{\rm phot} \lesssim 20$ by \citet{Yan2023}, we find no image counterparts at locations where they ought to be sufficiently magnified to be detectable.  Indeed, we find no image counterparts at redshifts down to the lowest values at which our lens model predicts these galaxies to be multiply lensed -- implying that they have $z_{\rm geo} \lesssim 1.6$--3.2.  In lieu of spectroscopy, the creation of reliable lens finder maps for cluster fields can therefore be used to test the claimed redshifts of candidate high-$z$ galaxies, and when not supported either correctly identify or place upper limits on their redshifts.  We urge the lensing community to create, after checking on the reliability of their lens models, their own lens finder maps for SMACS\,J0723.3–7327 to test the accuracy of our results, as well as lens finder maps for other clusters to check on the claimed redshifts of candidate high-$z$ galaxies behind those clusters.

\begin{acknowledgments}  
A. C., S. K. L., T. B., and J. L. acknowledges support from the Research Grants Council of Hong Kong through the Collaborative Research Fund C6017-20G.  M. C. A. L. is partially supported by Research Grants Council of Hong Kong through the Collaborative Research Fund C7015-19G. J. N. is partially supported by Research Grants Council of Hong Kong through the General Research Fund 201910159253. R. A. W. acknowledges support from NASA JWST Interdisciplinary Scientist grants NAG5-12460, NNX14AN10G and 80NSSC18K0200 from GSFC. This research employed observations made with the NASA/ESA/CSA James Webb Space Telescope as well as NASA/ESA Hubble Space Telescope, and made use of archival data from the Mikulski Archive for Space Telescope (MAST), which is a collaboration between the Space Telescope Science Institute (STScI/NASA), the Space Telescope European Coordinating Facility (ST-ECF/ESAC/ESA) and the Canadian Astronomy Data Centre (CADC/NRC/CSA). The specific JWST observations analyzed can be accessed via \dataset[https://doi.org/10.17909/txm7-h277]{https://doi.org/10.17909/txm7-h277}. This research has also made use of services of ESO Science Archive Facility. We thank the referee for constructive suggestions to improve the manuscript.
\end{acknowledgments}

\facilities{HST(ACS and WFC3), JWST (NIRCam), VLT (MUSE)}

\software{JWST calibration pipeline \citep{Bushouse2023},
           \texttt{BAGPIPES} \citep{Carnall2018},
           \texttt{imfit} \citep{Erwin2015},
           \texttt{glafic} \citep{oguri2010,Oguri2021},
           \texttt{NumPy} \citep{numpy},
           \texttt{Matplotlib} \citep{matplotlib},
           \texttt{Astropy} \citep{astropy:2013,astropy:2018,astropy:2022},
           \texttt{SciPy} \citep{scipy},
           \texttt{pandas} \citep{pandas},
           \texttt{PIL} \citep{PIL}
                    }

\bibliography{sample631}{}
\bibliographystyle{aasjournal}

\appendix
\section{Predicted vs. Observed Morphology of Image Counterparts} \label{sec:appen_A}

\begin{figure}
    \centering
        \gridline{\hspace{-1em} \fig{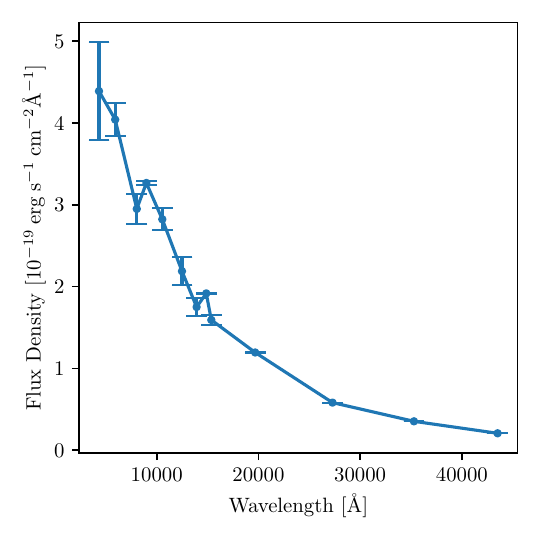}{0.14\textwidth}{} \hspace{-3.75em}
          \fig{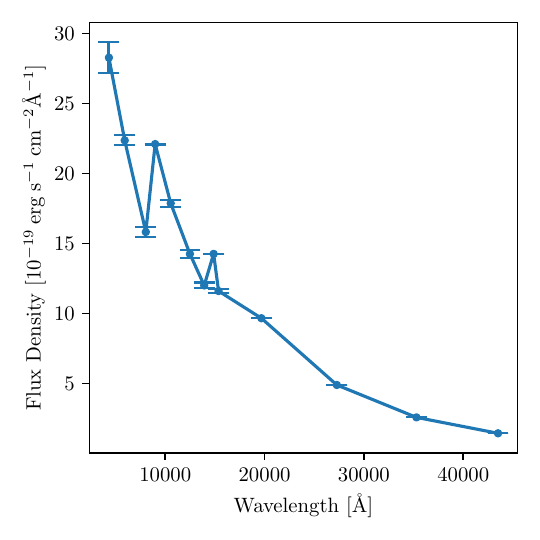}{0.14\textwidth}{} \hspace{-3.75em}
          \fig{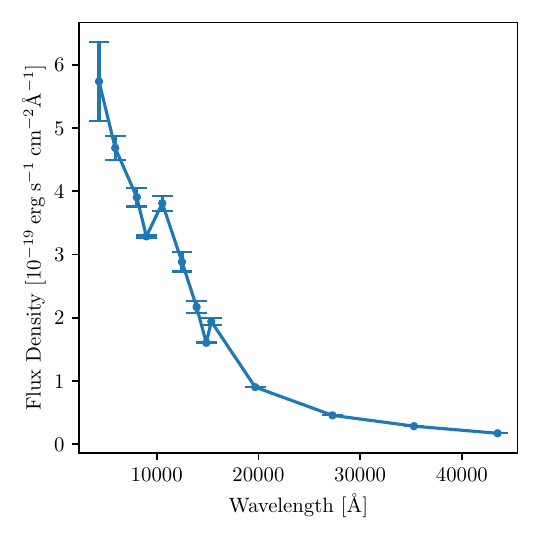}{0.14\textwidth}{}
          }
    \vspace{-3em}
    \gridline{\fig{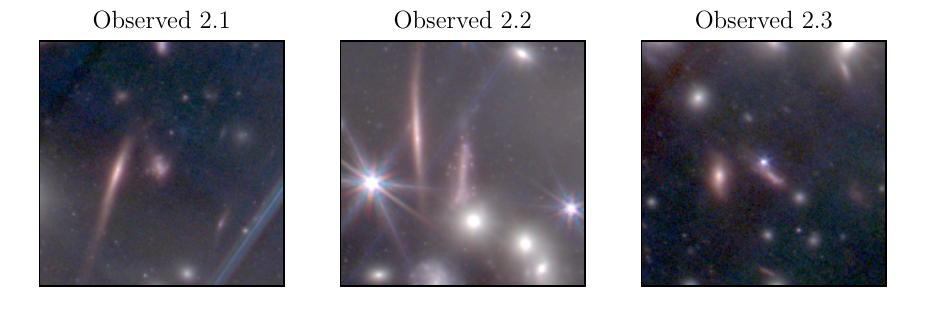}{0.45\textwidth}{}}
    \vspace{-3em}
    \gridline{\fig{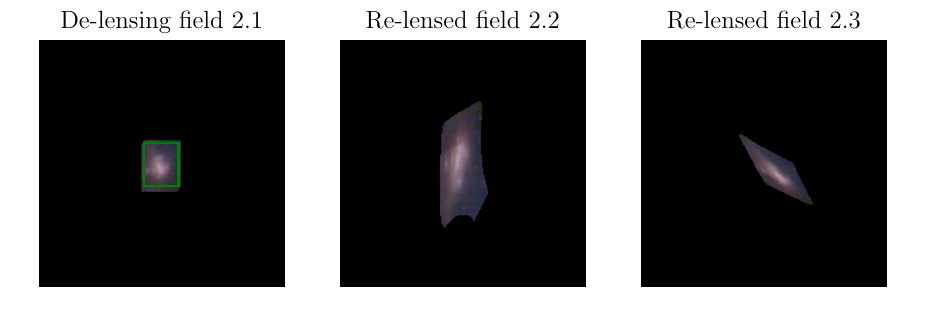}{0.45\textwidth}{}}
    \vspace{-3em}
    \gridline{\fig{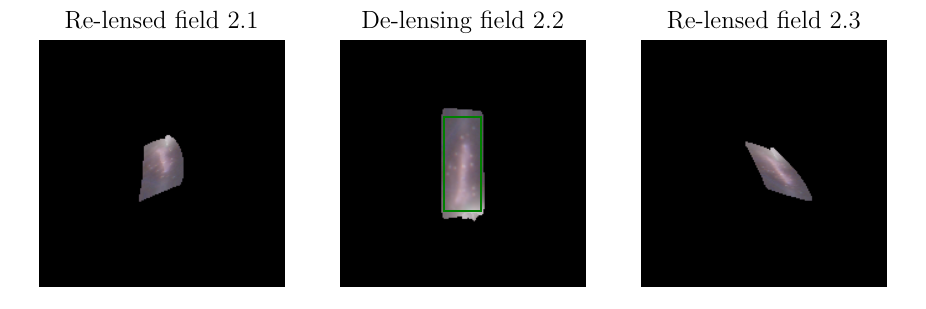}{0.45\textwidth}{}}
    \vspace{-3em}
    \gridline{\fig{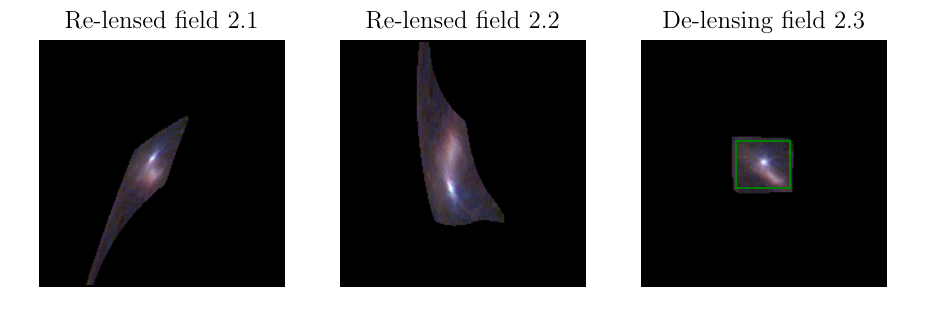}{0.45\textwidth}{}}
    \caption{Same as Fig.\,\ref{fig:relens_sys1}, but for system 2.}
    \label{fig:relens_sys2}
\end{figure}

\begin{figure}
    \centering
        \gridline{\hspace{-0.75em} \fig{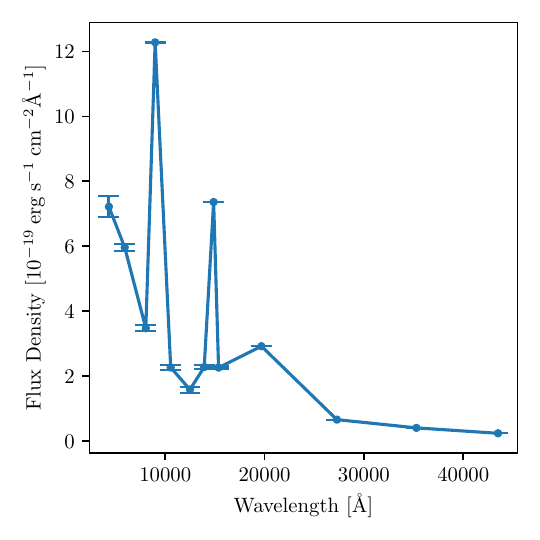}{0.105\textwidth}{} \hspace{-3.25em}
          \fig{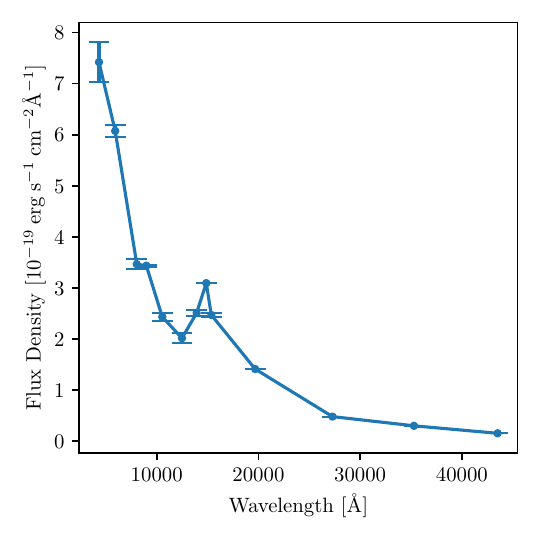}{0.105\textwidth}{} \hspace{-3.25em}
          \fig{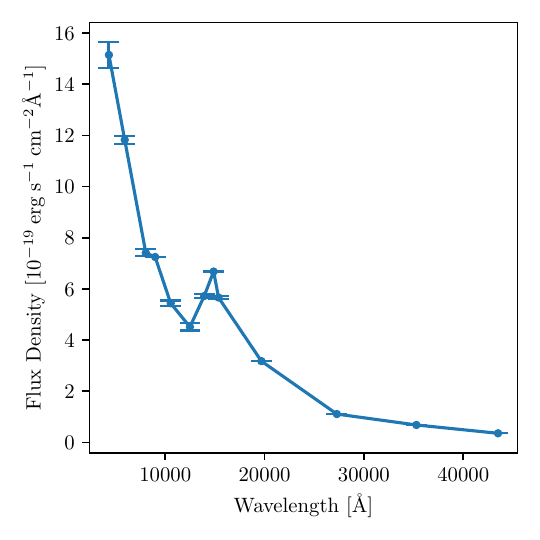}{0.105\textwidth}{} \hspace{-3.25em}
          \fig{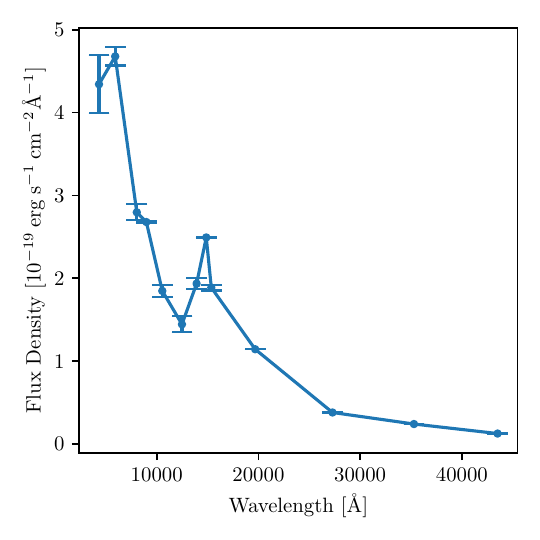}{0.105\textwidth}{}
          }
    \vspace{-3em}
    \gridline{\fig{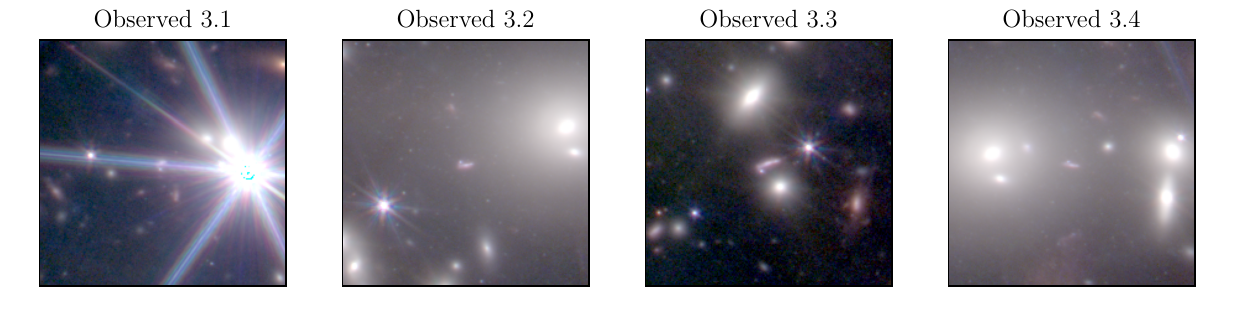}{0.45\textwidth}{}}
    \vspace{-3em}
    \gridline{\fig{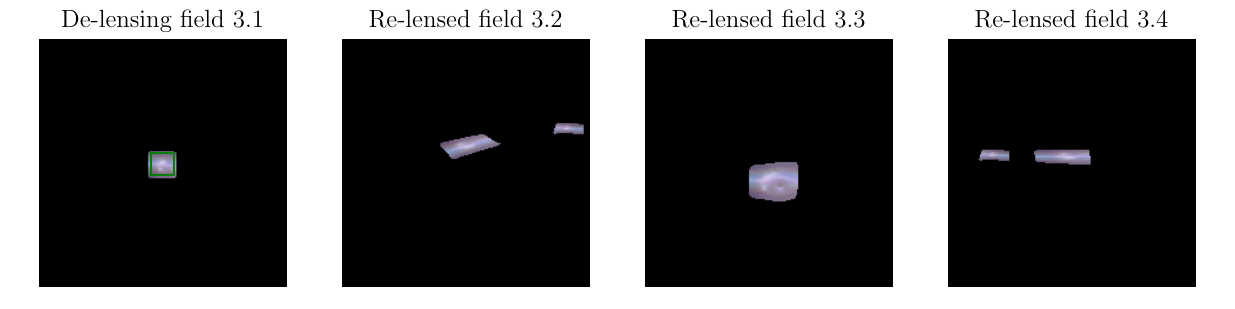}{0.45\textwidth}{}}
    \vspace{-3em}
    \gridline{\fig{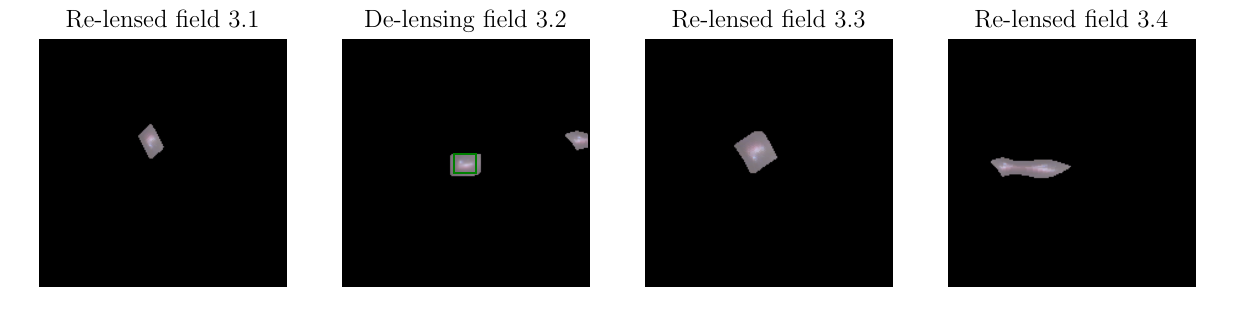}{0.45\textwidth}{}}
    \vspace{-3em}
    \gridline{\fig{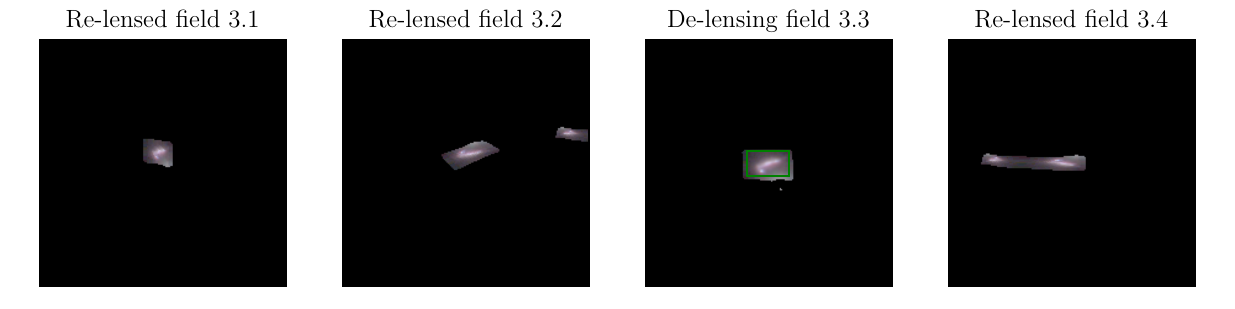}{0.45\textwidth}{}}
    \vspace{-3em}
    \gridline{\fig{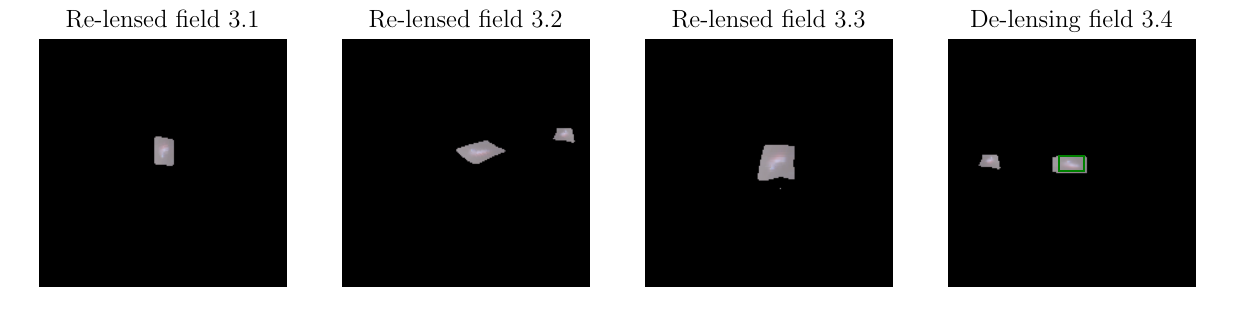}{0.45\textwidth}{}}
    \caption{Same as Fig.\,\ref{fig:relens_sys1}, but for system 3.}
    \label{fig:relens_sys3}
\end{figure}

\begin{figure}
    \centering
        \gridline{\hspace{-1em} \fig{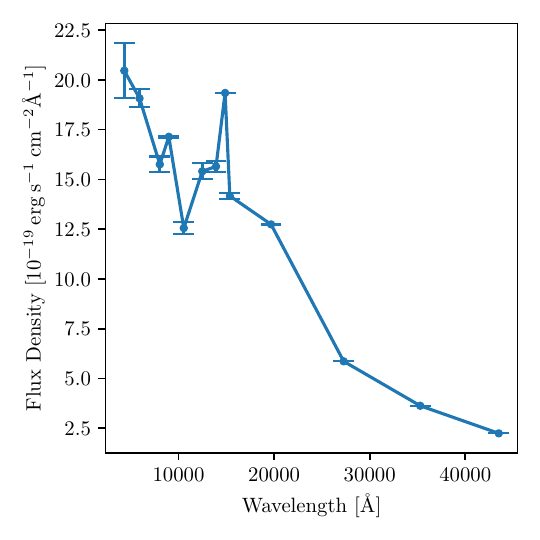}{0.14\textwidth}{} \hspace{-3.75em}
          \fig{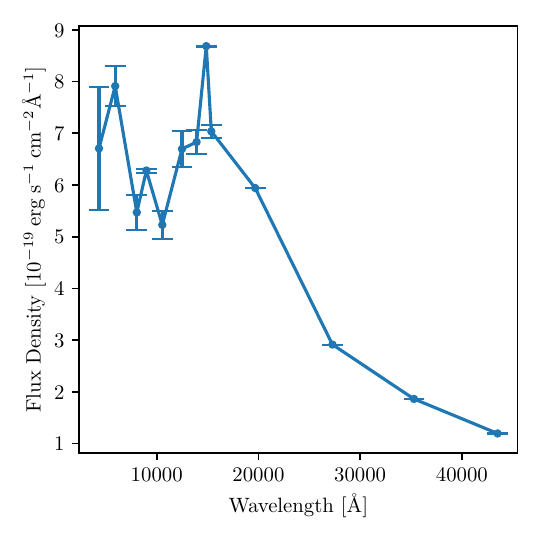}{0.14\textwidth}{} \hspace{-3.75em}
          \fig{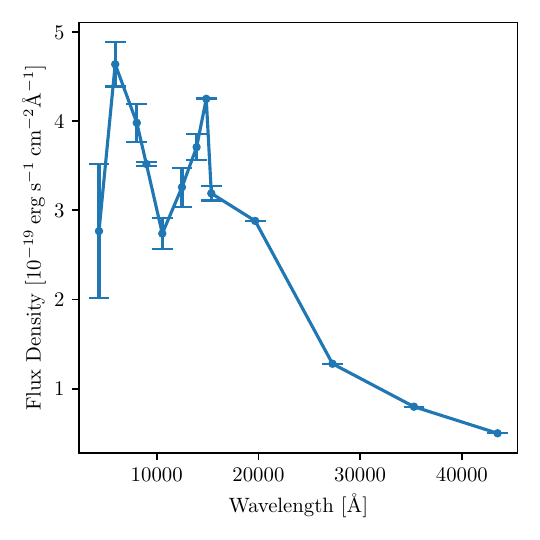}{0.14\textwidth}{}
          }
    \vspace{-3em}
    \gridline{\fig{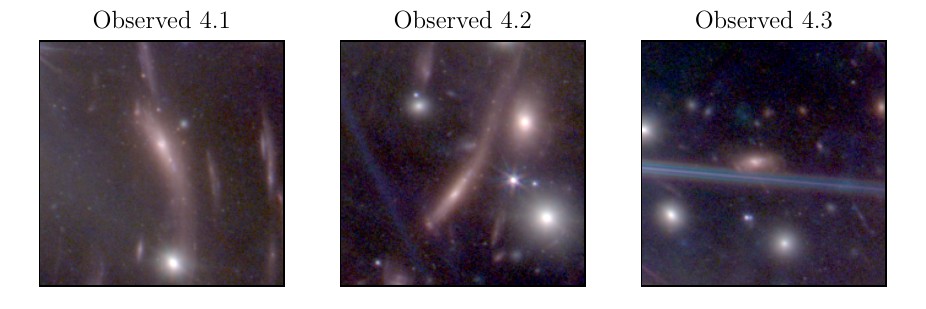}{0.45\textwidth}{}}
    \vspace{-3em}
    \gridline{\fig{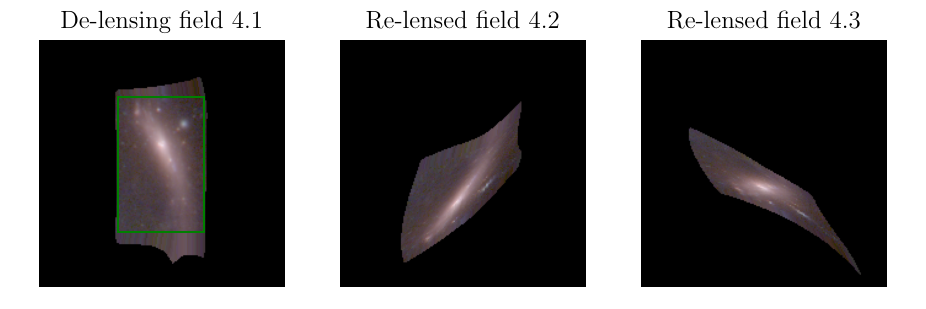}{0.45\textwidth}{}}
    \vspace{-3em}
    \gridline{\fig{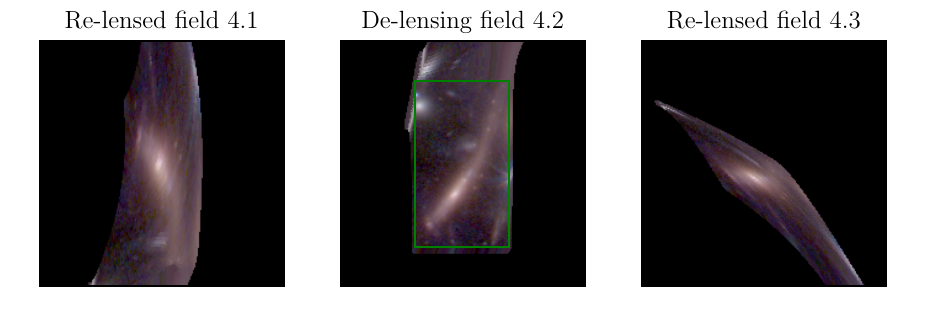}{0.45\textwidth}{}}
    \vspace{-3em}
    \gridline{\fig{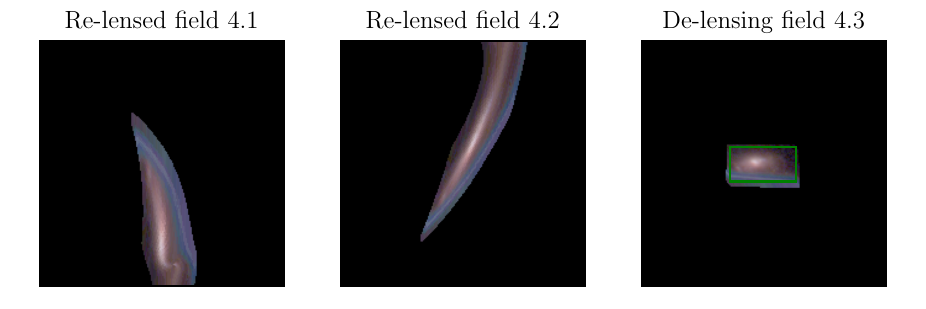}{0.45\textwidth}{}}
    \caption{Same as Fig.\,\ref{fig:relens_sys1}, but for system 4.}
    \label{fig:relens_sys4}
\end{figure}

\begin{figure}
    \centering
        \gridline{\hspace{-1em} \fig{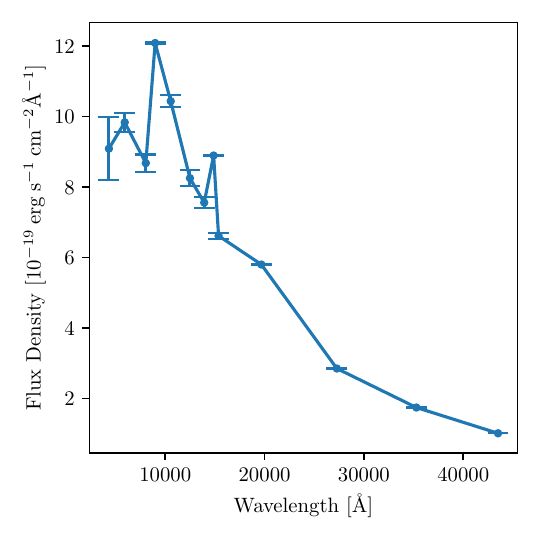}{0.14\textwidth}{} \hspace{-3.75em}
          \fig{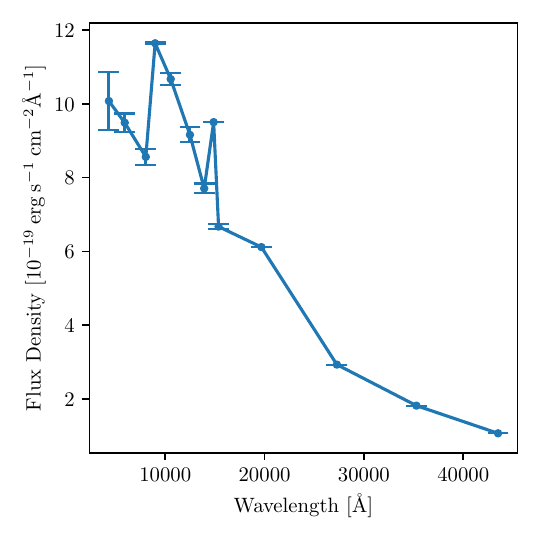}{0.14\textwidth}{} \hspace{-3.75em}
          \fig{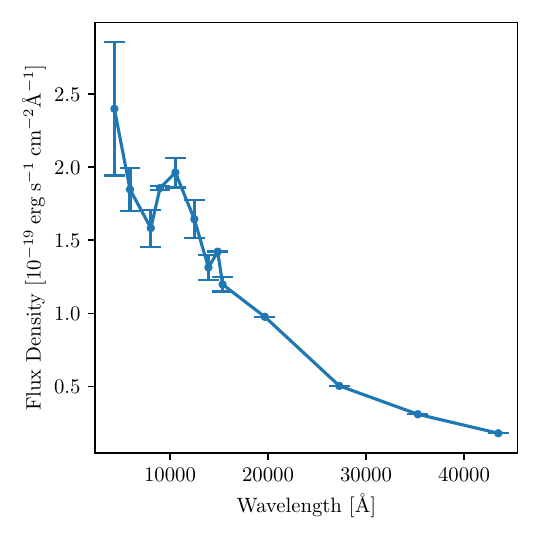}{0.14\textwidth}{}
          }
    \vspace{-3em}
    \gridline{\fig{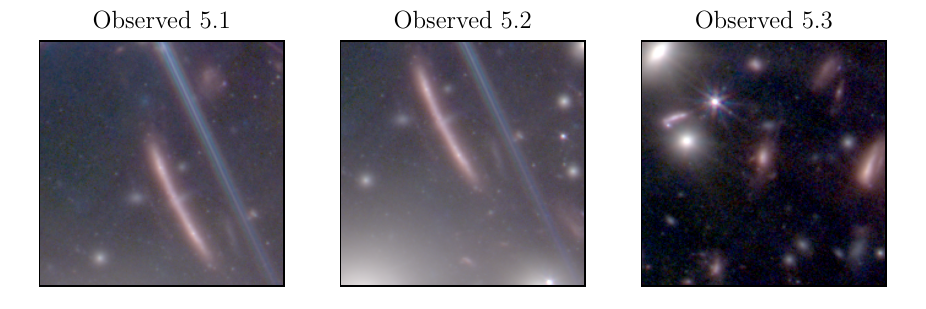}{0.45\textwidth}{}}
    \vspace{-3em}
    \gridline{\fig{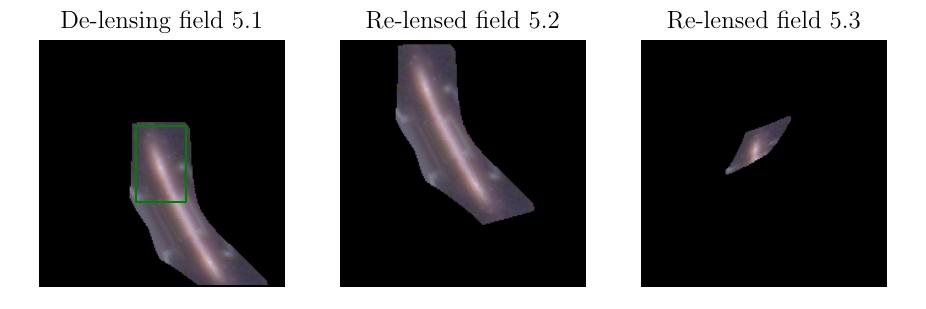}{0.45\textwidth}{}}
    \vspace{-3em}
    \gridline{\fig{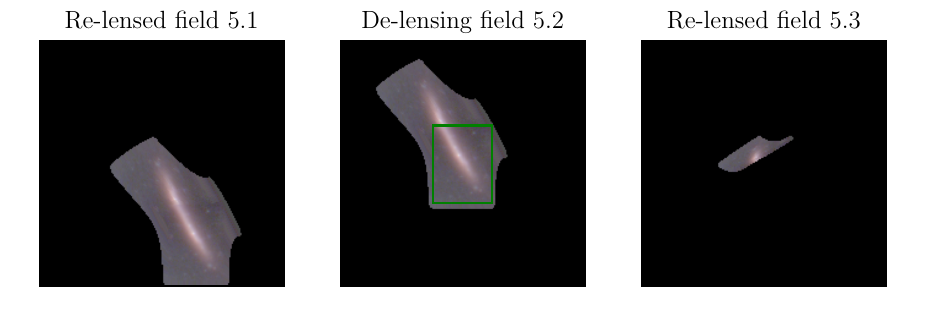}{0.45\textwidth}{}}
    \vspace{-3em}
    \gridline{\fig{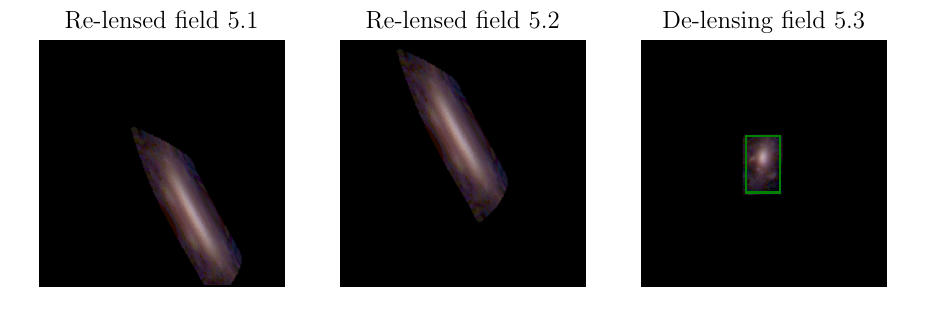}{0.45\textwidth}{}}
    \caption{Same as Fig.\,\ref{fig:relens_sys1}, but for system 5.}
    \label{fig:relens_sys5}
\end{figure}

\begin{figure}
    \centering
        \gridline{\hspace{-1em} \fig{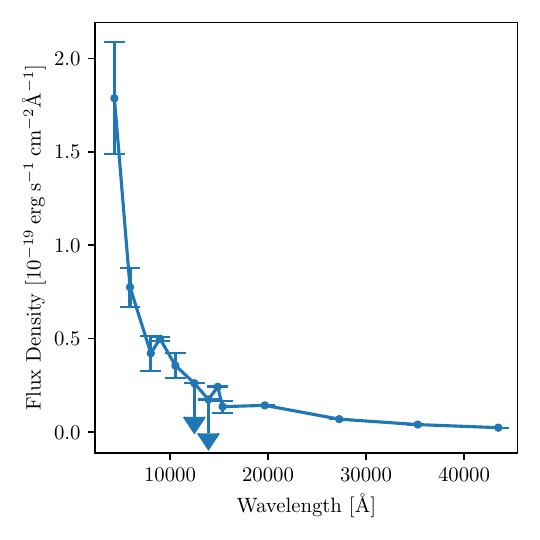}{0.14\textwidth}{} \hspace{-3.75em}
          \fig{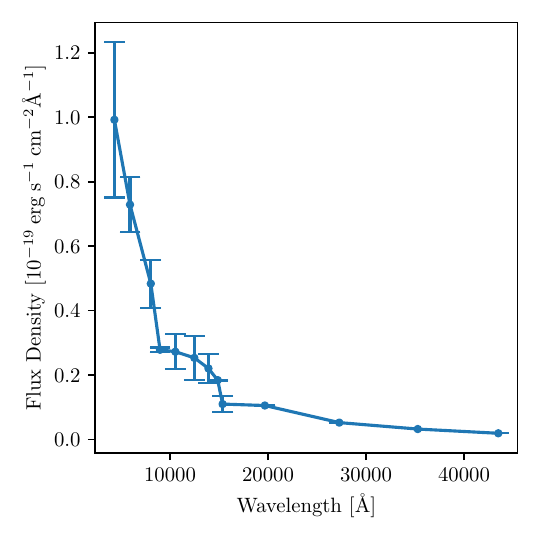}{0.14\textwidth}{} \hspace{-3.75em}
          \fig{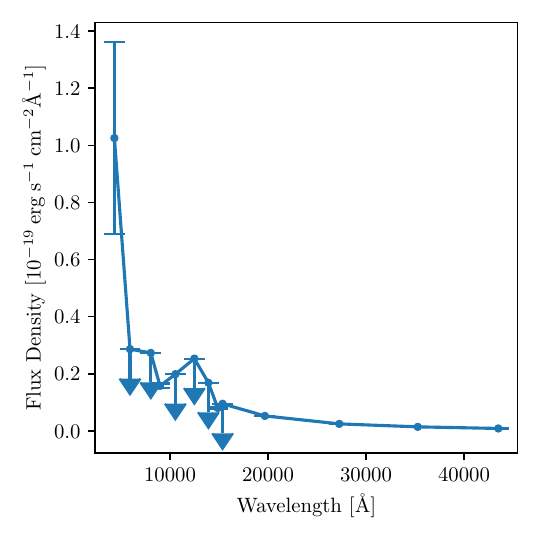}{0.14\textwidth}{}
          }
    \vspace{-3em}
    \gridline{\fig{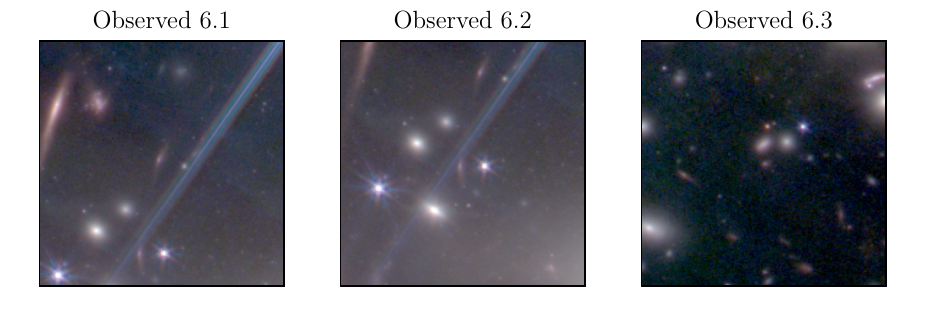}{0.45\textwidth}{}}
    \vspace{-3em}
    \gridline{\fig{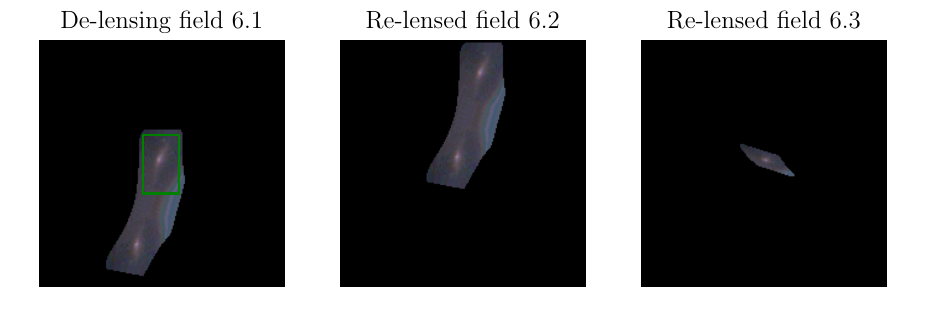}{0.45\textwidth}{}}
    \vspace{-3em}
    \gridline{\fig{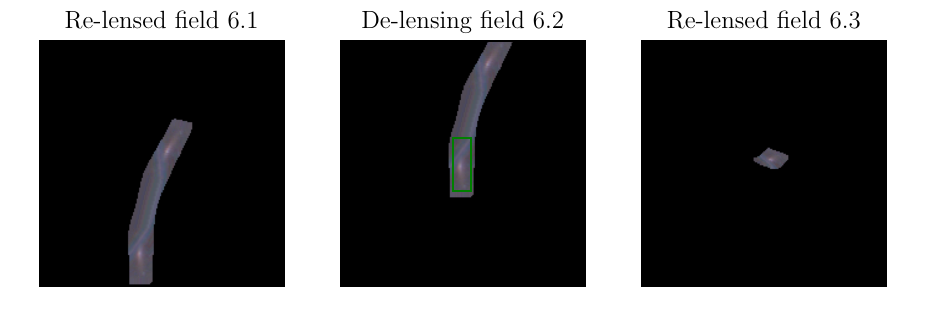}{0.45\textwidth}{}}
    \vspace{-3em}
    \gridline{\fig{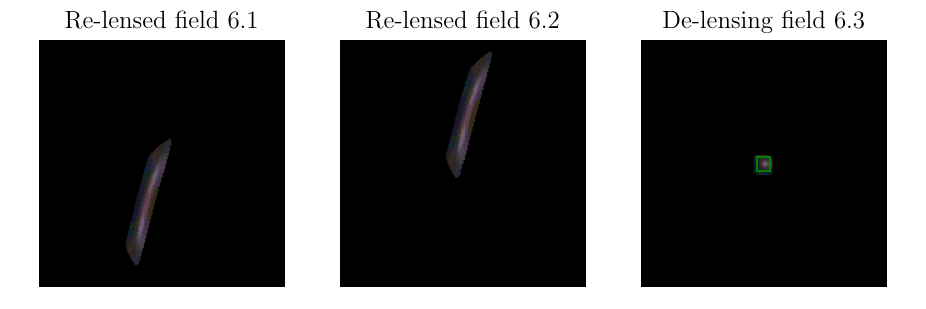}{0.45\textwidth}{}}
    \caption{Same as Fig.\,\ref{fig:relens_sys1}, but for system 6.}
    \label{fig:relens_sys6}
\end{figure}

\begin{figure}
    \centering
        \gridline{\hspace{-1em} \fig{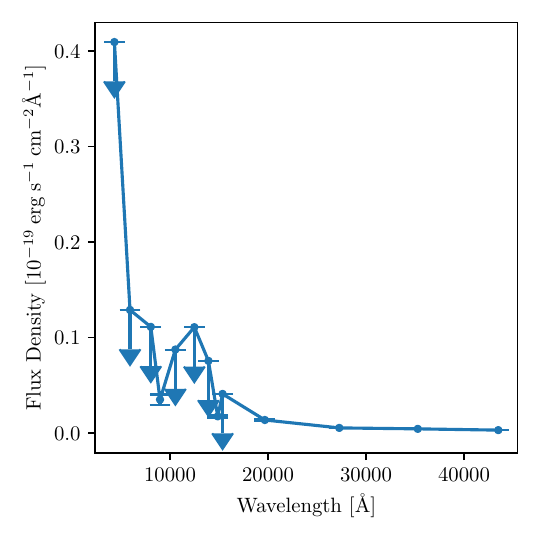}{0.14\textwidth}{} \hspace{-3.75em}
          \fig{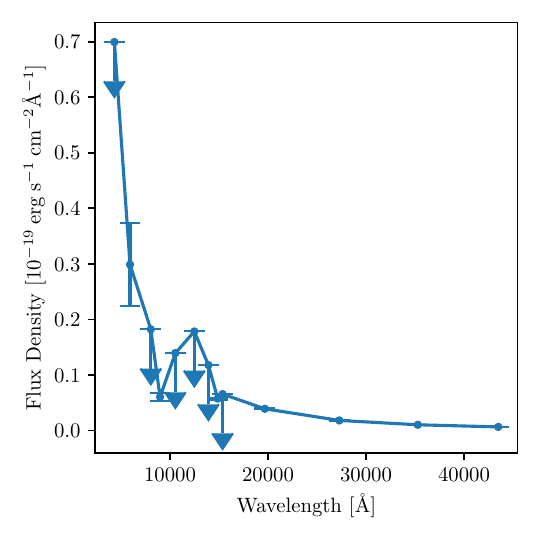}{0.14\textwidth}{} \hspace{-3.75em}
          \fig{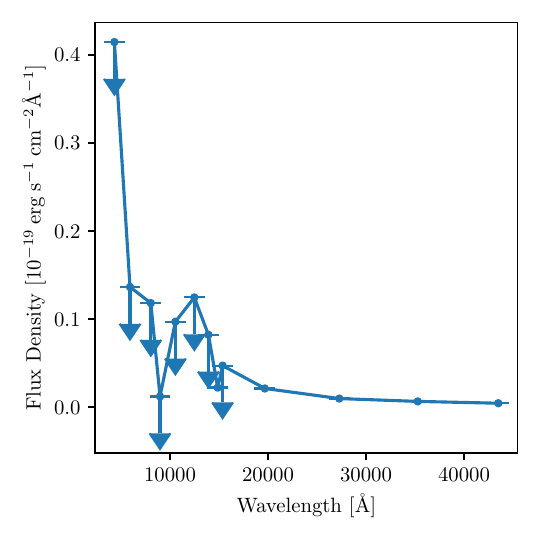}{0.14\textwidth}{}
          }
    \vspace{-3em}
    \gridline{\fig{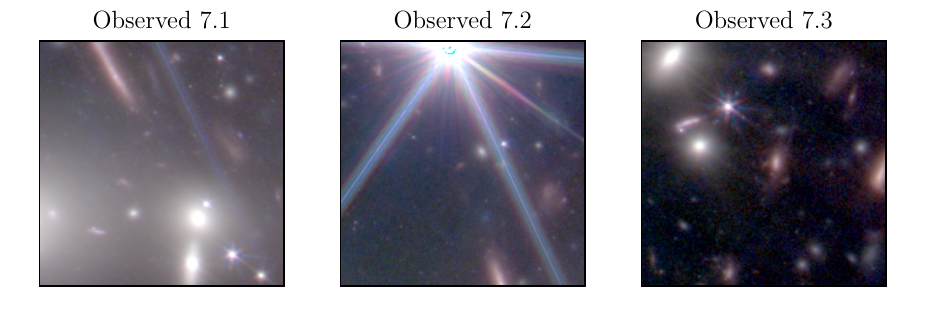}{0.45\textwidth}{}}
    \vspace{-3em}
    \gridline{\fig{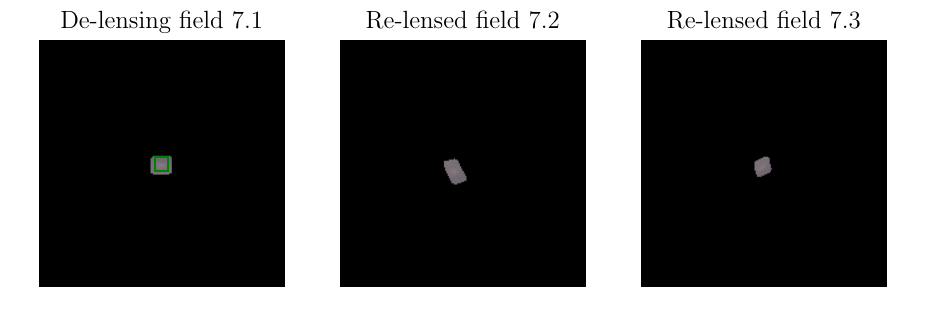}{0.45\textwidth}{}}
    \vspace{-3em}
    \gridline{\fig{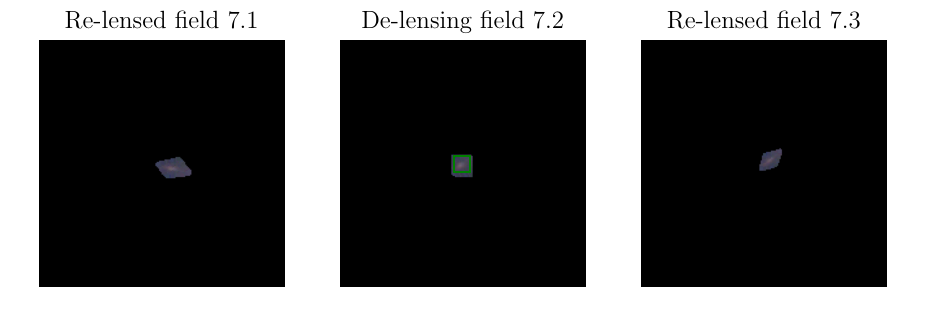}{0.45\textwidth}{}}
    \vspace{-3em}
    \gridline{\fig{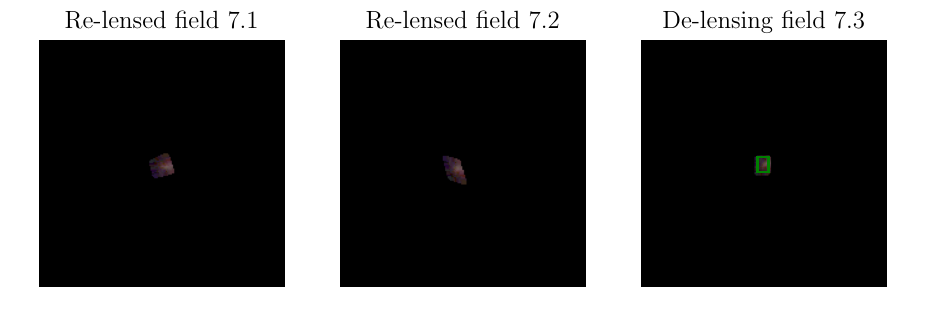}{0.45\textwidth}{}}
    \caption{Same as Fig.\,\ref{fig:relens_sys1}, but for system 7.}
    \label{fig:relens_sys7}
\end{figure}

\begin{figure}
    \centering
        \gridline{\hspace{-1em} \fig{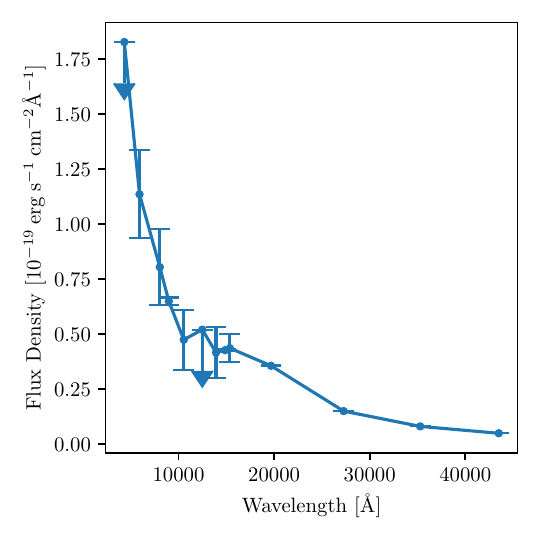}{0.14\textwidth}{} \hspace{-3.75em}
          \fig{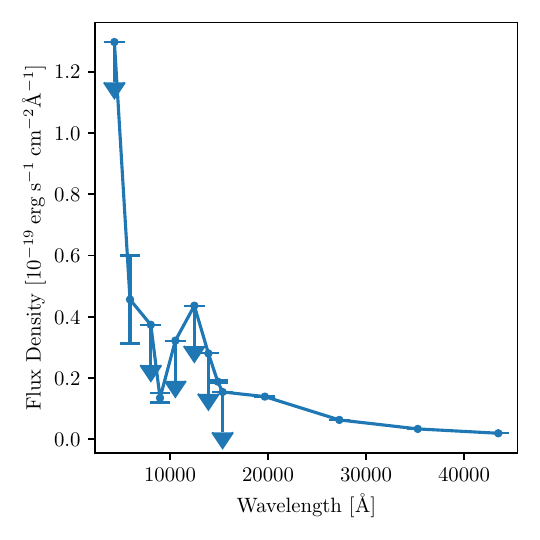}{0.14\textwidth}{} \hspace{-3.75em}
          \fig{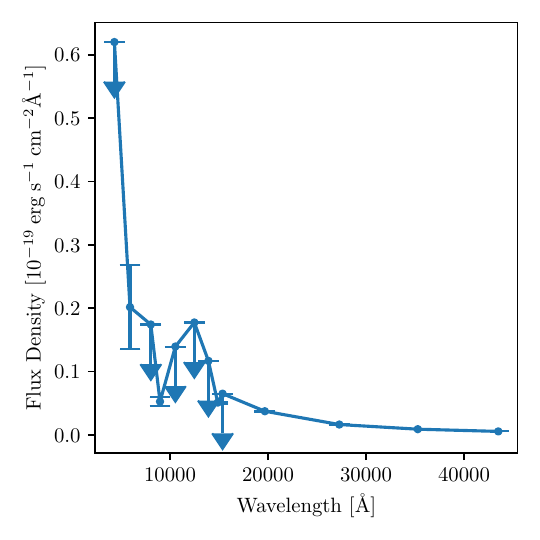}{0.14\textwidth}{}
          }
    \vspace{-3em}
    \gridline{\fig{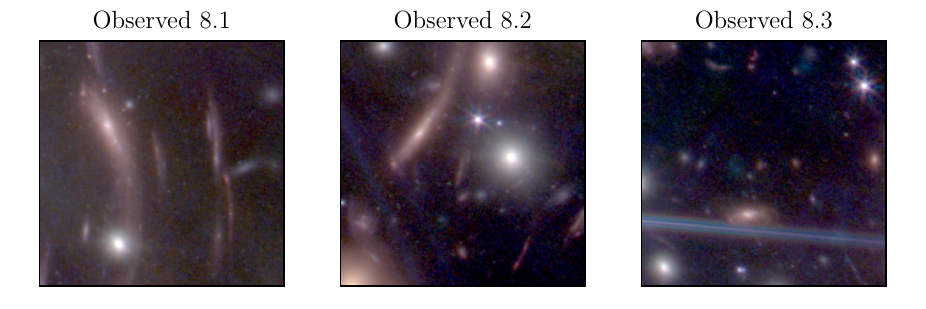}{0.45\textwidth}{}}
    \vspace{-3em}
    \gridline{\fig{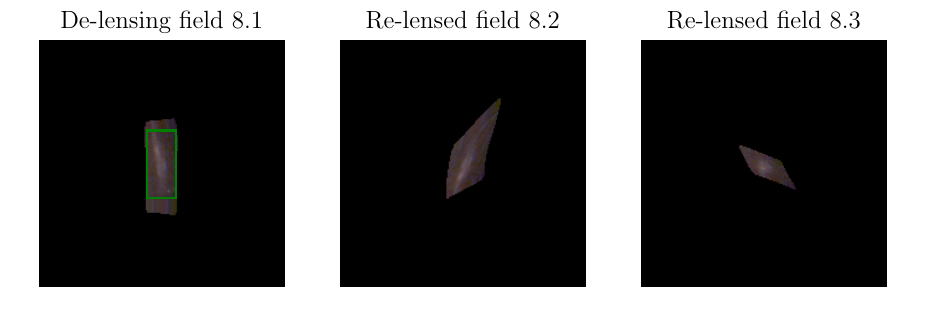}{0.45\textwidth}{}}
    \vspace{-3em}
    \gridline{\fig{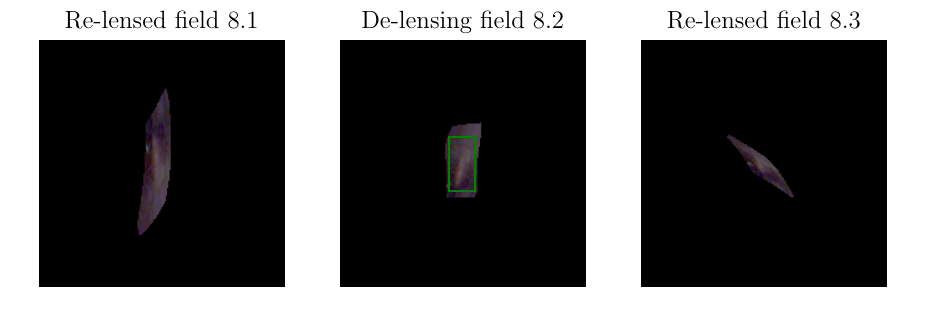}{0.45\textwidth}{}}
    \vspace{-3em}
    \gridline{\fig{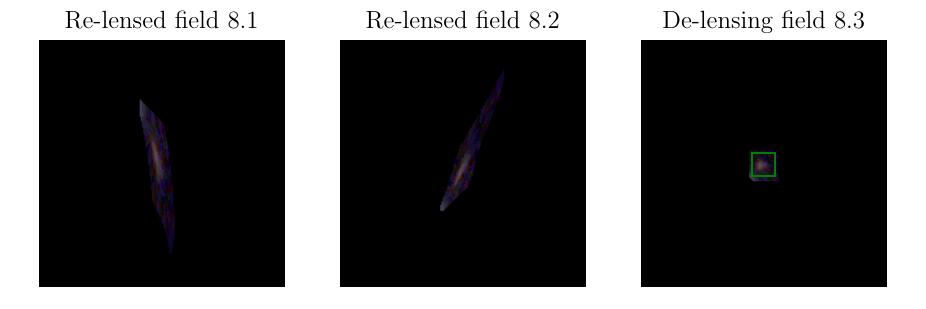}{0.45\textwidth}{}}
    \caption{Same as Fig.\,\ref{fig:relens_sys1}, but for system 8.}
    \label{fig:relens_sys8}
\end{figure}

\begin{figure}
    \centering
        \gridline{\hspace{-1em} \fig{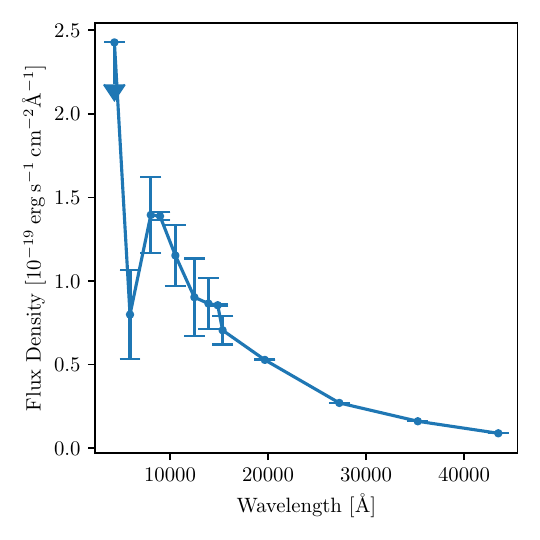}{0.14\textwidth}{} \hspace{-3.75em}
          \fig{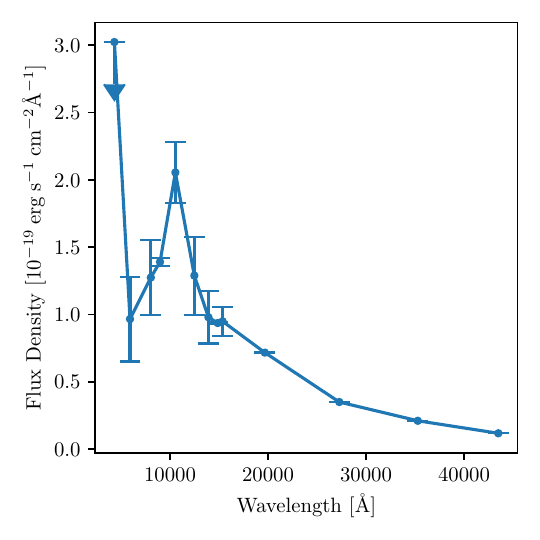}{0.14\textwidth}{} \hspace{-3.75em}
          \fig{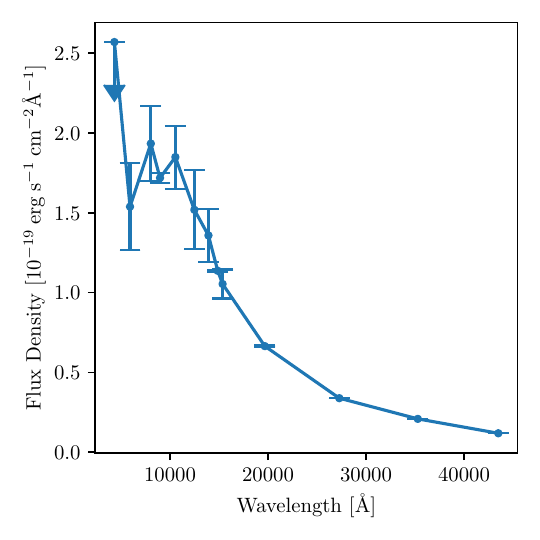}{0.14\textwidth}{}
          }
    \vspace{-3em}
    \gridline{\fig{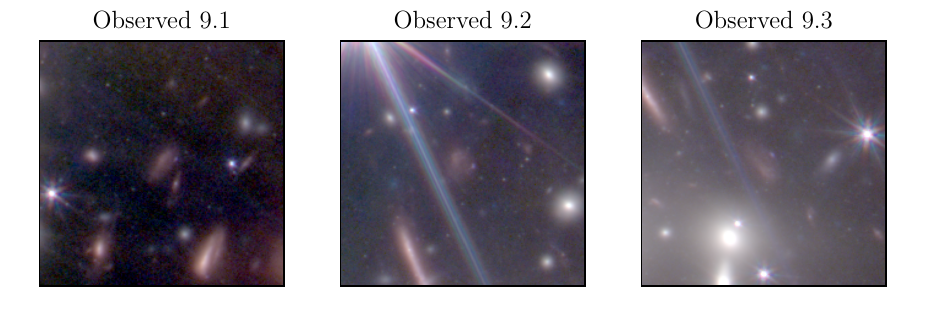}{0.45\textwidth}{}}
    \vspace{-3em}
    \gridline{\fig{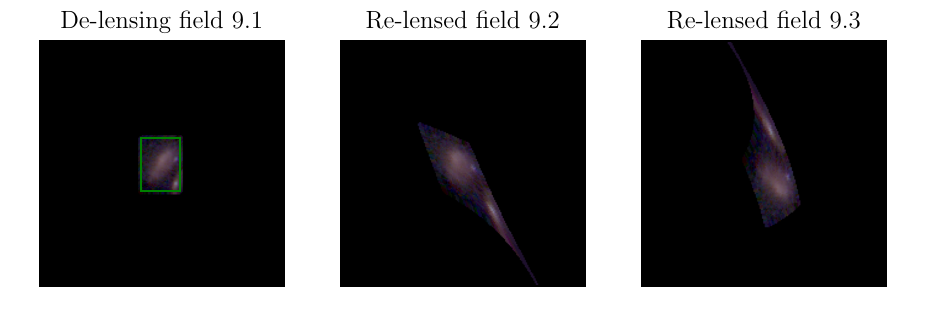}{0.45\textwidth}{}}
    \vspace{-3em}
    \gridline{\fig{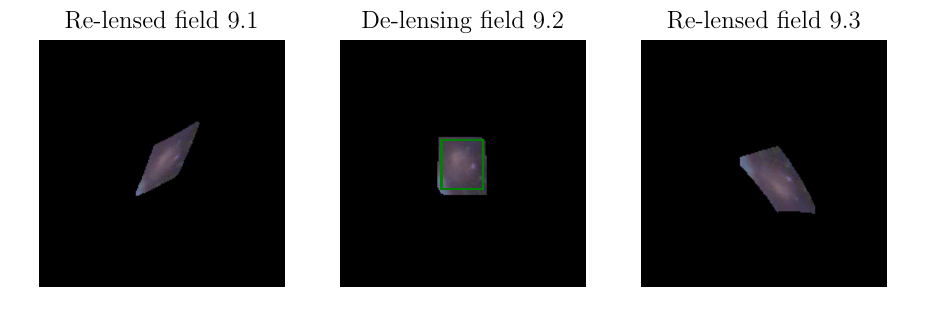}{0.45\textwidth}{}}
    \vspace{-3em}
    \gridline{\fig{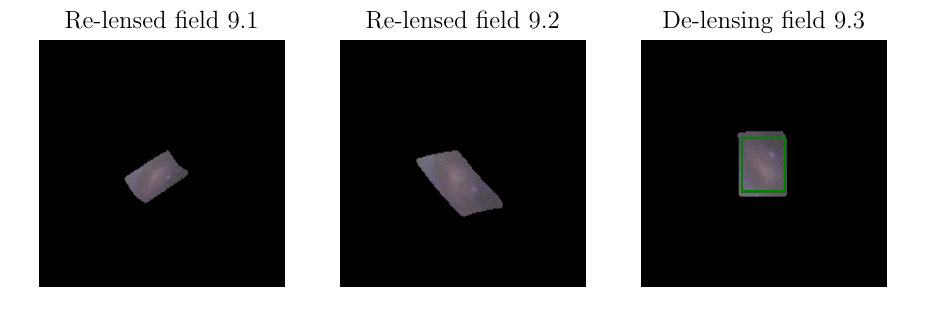}{0.45\textwidth}{}}
    \caption{Same as Fig.\,\ref{fig:relens_sys1}, but for system 9.}
    \label{fig:relens_sys9}
\end{figure}

\begin{figure}
    \centering
        \gridline{\hspace{-1em} \fig{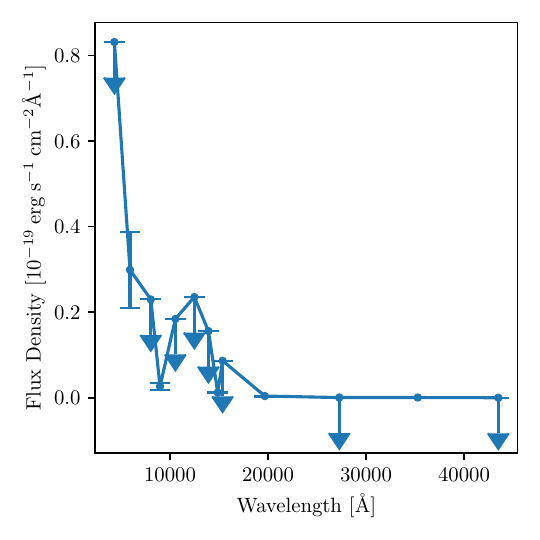}{0.14\textwidth}{} \hspace{-3.75em}
          \fig{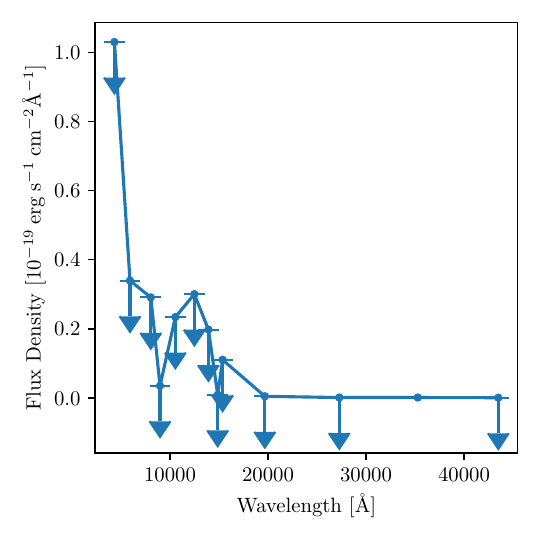}{0.14\textwidth}{} \hspace{-3.75em}
          \fig{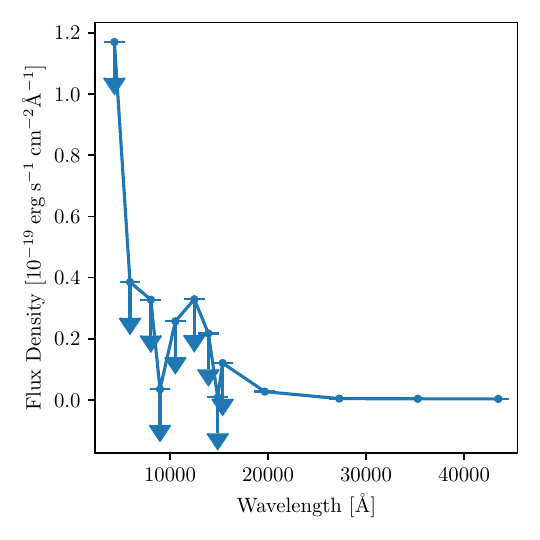}{0.14\textwidth}{}
          }
    \vspace{-3em}
    \gridline{\fig{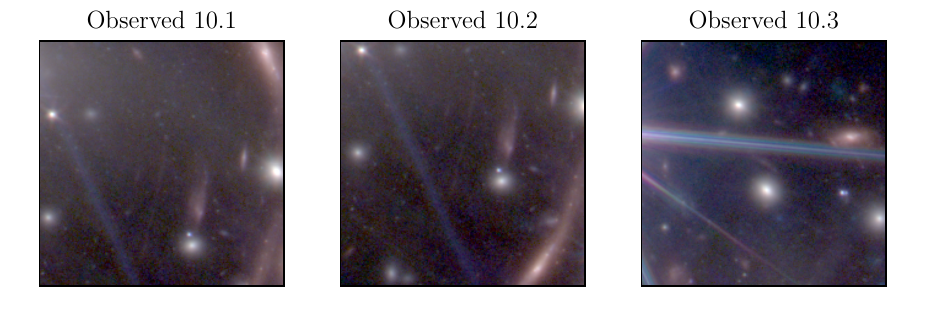}{0.45\textwidth}{}}
    \vspace{-3em}
    \gridline{\fig{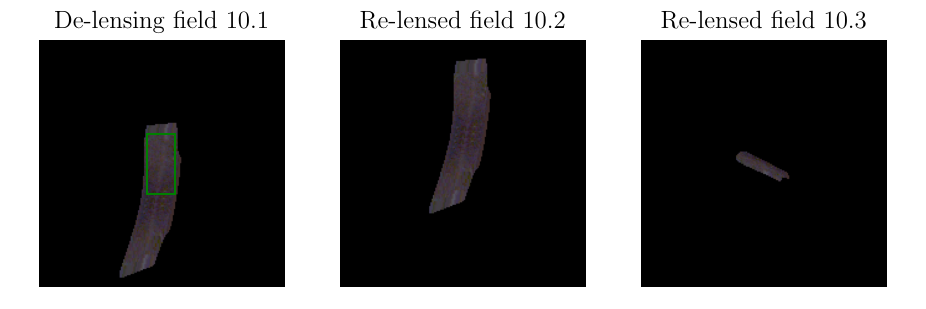}{0.45\textwidth}{}}
    \vspace{-3em}
    \gridline{\fig{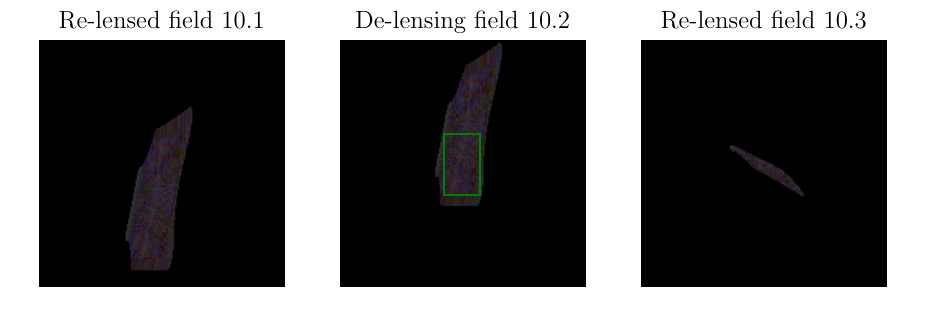}{0.45\textwidth}{}}
    \vspace{-3em}
    \gridline{\fig{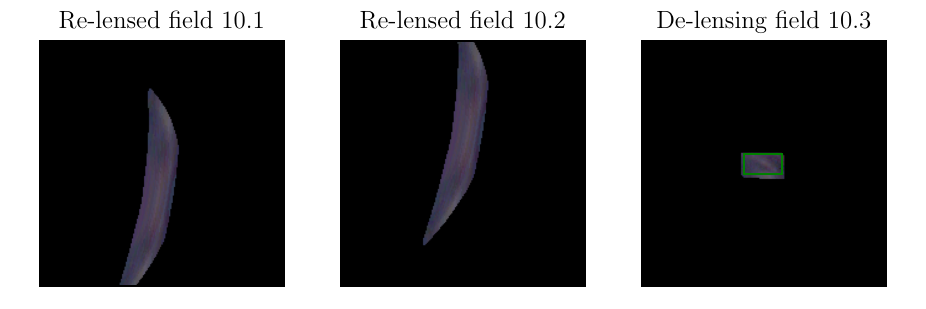}{0.45\textwidth}{}}
    \caption{Same as Fig.\,\ref{fig:relens_sys1}, but for system 10.}
    \label{fig:relens_sys10}
\end{figure}

\begin{figure}
    \centering
        \gridline{\hspace{-1em} \fig{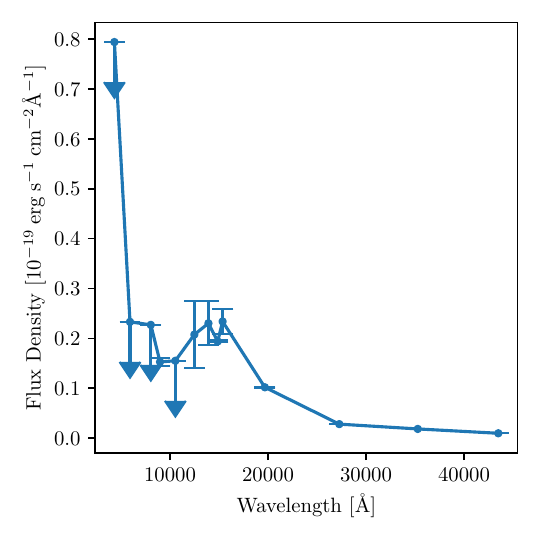}{0.14\textwidth}{} \hspace{-3.75em}
          \fig{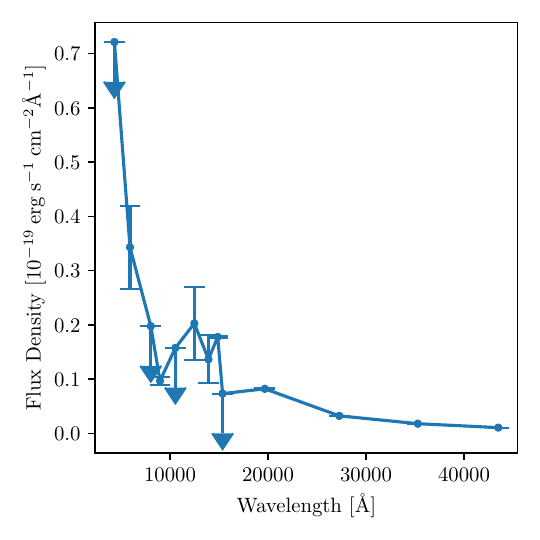}{0.14\textwidth}{} \hspace{-3.75em}
          \fig{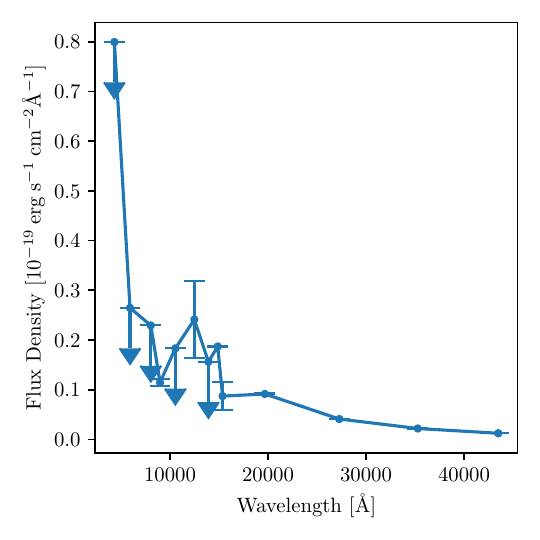}{0.14\textwidth}{}
          }
    \vspace{-3em}
    \gridline{\fig{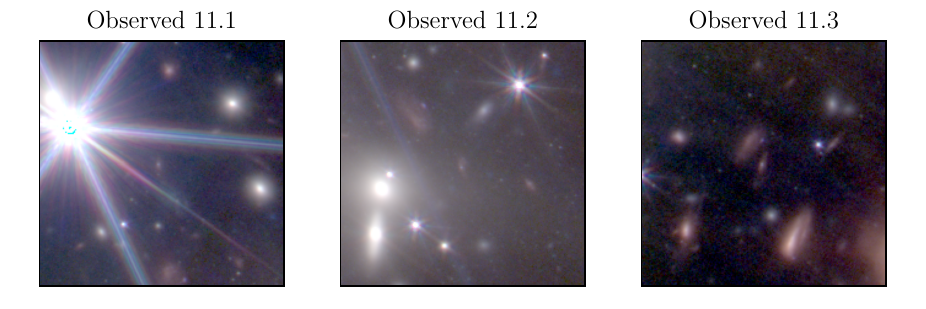}{0.45\textwidth}{}}
    \vspace{-3em}
    \gridline{\fig{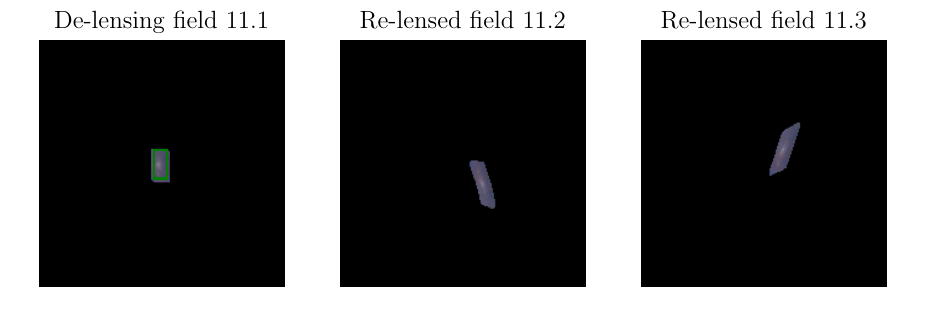}{0.45\textwidth}{}}
    \vspace{-3em}
    \gridline{\fig{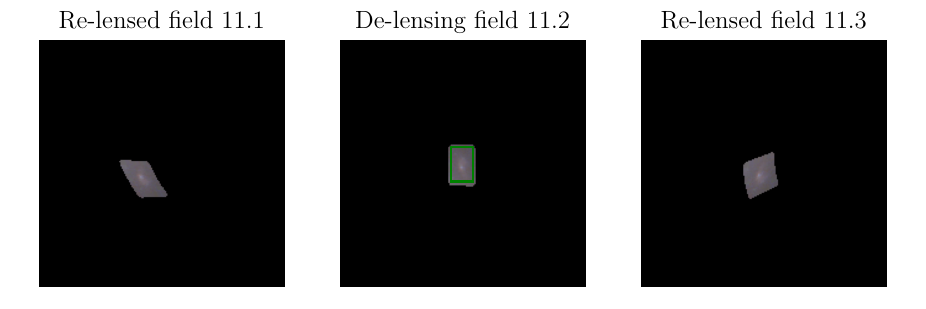}{0.45\textwidth}{}}
    \vspace{-3em}
    \gridline{\fig{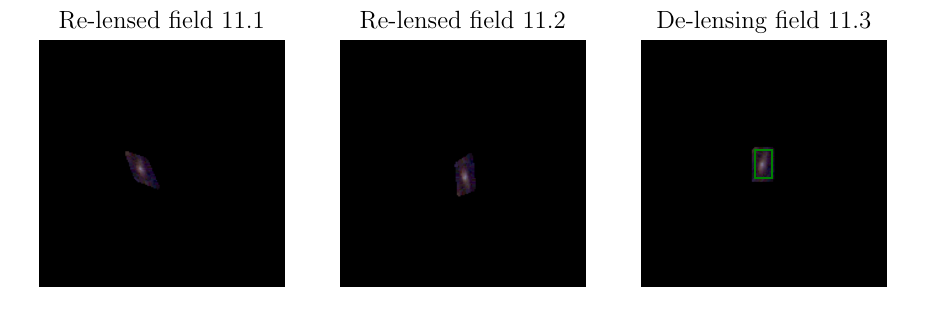}{0.45\textwidth}{}}
    \caption{Same as Fig.\,\ref{fig:relens_sys1}, but for system 11.}
    \label{fig:relens_sys11}
\end{figure}

\begin{figure}
    \centering
        \gridline{\hspace{-0.75em} \fig{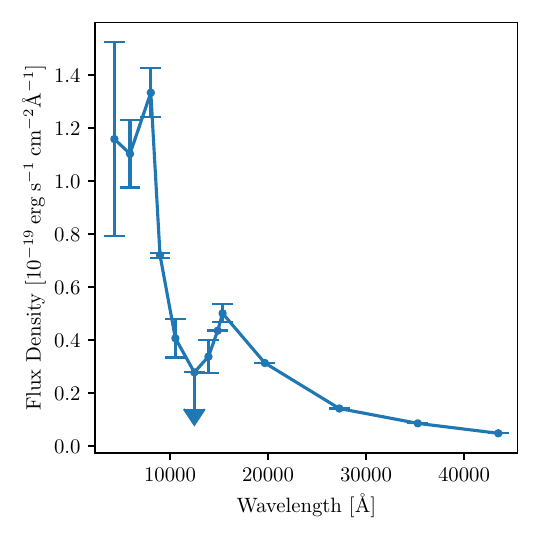}{0.105\textwidth}{} \hspace{-3.25em}
          \fig{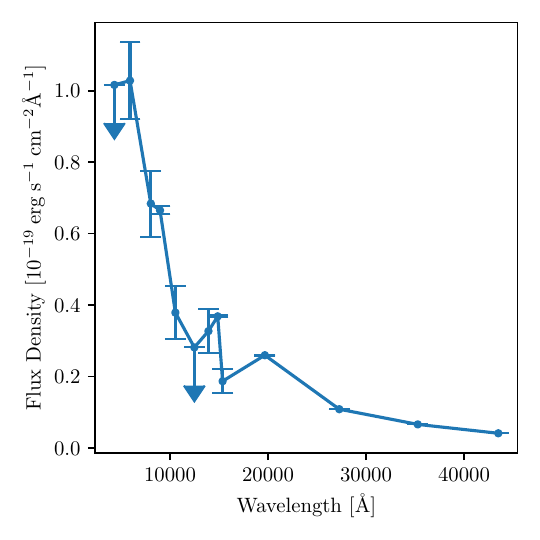}{0.105\textwidth}{} \hspace{-3.25em}
          \fig{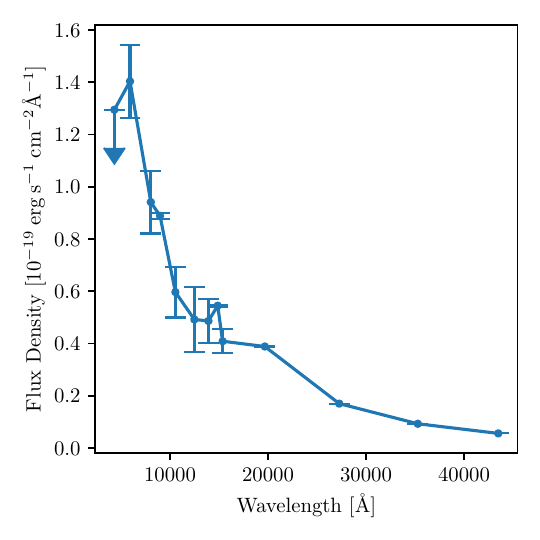}{0.105\textwidth}{} \hspace{-3.25em}
          \fig{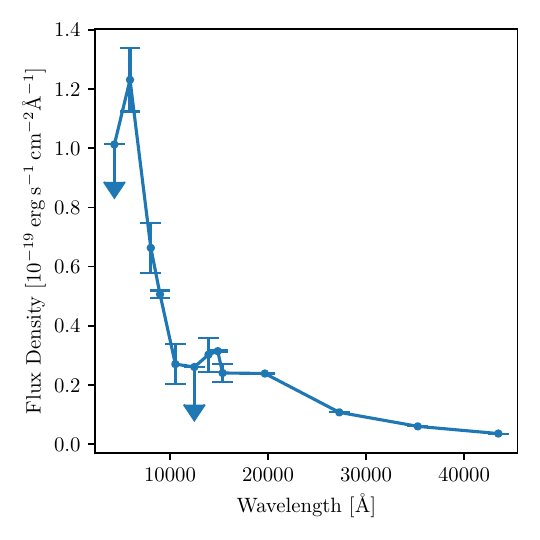}{0.105\textwidth}{}
          }
    \vspace{-3em}
    \gridline{\fig{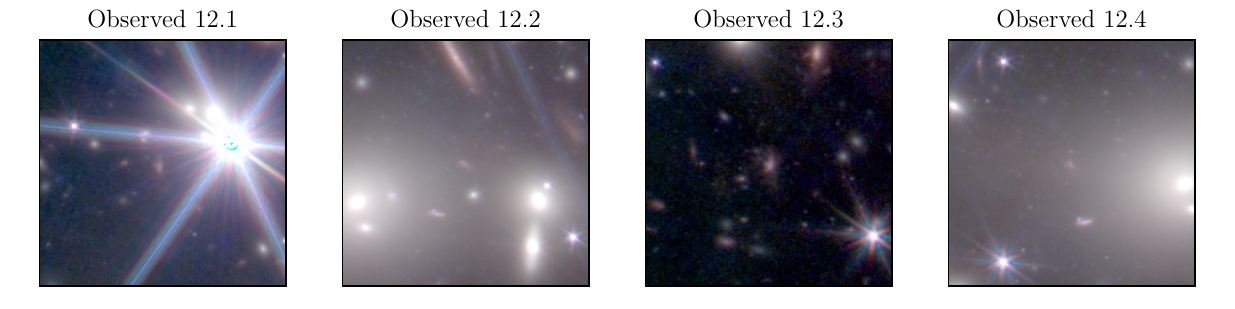}{0.45\textwidth}{}}
    \vspace{-3em}
    \gridline{\fig{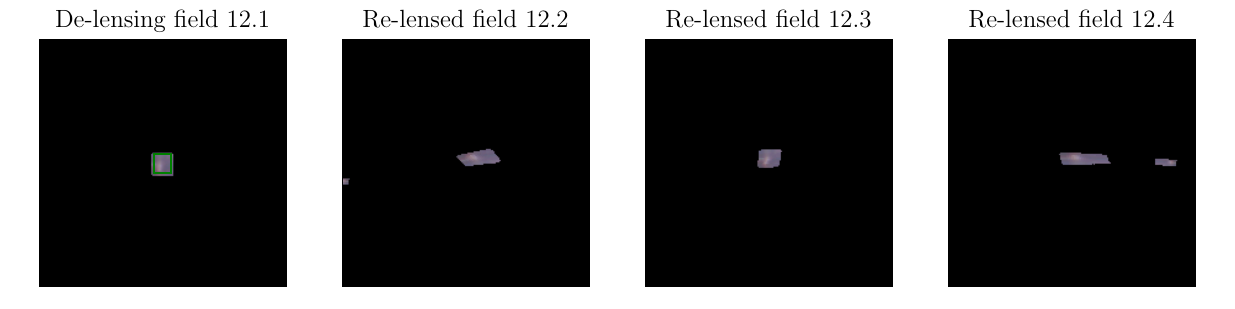}{0.45\textwidth}{}}
    \vspace{-3em}
    \gridline{\fig{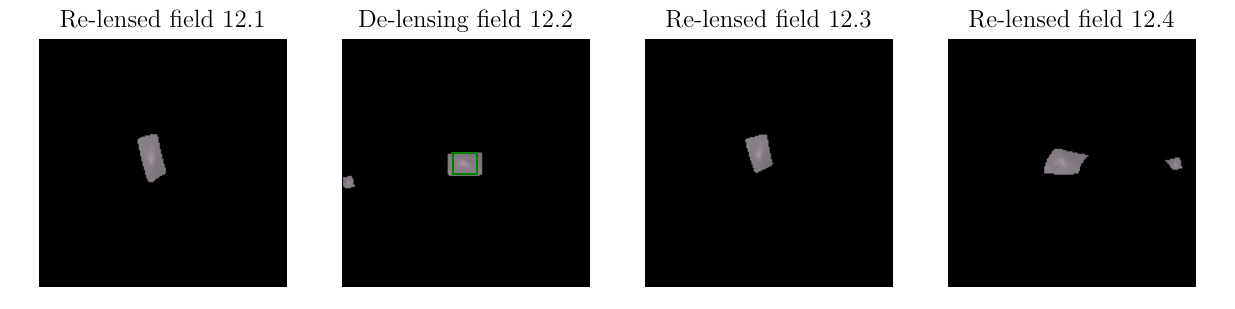}{0.45\textwidth}{}}
    \vspace{-3em}
    \gridline{\fig{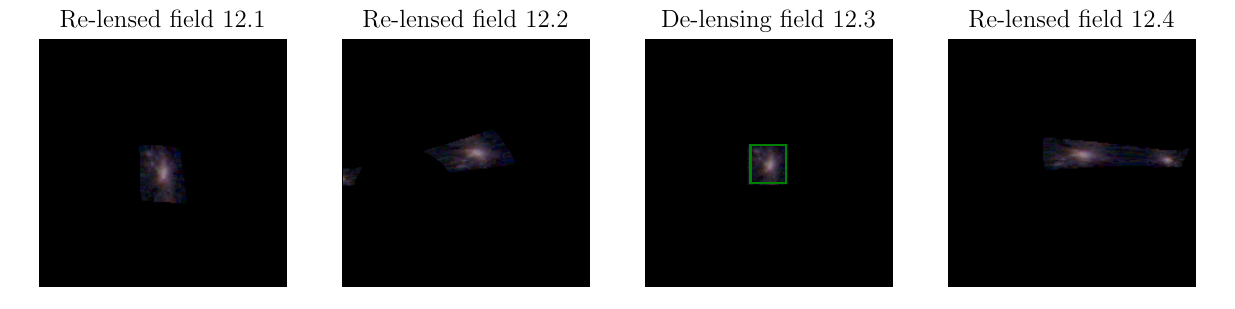}{0.45\textwidth}{}}
    \vspace{-3em}
    \gridline{\fig{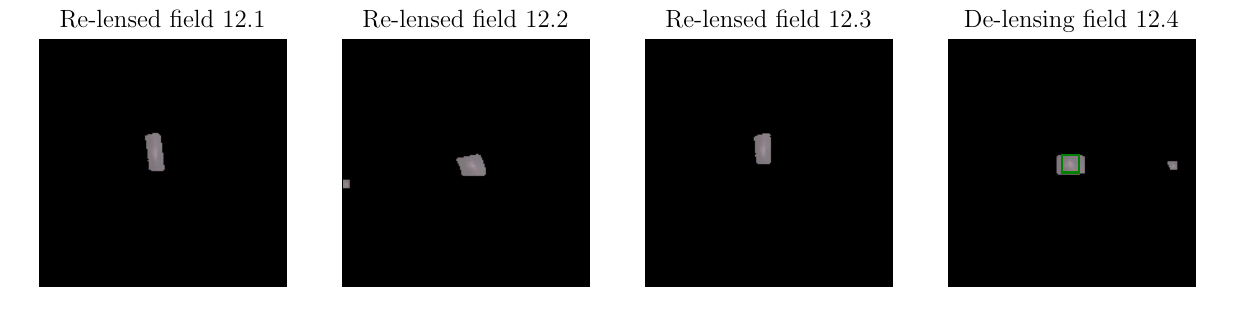}{0.45\textwidth}{}}
    \caption{Same as Fig.\,\ref{fig:relens_sys1}, but for system 12.}
    \label{fig:relens_sys12}
\end{figure}

\begin{figure}
    \centering
        \gridline{\hspace{-1em} \fig{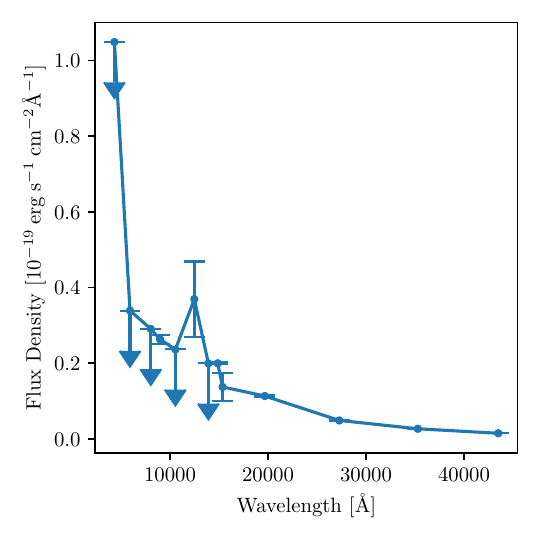}{0.14\textwidth}{} \hspace{-3.75em}
          \fig{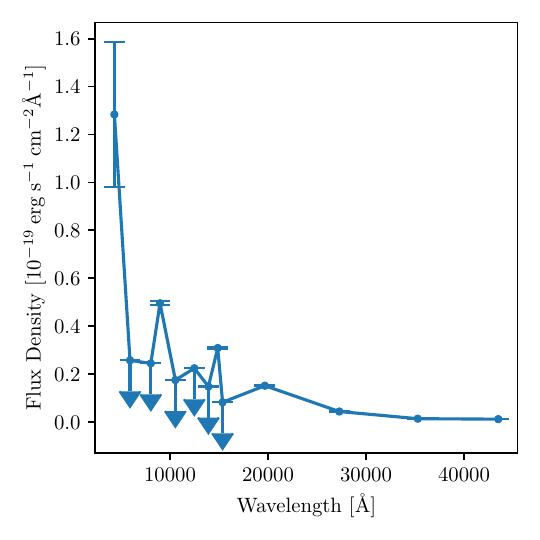}{0.14\textwidth}{} \hspace{-3.75em}
          \fig{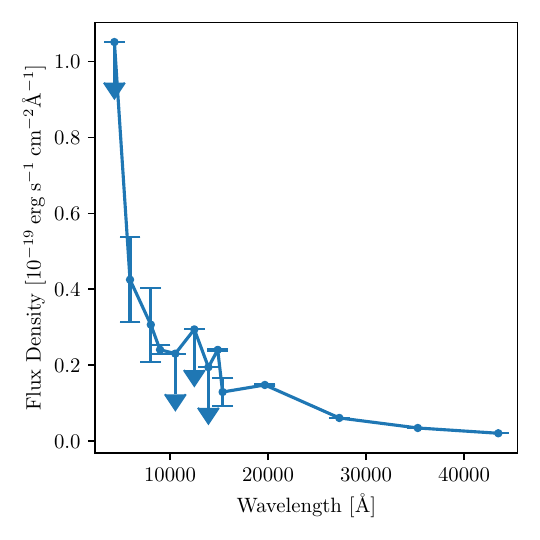}{0.14\textwidth}{}
          }
    \vspace{-3em}
    \gridline{\fig{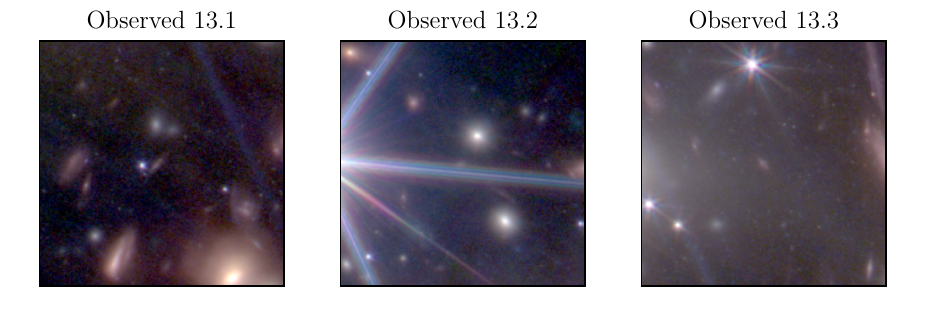}{0.45\textwidth}{}}
    \vspace{-3em}
    \gridline{\fig{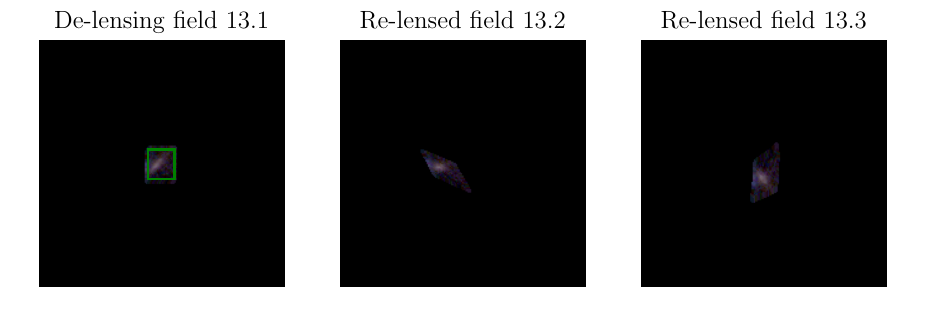}{0.45\textwidth}{}}
    \vspace{-3em}
    \gridline{\fig{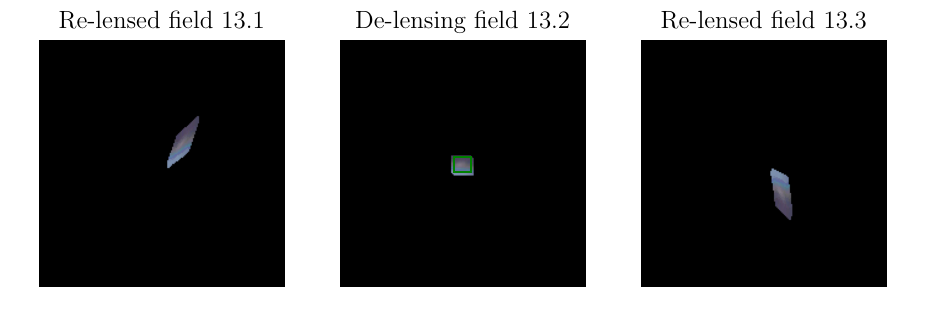}{0.45\textwidth}{}}
    \vspace{-3em}
    \gridline{\fig{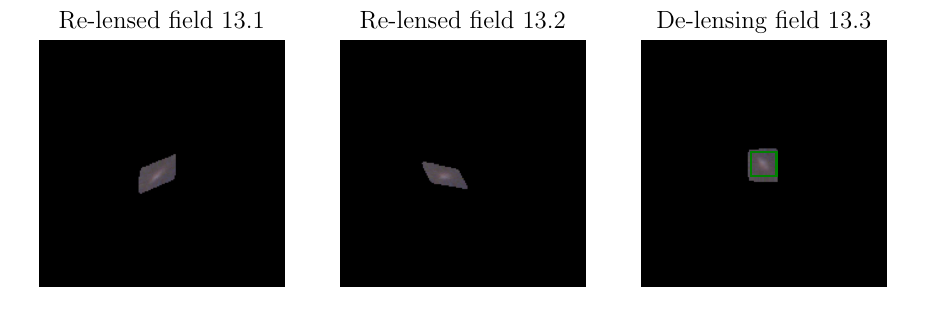}{0.45\textwidth}{}}
    \caption{Same as Fig.\,\ref{fig:relens_sys1}, but for system 13.}
    \label{fig:relens_sys13}
\end{figure}

\begin{figure}
    \centering
        \gridline{\hspace{-1em} \fig{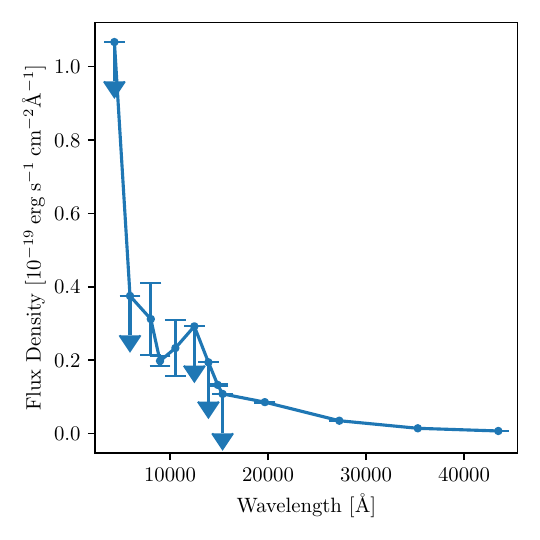}{0.14\textwidth}{} \hspace{-3.75em}
          \fig{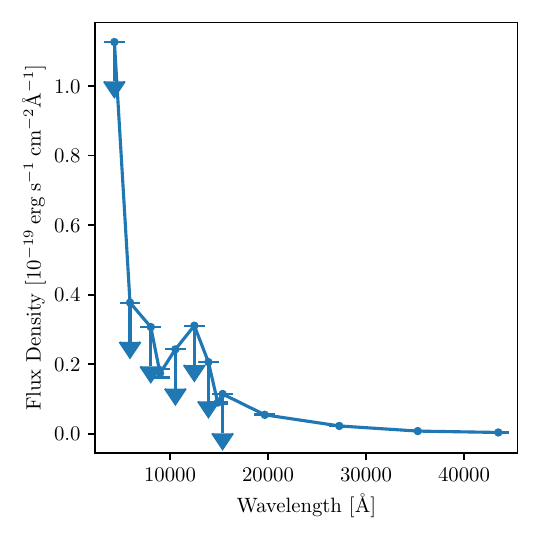}{0.14\textwidth}{} \hspace{-3.75em}
          \fig{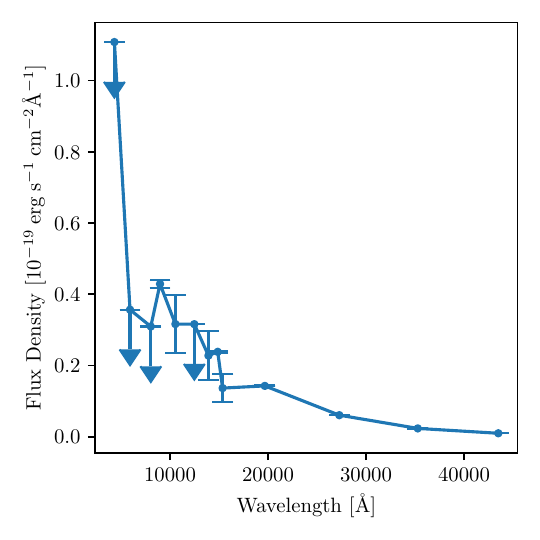}{0.14\textwidth}{}
          }
    \vspace{-3em}
    \gridline{\fig{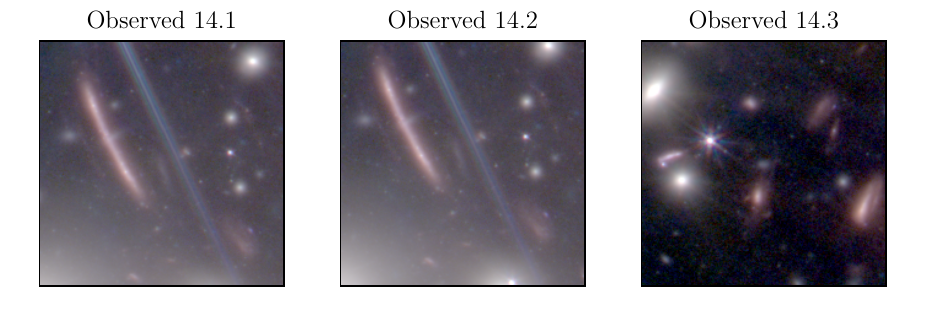}{0.45\textwidth}{}}
    \vspace{-3em}
    \gridline{\fig{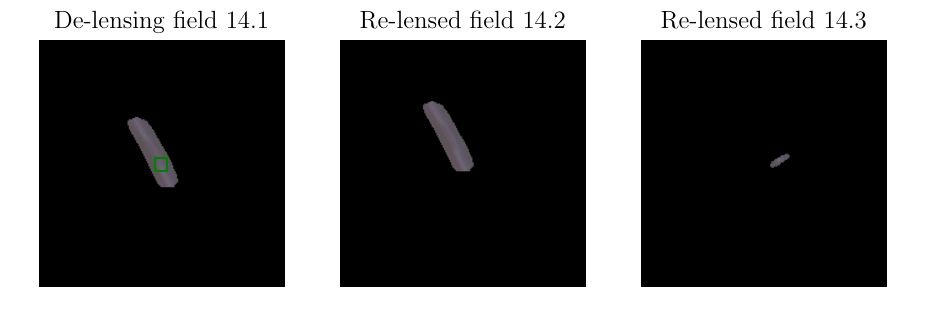}{0.45\textwidth}{}}
    \vspace{-3em}
    \gridline{\fig{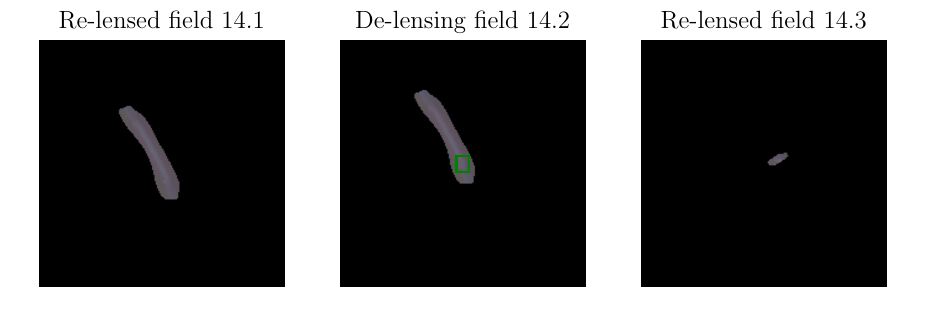}{0.45\textwidth}{}}
    \vspace{-3em}
    \gridline{\fig{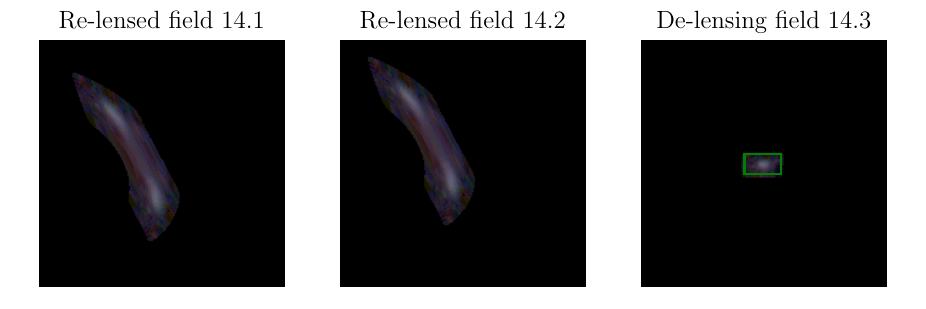}{0.45\textwidth}{}}
    \caption{Same as Fig.\,\ref{fig:relens_sys1}, but for system 14.}
    \label{fig:relens_sys14}
\end{figure}

\begin{figure}
    \centering
        \gridline{\hspace{-1em} \fig{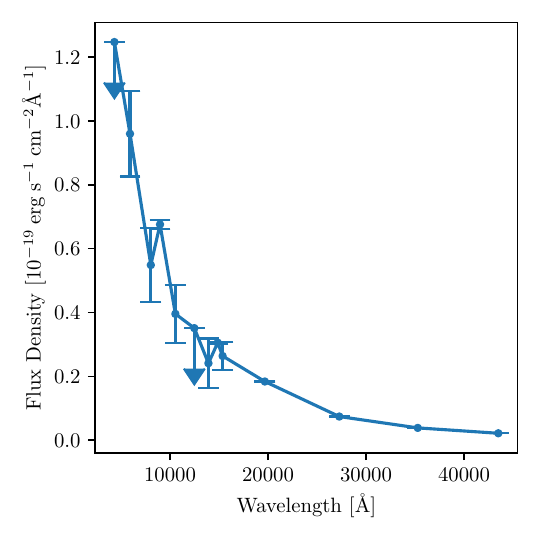}{0.14\textwidth}{} \hspace{-3.75em}
          \fig{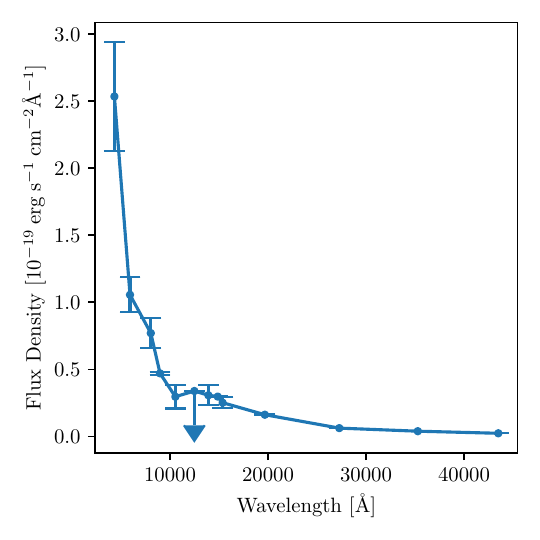}{0.14\textwidth}{} \hspace{-3.75em}
          \fig{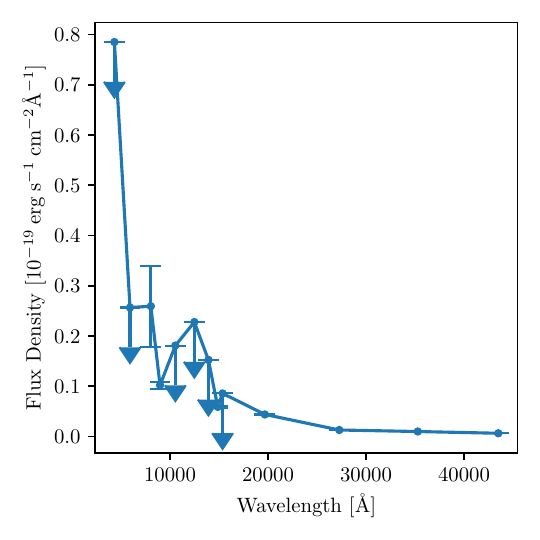}{0.14\textwidth}{}
          }
    \vspace{-3em}
    \gridline{\fig{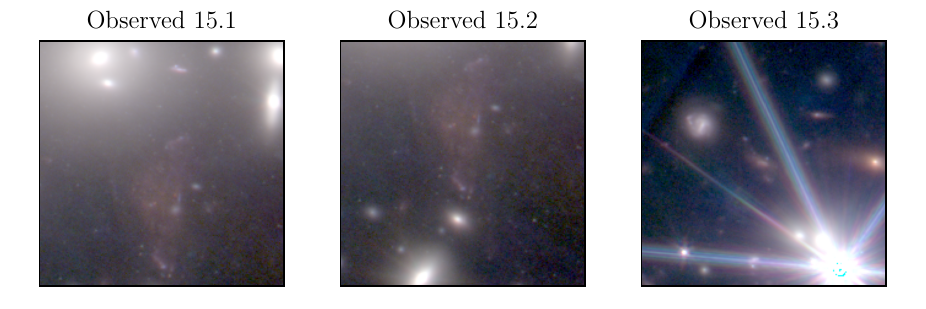}{0.45\textwidth}{}}
    \vspace{-3em}
    \gridline{\fig{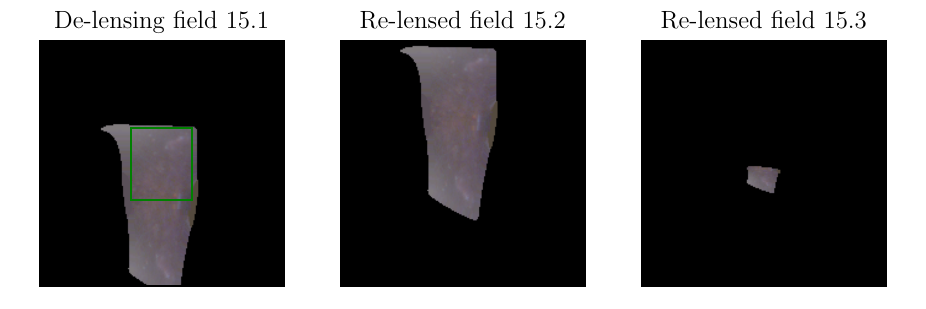}{0.45\textwidth}{}}
    \vspace{-3em}
    \gridline{\fig{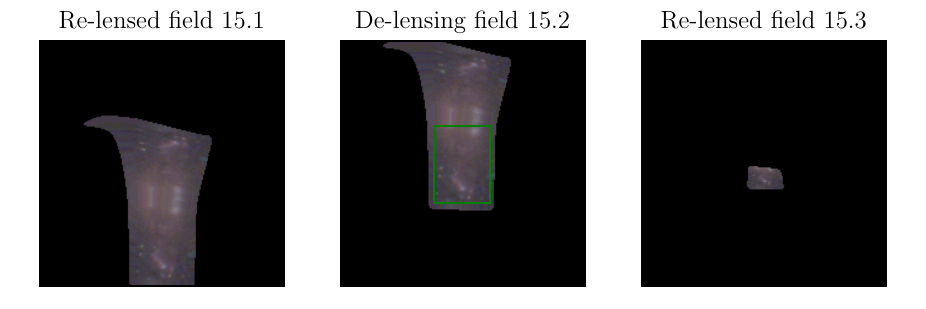}{0.45\textwidth}{}}
    \vspace{-3em}
    \gridline{\fig{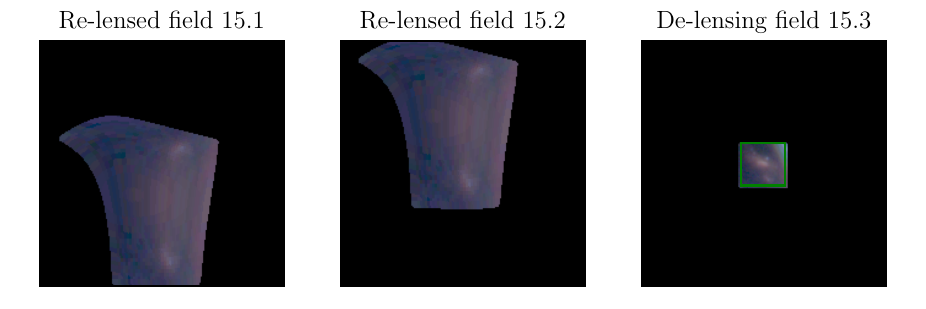}{0.45\textwidth}{}}
    \caption{Same as Fig.\,\ref{fig:relens_sys1}, but for system 15.}
    \label{fig:relens_sys15}
\end{figure}

\begin{figure}
    \centering
        \gridline{\hspace{-1em} \fig{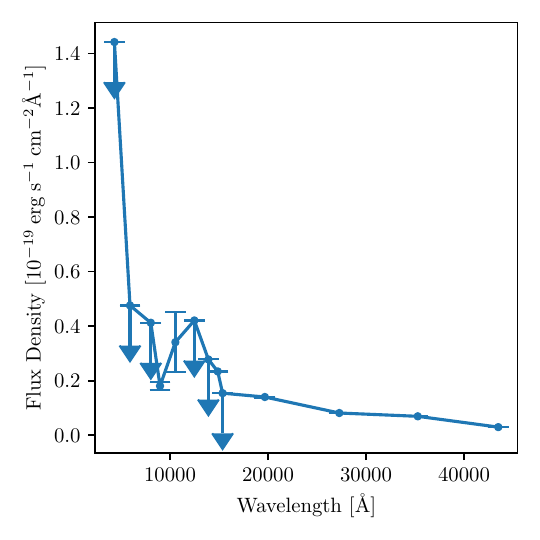}{0.14\textwidth}{} \hspace{-3.75em}
          \fig{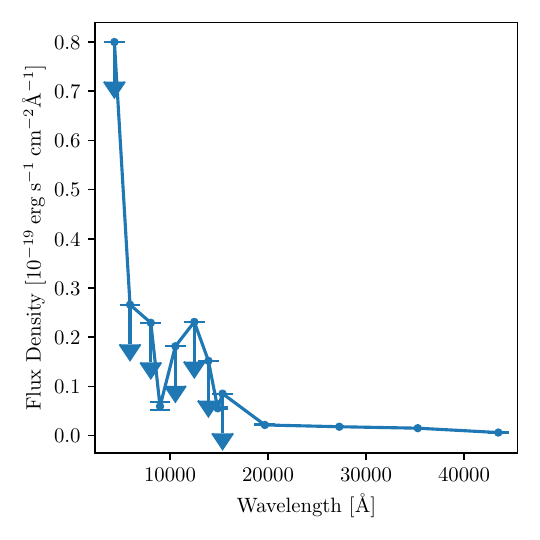}{0.14\textwidth}{} \hspace{-3.75em}
          \fig{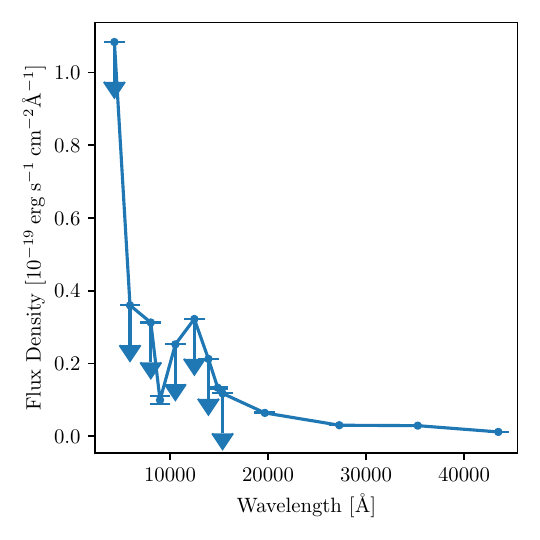}{0.14\textwidth}{}
          }
    \vspace{-3em}
    \gridline{\fig{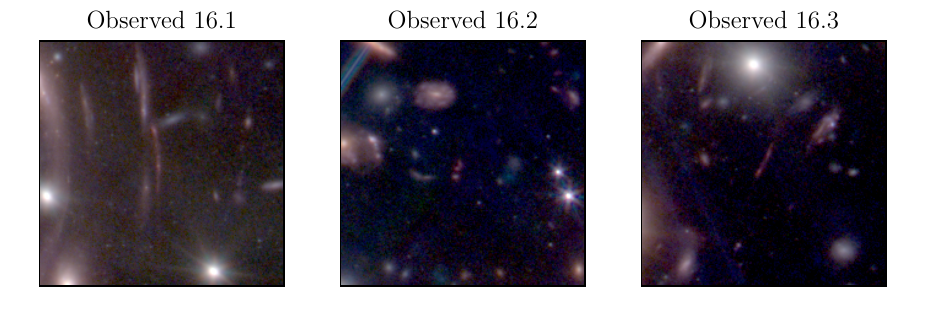}{0.45\textwidth}{}}
    \vspace{-3em}
    \gridline{\fig{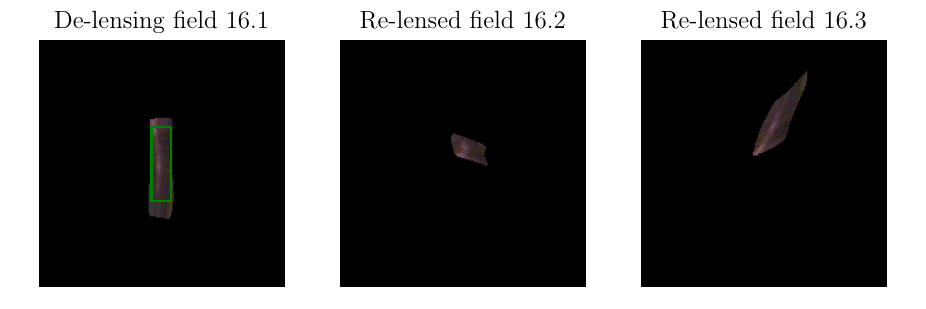}{0.45\textwidth}{}}
    \vspace{-3em}
    \gridline{\fig{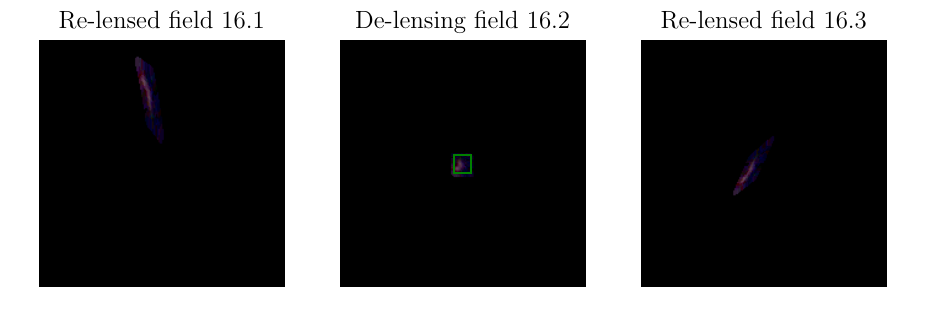}{0.45\textwidth}{}}
    \vspace{-3em}
    \gridline{\fig{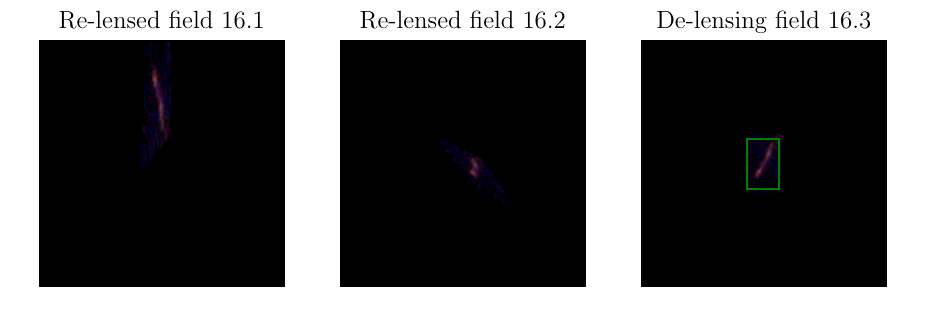}{0.45\textwidth}{}}
    \caption{Same as Fig.\,\ref{fig:relens_sys1}, but for system 16.}
    \label{fig:relens_sys16}
\end{figure}

\begin{figure}
    \centering
        \gridline{\hspace{-1em} \fig{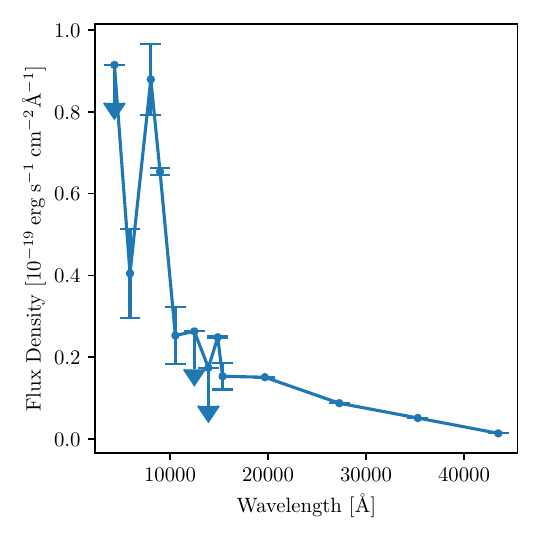}{0.14\textwidth}{} \hspace{-3.75em}
          \fig{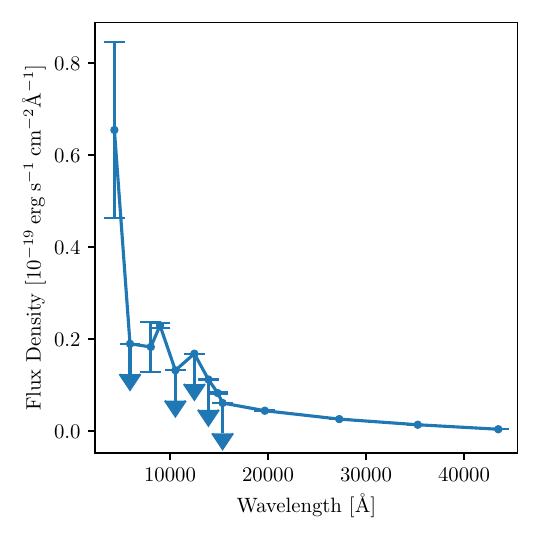}{0.14\textwidth}{} \hspace{-3.75em}
          \fig{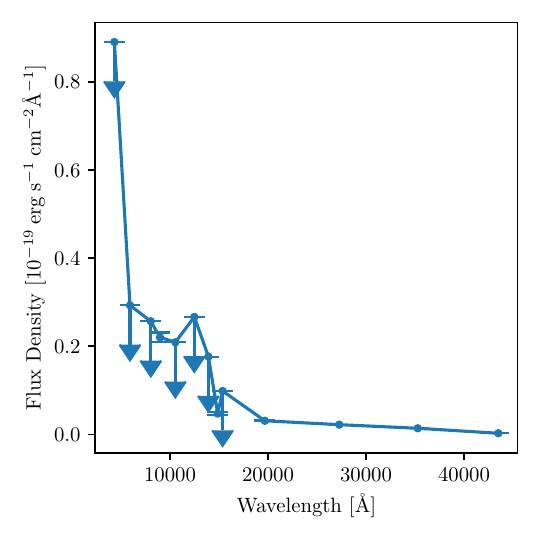}{0.14\textwidth}{}
          }
    \vspace{-3em}
    \gridline{\fig{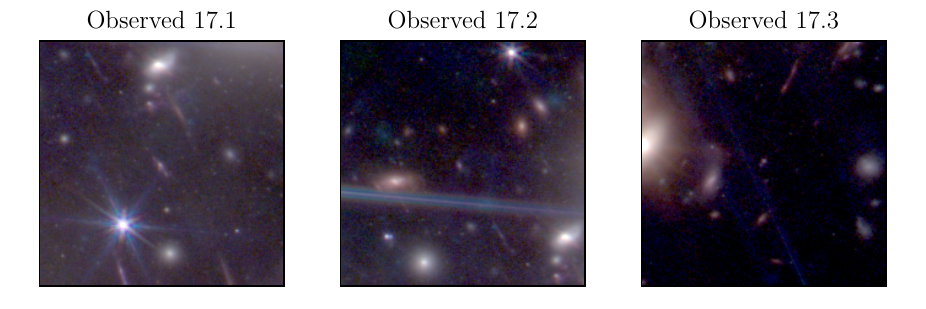}{0.45\textwidth}{}}
    \vspace{-3em}
    \gridline{\fig{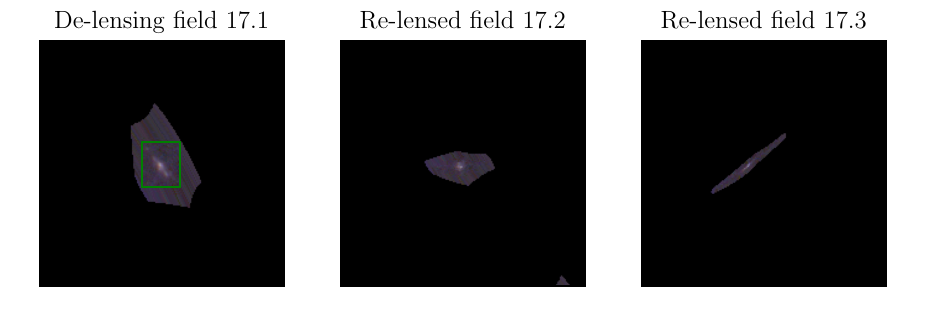}{0.45\textwidth}{}}
    \vspace{-3em}
    \gridline{\fig{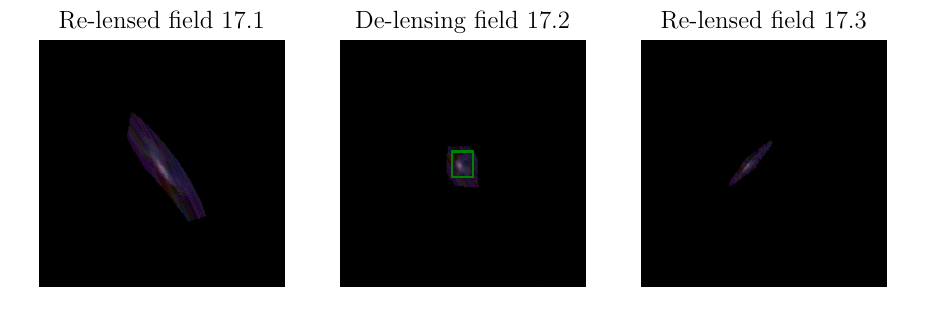}{0.45\textwidth}{}}
    \vspace{-3em}
    \gridline{\fig{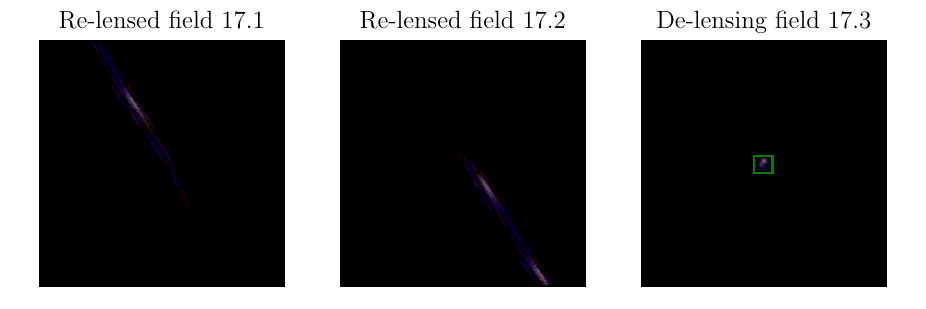}{0.45\textwidth}{}}
    \caption{Same as Fig.\,\ref{fig:relens_sys1}, but for system 17.}
    \label{fig:relens_sys17}
\end{figure}

As a test of the predictability of our lens model, we de-lens each image counterpart of a particular system used to constrain our lens model to the source plane, and then re-lens it back to the image plane at the locations of its other image counterparts.  The results for system 1 are shown in Figure \ref{fig:relens_sys1}, and for the remaining systems in Figures\,\ref{fig:relens_sys2}--\ref{fig:relens_sys17}.  A good agreement between the predicted and observed lensed morphologies, as is found in all cases albeit poorest in system 14, serves as a test of the predictability and hence reliability of our lens model.  In the case of system 14, which is closely adjacent to system 5, the lens model is better optimised for system 5, such that the critical curve at the redshift of system 14 does not cut through the middle of the arc comprising this system.

\end{document}